\documentclass[aip,jcp,showpacs,floatfix,nofootinbib,preprintnumbers,superscriptaddress,amsmath,amssymb]{revtex4-2}

\usepackage{graphicx}
\usepackage{appendix}
\usepackage{mciteplus}
\usepackage{multirow}
\usepackage{amsfonts,calrsfs}
\usepackage{amsmath}
\usepackage{verbatim}
\usepackage{enumerate}
\usepackage[T1]{fontenc}
\usepackage{bm}
\usepackage{siunitx}
\usepackage{orcidlink}

\usepackage[utf8]{inputenc}
\usepackage{hyperref}
\hypersetup{breaklinks=true,colorlinks=true,linkcolor=blue,citecolor=blue,filecolor=magenta,urlcolor=blue}
\usepackage{siunitx}
\usepackage{orcidlink}

\newcommand{\ud}{\mathrm{d}}

\draft 

\begin{document}


\title{Barlow Packings All Have Maximum Packing Fraction: Why it is so Hard to Find Barlow Structures Other than fcc or hcp?}

\author{Andres Robles-Navarro\,\orcidlink{0000-0001-9586-1003}}\email{andres.robles.n@gmail.com}
\affiliation{Centre for Theoretical Chemistry and Physics, The New Zealand Institute for Advanced Study (NZIAS), Massey University Albany, Private Bag 102904, Auckland 0745, New Zealand}
\author{Shaun Cooper\,\orcidlink{0000-0001-5103-0400}}\email{s.cooper@massey.ac.nz}
\affiliation{School of Natural and Computational Sciences, Massey University Albany, Private Bag 102904, Auckland 0745, New Zealand.}
\author{Alex Travesset\,\orcidlink{0000-0001-7030-9570}}\email{trvsst@ameslab.gov}
\affiliation{Ames Laboratory and Department of Physics and Astronomy, Iowa State University, Ames, IA 50011, USA}
\author{Peter Schwerdtfeger\,\orcidlink{0000-0003-4845-686X}}\email{peter.schwerdtfeger@gmail.com}
\affiliation{%
Centre for Theoretical Chemistry and Physics, The New Zealand Institute for Advanced Study (NZIAS), Massey University Albany, Private Bag 102904, Auckland 0745, New Zealand.}


\date{\today}

\begin{abstract}
Abstract: Lattice theory for periodic Barlow packings, $(X_1X_2\ldots X_N)_\infty$ with $X_i\in{A,B,C}$ and $X_i\neq X_{i+1}$, consisting of stacked two-dimensional hexagonal close-packed layers, is applied to generalized $(n,m)$-Lennard-Jones potentials. 
The results show that the cohesive energies of all Barlow packings lie between those of the hexagonal close-packed (hcp) $(AB)$ and face-centered cubic (fcc) $(ABC)$ structures and exhibit an almost perfectly linear dependence on the fcc stacking fraction. A similarly linear dependence is found for other properties, including the equilibrium lattice constant, bulk modulus, Einstein frequency, and vibrational entropy. For the commonly used $(12,6)$-Lennard-Jones potential, the fcc structure has the highest vibrational entropy and the hcp structure the lowest, with all other Barlow packings lying in between, in agreement with previous sticky hard-sphere Monte Carlo simulations.
This is supported by phonon dispersion calculations. Since all major energetic and vibrational properties exhibit this approximately linear behavior, the free energy of a Barlow packing is likewise expected to depend nearly linearly on the fcc stacking fraction, with hcp and fcc representing the two limiting cases. This simple model suggests that the rarity of Barlow packings in nature, beyond the common hcp and fcc structures, originates mainly from this near-linear dependence, which provides little thermodynamic driving force for intermediate stackings. Exceptions to this general behavior are expected only when the hcp and fcc structures are nearly degenerate in energy or if many-body forces not considered here lead to a slight preference of some Barlow packings other than fcc or hcp. Density functional calculations including dispersion forces for argon, performed for Barlow packings with periodicities $N=2$, 3, 4, and 12, support the predicted near linear dependence of the cohesive energy.
\end{abstract}


\maketitle 



\section{Introduction}
\label{sec:intro}
 
\begin{figure}[htb]
\centering
\includegraphics[width=.45\columnwidth]{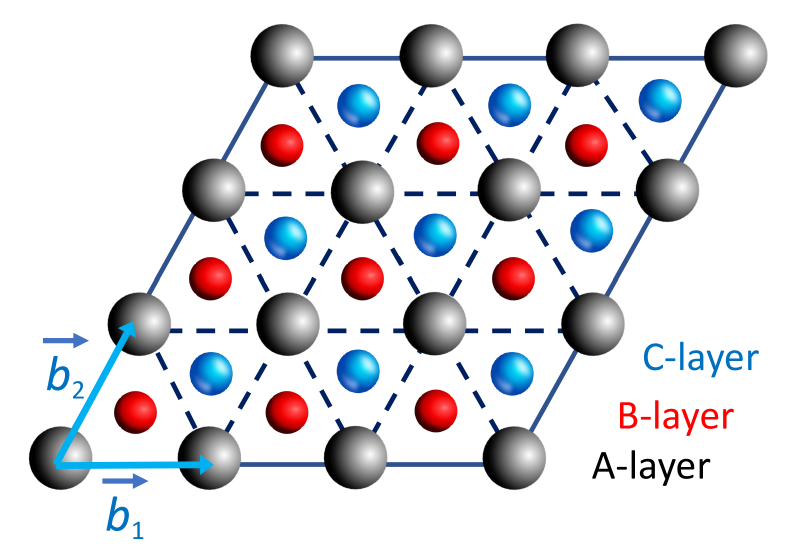} \quad
\includegraphics[width=.22\columnwidth]{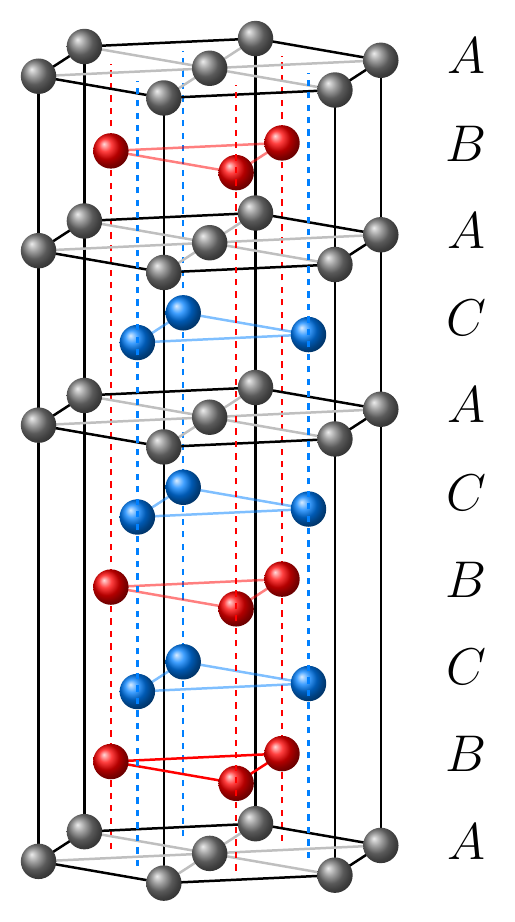}
\caption{Left: The three possible layers $A$, $B$ and $C$ in Barlow packings. Right: The 9R structure with a sequence of $(ABCBCACAB)_\infty$ observed for example for samarium at room temperature\cite{ellinger1953,Daane1954} and metallic lithium at 75K\cite{Smith1987}. The last layer at the top of the R9 structure belongs to the next unit cell with the same layer sequence.}
\label{fig:Barlow}
\end{figure}

A Barlow packing is described by an infinite stacking sequence $(X_1X_2X_3\ldots)$, where $X_i\in\{A,B,C\}$ denotes one of the three possible densely packed hexagonal layers, distinguished by their lateral positions. Aa adjacent layers cannot occupy the same position we have $X_i\neq X_{i+1}$ for all $i\in\mathbb{N}$, thereby excluding the pairs $AA$, $BB$, and $CC$ (see Figure~\ref{fig:Barlow}). There are infinitely many distinct Barlow packings, all of which attain the same maximum packing fraction, $\rho=\pi/3\sqrt{2}$=0.74048...,
which is the highest possible density for packings of identical non-overlapping spheres\cite{Barlow1883,conway1999recent,Osang2021} as proven by Hales\cite{Hales2011,hales-2005}. General aspects of the lattice theory of Barlow packings are discussed in Refs.~\onlinecite{Thompson2001,Torquato2010,Osang2021,Cooper2025,Jaskolski2025}. For a recent review on sphere packing problems see Kumar and Deep\cite{Kumar2026}.

Apart from the two most common close-packed structures, namely hexagonal close packing (hcp), $(AB)_\infty$, and face-centered cubic (fcc), $(ABC)_\infty$, other Barlow packings are exceedingly rare\cite{ellinger1953,Daane1954,Smith1987,wells1984,Overhauser1984,Ackland2017,Stillinger2019}. They are often regarded as close-packed structures containing stacking faults\cite{Warnes1990}. For example, the low-temperature phase of lithium was proposed to adopt the close-packed rhombohedral (9R) structure\cite{Overhauser1984,Smith1987} with the stacking sequence $(ABABCBCAC)_\infty$ of hexagonal layers (using the natural notation introduced in Ref.\onlinecite{Cooper2025}; see Figure~\ref{fig:Barlow}). More recently, however, this assignment was challenged by Ackland, who argued that the phase is instead fcc\cite{Ackland2017}, a conclusion supported by recent density-functional calculations of the Burgers phase transition\cite{RoblesNavarro2025}.

The prototypical example of a non-conventional Barlow packing is the paramagnetic $\alpha$-Sm 9R phase, which is stable at room temperature\cite{ellinger1953,Daane1954} and transforms to hcp at $731^\circ$C.\cite{SHI1985} In samarium, the double hexagonal close-packed (dhcp) structure with stacking sequence $(ABAC)_\infty$ has also been observed,\cite{SHI1985} highlighting the transient character of the various close-packed polytypes in this element. Only a few other materials are known to adopt unusual Barlow packings\cite{wells1984}. The 9R structure also occurs as a metastable transitional phase in several metals and high-entropy alloys\cite{Gooding1988,Wang2022}.
Despite the existence of infinitely many Barlow packings, it remains unclear why the fcc and hcp structures are overwhelmingly the most prevalent. Preliminary work from our group suggests that these two structures represent the extremal cases in the energetic stability sequence of all Barlow packings.\cite{Cooper2024}

The theory of Barlow packings has been investigated extensively over the past decades\cite{conway1999recent,Thompson2001,Ackland2017a,Stillinger2019,Jaskolski2025,Cooper2025}. In our recent work, we presented a rigorous mathematical framework for periodically repeating Barlow sequences of the form $(X_1X_2\ldots X_N)_\infty$ and introduced an efficient algorithm together with a Fortran program\cite{Cooper2025} for generating all such stacking sequences up to a computationally feasible periodicity $N$. In the present paper, we extend this theory to account for symmetry-breaking effects and investigate in detail the energetics and thermodynamics of these structures for a general $(n,m)$-Lennard-Jones (LJ) potential\cite{Jones-1924b,Jones-1925,Wales2024} (also known as the Mie-Gr\"uneisen potential\cite{Mie-1903,Gruneisen1912}),
\begin{equation} \label{eq:VLJ}
V_{\textrm{LJ}}(r)=\frac{\epsilon nm}{n-m} \;  \left[ \frac{1}{n}\left( \frac{r_e}{r} \right)^{n} - \frac{1}{m} \left( \frac{r_e}{r} \right)^{m} \right] \,,
\end{equation}
with $n>m$. Here, $\epsilon$ and $r_e$ denote the binding energy and equilibrium bond length of the diatomic molecule, respectively. Using dimensionless units, both quantities are set to unity without loss of generality. Since both free and internal energy differences between different Barlow packings are expected to be extremely small ($\sim10^{-5}$ eV)\cite{Schwerdtfeger-2016,Cooper2025}, we derive rapidly convergent lattice sums based on Bessel-function expansions to evaluate the long-range interactions accurately, thereby enabling the cohesive energies to be computed to machine precision\cite{borwein-2013,burrows-2020,Smits2021}. We also report density functional theory (DFT) calculations, including dispersion interactions, for selected Barlow packings of solid argon, since its cohesive energy is dominated by two-body dispersion interactions, while many-body effects may play an important role in accurately determining the extremely small energy differences between these densely packed structures.

\section{Theory}

\subsection{Barlow packings}

A detailed account on the theory of Barlow packings can be found in our previous work \cite{Cooper2025}. We only repeat the most important features required for this paper.

A Barlow packing is defined as a bi-infinite sequence $(\ldots,X_{-2},X_{-1},X_0,X_1,X_2,X_3,\ldots)$ with the condition that 
\begin{equation}
\label{barlowdefinition}
X_i\in\left\{A,B,C\right\} \quad\text{and}\quad X_i\neq X_{i+1} \quad \text{for all} \quad i\in\mathbb{Z}.
\end{equation}
The infinite Barlow sequence is said periodic with period~$N\geq 2$ if 
\begin{equation}
\label{periodicity}
X_{i+N}=X_i \quad \text{for all $i\in\mathbb{Z}$}.
\end{equation}
Without loss of generality, we can set $X_1=A$, $X_2=B$, and $X_N=C$\cite{Cooper2025}. Further, we assume that the period~$N$ is minimal, i.e. $N$ is the smallest positive integer for which \eqref{periodicity} holds. The most commonly observed Barlow packings are the hexagonal close packed structure of period 2, $(AB)$, and the face centred cubic packing lattice of period 3, $(ABC)$, and we understand that the letters in parentheses are periodically continued at the left and right to infinity, i.e. $(ABC)\equiv(ABC)_\infty$.

We define the stacking fraction of hexagonal layers $\alpha(\ell)$ by
\begin{equation}
\label{pfc}
\alpha(\ell) = \frac{1}{N} \sum_{i=1}^N \left[ 1- p(i,\ell) \right] .
\end{equation}
with $1\leq i,\ell \leq N$, which measures the fcc stacking fraction (also called fcc packing fraction in certain contexts) identical to the one used by Middlemas et al \cite{Stillinger2019}. Here, 
$p(i,\ell)$ are the stacking sequence numbers defined by
\begin{equation}
\label{eq:packfrac}
p(i,\ell) = \begin{cases}
1 & \text{if} \quad X_i=X_{i+\ell}, \\
0 & \text{if} \quad X_i \neq X_{i+\ell}
\end{cases}
\end{equation}  
Next we introduce the natural sequence by
\begin{equation}
\label{eq:cond}
\{ p_i^{'} \} = \{ p_i \}~ \text{with} \quad \text{max} \{  N_c=\sum_{i=0}^N 2^{i} p(i,2)  \quad \forall ~\text{left or right cycle}~ \{p(i,2)\}  \}
\end{equation}
Note that for hcp we have $\alpha_(2)=0$ and for fcc $\alpha(2)=1$, with all other Barlow packings in between. For example, for the 9R structure\cite{Overhauser1984} ($N=9$) we get the sequence $(ABABCBCAC)$ with stacking fraction $\alpha(2)=\frac{1}{3}$. An efficient algorithm to produce all non-isomorphic Barlow structures for the exponentially growing numbers of sequences can be found in Ref.\onlinecite{Cooper2025}.

\subsection{Lattice constants and sums}

The generator matrix for the underlying hexagonal Bravais lattice of a Barlow packing of period $N$ may be defined as\cite{Cooper2025}, 
\begin{equation}
B = \begin{pmatrix}
\vec{b}_1^\top \\
\vec{b}_2^\top \\
\vec{b}_3^\top 
\end{pmatrix}
= a
\begin{pmatrix}
1  & 0  &  0 \\
\frac{1}{2} &\frac{\sqrt{3}}{2} &  0 \\
0 & 0  &  \gamma_T
\end{pmatrix}
\end{equation}
where $\gamma_T=\frac{c}{a}$. Here the lattice basis vectors and parameters are defined by $|\vec{b}_1|=|\vec{b}_2|=a$ and $|\vec{b}_3|=c$, and $\angle(\vec{b}_1,\vec{b}_2)=60^o$ and $\angle(\vec{b}_1,\vec{b}_3)=\angle(\vec{b}_2,\vec{b}_3)=90^o$. The corresponding positive definite and symmetric Gram matrix then becomes, 
\begin{equation}
G = BB^\top = a^2
\begin{pmatrix}
1  & \frac{1}{2}  &  0 \\
\frac{1}{2} &1 &  0 \\
0 & 0  & \gamma_T^2 
\end{pmatrix}
\end{equation}
with det$(G)=\frac{3}{4}a^4c^2=\frac{3}{4}a^6\gamma_T^2>0$. From an arbitrarily chosen lattice point as the origin we obtain all other lattice points in the hexagonal lattice from, 
\begin{equation}
\vec{r}(\vec{i}) = B^\top \vec{i} = i_1\vec{b}_1+ i_2\vec{b}_2+ i_3\vec{b}_3 \\
\end{equation}
with $\vec{i}\in\mathbb{Z}^3$ with their distances from the origin given by the quadratic form 
\begin{equation}
|\vec{r}(\vec{i})| = \left(\vec{i}^\top G \vec{i} \right)^\frac{1}{2} = a \left(i_1^2 + i_1i_2 + i_2^2 +\gamma_T^2 i_3^2\right)^\frac{1}{2} \,.
\end{equation}
The volume for the unit cell is determined through the generator or Gram matrix
\begin{equation}
V(a,\gamma)={\textrm{det}}B=\sqrt{{\textrm{det}}G}= \frac{\sqrt{3}}{2} a^2 c =\frac{\sqrt{3}}{2} a^3 \gamma_T
\label{eq:vol} 
\end{equation}

The Barlow packing is a multi-lattice, and the atoms at different Wyckoff positions in the hexagonal layers $X_1,X_2,...,X_N$ within the unit cell are defined by the shift vectors,
\begin{align}
\label{eq:shift}
\vec{r}^{X_k}(\vec{i}) &=  B^\top \vec{i} +\vec{v}^{X_k} \;.
\end{align}
The set of all vectors $\{\vec{r}^{X_k}(\vec{i})\}$ produce all points (atomic positions) in 3D space for a Barlow multi-lattice. We use the following shift vectors ($1\le k\le N$),
\begin{align}
\left(\vec{v}^{X_k}\right)^\top
&= \begin{cases}
a\left(0,0,z_k\right) & \text{if} ~X_k=\text{A}, \\
a\left(\frac{1}{2},\frac{1}{2\sqrt{3}}, z_k\right) & \text{if} ~X_k=\text{B}, \\
a\left(\frac{1}{2},-\frac{1}{2\sqrt{3}}, z_k\right) & \text{if} ~X_k=\text{C}.
\end{cases}
\end{align}
with $\left(\vec{v}^{X_1}\right)^\top=\left(0,0,0\right)$. These shift vectors can be derived from the fact that an atom in the subsequent layer sits above the centroid of a triangle of neigboring lattice points in the $A$ layer as seen in Figure \ref{fig:Barlow}. The kissing number, defined by the number of minimum distances around a hard sphere, for a Barlow packing is $\kappa=12$, although for real crystals such as samarium there might be distortions reducing the kissing number\cite{ellinger1953,Daane1954}.

For the vertical distance $z_k$ between one of the hexagonal layers $X_k$ and the base layer $X_1$, consider an ideal close-packed Barlow structure for which we have $\gamma_T=N\gamma_0$ with $\gamma_0=\sqrt{\frac{2}{3}}$. For example, for $N=2$ we have the hcp structure with the well known ratio of $c/a=\gamma_T=2\gamma_0=\sqrt{\frac{8}{3}}$. For equally distributed layers in the hexagonal unit cell we then have $z_k=\frac{k-1}{N}\gamma_T=(k-1)\gamma_0$. This neglects small distortions between the individual hexagonal layers along the $c$-axis. For example, already for the hcp structure using a (12,6)-LJ potential one gets\cite{Schwerdtfeger2023} $\gamma_T=\sqrt{\frac{8}{3}}-\delta$ with $\delta=2.298569\times 10^{-4}$.

Let the distance (in units of $a$) between two neighboring hexagonal planes with the cardinal numbers for the individual layers $k_1$ and $k_2>k_1$ ($k_1,k_2=1,...,N$) be
\begin{align} \label{eq:distancehexplanes}
d_{k_1k_2}=z_{k_2}-z_{k_1}=\sum_{i=k_1}^{k_2-1} \gamma_i \quad \text{with}\quad \gamma_T \equiv d_{1N}=\sum_{i=1}^{N} \gamma_i
\end{align}
For the case that $\gamma_i=\gamma_0$ for all $i=1,...,N$ we simply have $d_{k_1k_2}=(k_2-k_1)\gamma_T/N$.

Using the definitions in \eqref{eq:shift}, we get for the distances between the different lattice points in layers $X_{k_1}$ and $X_{k_2}$ in terms of the generator matrix $B$ and shift vectors $\vec{v}^{X_{k}}$,
\begin{equation}
r(N,B,\vec{i},\vec{j},X_{k_1},X_{k_2})=\left|\vec{r}^{X_{k_2}}(\vec{i})-\vec{r}^{X_{k_1}}(\vec{j})\right| = \left| B^\top \left( \vec{i} - \vec{j} \right) +\left( \vec{v}^{X_{k_2}} -\vec{v}^{X_{k_1}} \right)  \right| 
\end{equation}
We need to consider 6 different cases for these distances, i.e. the cases $(X_{k_1}X_{k_2})=(AA), (BB), (CC), (AB), (AC)$ and $(BC)$\cite{},
\begin{align}
r(N,B,\vec{i},\vec{j},X_{k_1},X_{k_2}) = \hspace{11cm}\\
\nonumber
\begin{cases}
 \left| {B^\top \left( \vec{i} - \vec{j} \right) + a\left( 0,0, d_{k_1,k_2}\right)} \right| 
 & \text{if} \quad (X_{k_1}X_{k_2})=\text{(AA), (BB) or (CC)} \\\\
 \left| {B^\top \left( \vec{i} - \vec{j} \right) + a\left( \frac{1}{2},\frac{1}{2\sqrt{3}}, d_{k_1,k_2}\right)} \right| 
 & \text{if} \quad (X_{k_1}X_{k_2})=\text{(AB) } \\\\
 \left| {B^\top \left( \vec{i} - \vec{j} \right) + a\left( \frac{1}{2},-\frac{1}{2\sqrt{3}}, d_{k_1,k_2}\right)} \right| 
 & \text{if} \quad (X_{k_1}X_{k_2})=\text{(AC)  }\\\\
  \left| {B^\top \left( \vec{i} - \vec{j} \right) + a\left( 0,\frac{1}{\sqrt{3}}, d_{k_1,k_2} \right)} \right| 
 & \text{if} \quad (X_{k_1}X_{k_2})=\text{(BC)  }
\end{cases}
\label{eq:distdiff}
\end{align}
To derive the lattice sums we consider an inverse power potential of the form
\begin{equation}
V(r)=\alpha r^{-n}.
\end{equation}
with $n>3, n\in\mathbb{R}$ \cite{Schwerdtfeger-2006}. The cohesive energy per atom in the unit cell for such a potential becomes\cite{Cooper2025},
\begin{equation}
E_{\textrm{coh}} = \frac{\alpha}{2N}  \left( N{ \sum_{\vec{i}\in \mathbb{Z}^3}}^{'} Q(B,\vec{i})^{-s} 
+ 2 \sum_{k_1<k_2}^N \;\;  \sum_{\vec{i}\in \mathbb{Z}^3} q(N,B,\vec{i},X_{k_1},X_{k_2})^{-s} \right)
\end{equation}
with exponent $s=n/2$. The prime notation for the sum indicates that we avoid the case $\vec{i}=\vec{0}$~ in our summation. The quadratic form $Q$ originates from the unit cell and is given by,
\begin{equation}
 Q(B,\vec{i})= \vec{i}^\top BB^\top \vec{i} = a^2 \left(i_1^2 + i_1i_2 + i_2^2 +\gamma_T^2 i_3^2\right)
\end{equation}
while the quadratic function $q$ for the interaction between the hexagonal layers becomes
\begin{align}
q(N,B,\vec{i},X_{k_1},X_{k_2}) &= \left( \vec{i}^\top B  + \left(\vec{w}^{X_{k_1}X_{k_2}}\right)^\top \right) \left({B}^\top \vec{i} + \vec{w}^{X_{k_1}X_{k_2}}\right) \\
\nonumber
		&=  \left[  \vec{i}^\top G \vec{i} + 2 \vec{i}^\top B \vec{w}^{X_{k_1}X_{k_2}} + |\vec{w}^{X_{k_1}X_{k_2}}|^2\right] \\
\nonumber
		&=a^2 \left[ \left( i_1 +\frac{1}{2}i_2+\beta_1\right)^2 + \left( \frac{\sqrt{3}}{2} i_2 + \beta_2 \right)^2 + \left( \gamma_T i_3 + d_{k_1k_2} \right)^2 \right]
\end{align}
where $\vec{w}^{X_{k_1}X_{k_2}}=\vec{v}^{X_{k_2}} -\vec{v}^{X_{k_1}}$ and $\beta_1=0$ for the $(AA)$, $(BB)$, $(CC)$ or $(BC)$ sequence, $\beta_1=\frac{1}{2}$ for $(AB)$ or $(AC)$,  $\beta_2=0$ for $(AA)$, $(BB)$, $(CC)$, $\beta_2=\frac{1}{2\sqrt{3}}$ for $(AB)$, $\beta_2=-\frac{1}{2\sqrt{3}}$ for $(AC)$, and $\beta_2=\frac{1}{\sqrt{3}}$ for $(BC)$. As shown in our first paper\cite{Cooper2025}, the lattice sums for the cases (AA), (BB) and (CC) are formally identical, and so are the cases (AB), (AC) and (BC).

We now rewrite the cohesive energy in the following form,
\begin{equation}
\label{eq:Ecoh}
E_{\textrm{coh}} =  \frac{\alpha}{2a^{2s}} L^\textbf{\textit{X}}(N,s,\{\gamma_i\}) = \frac{\alpha}{2a^{2s}}  \left\{   L^A(s,\gamma_T) 
+ \frac{2}{N} \sum_{1\le k_1<k_2\le N} \;\; L_q^{X_{k_1}X_{k_2}}(s,\{\gamma_i\}) \right\}
\end{equation}
with the Barlow lattice sums $L$ for a specific period $N$ defined as,
\begin{align}
\label{eq:LA}
L^{A}(s,\gamma_T)& ={\sum_{\vec{i}\in \mathbb{Z}^3}}^{'} S^{A}(\vec{i},\gamma_T)^{-s} = {\sum_{\vec{i}\in \mathbb{Z}^3}}^{'} \left[ i_1^2 + i_1i_2 + i_2^2 + \gamma_T^2 i_3^2 \right]^{-s}\;,
\end{align}
\begin{align}
\label{eq:LAA}
\nonumber
L^{AA}&(s,\gamma_T,\{\gamma_{k_1},\dots,\gamma_{k_2}\}) = {\sum_{\vec{i}\in \mathbb{Z}^3}} S^{AA}(\vec{i},k_1,k_2,\{\gamma_i\})^{-s} \\
&= {\sum_{\vec{i}\in \mathbb{Z}^3}} \left[ i_1^2 + i_1i_2 + i_2^2 +  \left( \gamma_T i_3 +d_{k_1k_2} \right)^2 \right]^{-s}
\end{align}
and
\begin{align}
\label{eq:LAB}
\nonumber
L^{AB}&(s,\gamma_T,\{\gamma_{k_1},\dots,\gamma_{k_2}\}) = {\sum_{\vec{i}\in \mathbb{Z}^3}} S^{AB}(\vec{i},k_1,k_2,\{\gamma_i\})^{-s} \\
&= {\sum_{\vec{i}\in \mathbb{Z}^3}} \left[ 
\left( i_1 +\tfrac{1}{3}\right)^2 
+ \left( i_1 +\tfrac{1}{3}\right)\left( i_2 +\tfrac{1}{3}\right)
+ \left( i_2 +\tfrac{1}{3}\right)^2 + \left( \gamma_T i_3 +d_{k_1k_2} \right)^2 \right]^{-s}
\end{align}
where $\{\gamma_i \}$ stands for the set of all $N$ parameters $\gamma_i$ which need to be optimized. These lattice sums are special cases of the more general Epstein zeta function, for which alternative treatments exist\cite{crandall1998fast,Buchheit2026}.

Rapidly converging series in terms of Bessel functions have already been provided for the special case of equally distributed hexagonal layers in the unit cell\cite{Cooper2025}. For the more general case including symmetry breaking along the $c$-axis we refer to Appendix \ref{app:Bessel}. However, we expect small deviations from the ideal value of $\gamma_0=\sqrt{\frac{2}{3}}$ for each $\gamma_i$.\cite{Cooper2024} For some specific Barlow packings one has additional symmetry relations between the parameters $\gamma_i$, e.g. in the case of a Barlow sequence (ABCB) we have $\gamma_1=\gamma_4$ and $\gamma_2=\gamma_3$.

\subsection{Properties for a general Lennard-Jones potential} 
The cohesive energy for periodic Barlow packings of length $N$, with the lattice points interacting through a general $(n,m)$-LJ potential, expressed in terms of lattice sums, is given by
\begin{equation} \label{eq:cohLJhcp}
E^\textbf{\textit{X}}_{\textrm{coh}}(n,m,a,\{\gamma_i\})= \frac{\epsilon nm}{2(n-m)} \; \left[ \frac{L^\textbf{\textit{X}}(\tfrac{n}{2},\{\gamma_i\})}{n}\left( \frac{r_e}{a} \right)^{n} - \frac{L^\textbf{\textit{X}}(\tfrac{m}{2},\{\gamma_i\})}{m} \left( \frac{r_e}{a} \right)^{m} \right] 
\end{equation}
with $n>m>3$ and the lattice sums $L^\textbf{\textit{X}}(\tfrac{n}{2},\{\gamma_i\})$ as defined in \eqref{eq:LA}, \eqref{eq:LAA} and \eqref{eq:LAB}. Here we write the cohesive energy in terms of lattice sums in a special form by taking (arbitrarily) the lattice parameter $a$ out of the lattice sum so we can more easily see the link to a Lennard-Jones potential for two interacting atoms. This allows the lattice sums $L$ to become dimensionless which is preferred for mathematical functions such as the Epstein zeta function. Alternatively, we could have taken the inverse power term $a^{-n}$ into the lattice sum. Nevertheless, the former approach has the advantage that for the minimum cohesive energy at fixed values of $\gamma_i$  we get the lattice parameter $a$ directly from\cite{Cooper2025},
\begin{equation} \label{eq:RLJnmmin}
 a^*_{\textrm{min}}(n,m,\{\gamma_i\})=\left( \frac{L^\textbf{\textit{X}}(\tfrac{n}{2},\{\gamma_i\})}{L^\textbf{\textit{X}}(\tfrac{m}{2},\{\gamma_i\})}\right)^{\tfrac{1}{n-m}} \,.
\end{equation}
when minimizing the Barlow structures. The * notation implies that reduced (or dimensionless) units are used ($E^*=E/ \epsilon, a^*=a/r_e$). This results in the minimum cohesive energy \cite{burrows2021b},
 \begin{equation} \label{eq:ELJnmmin}
{E_{\textrm{coh}}^\textbf{\textit{X}}}^*(n,m,a^*_{\textrm{min}},\{\gamma_i\})=-\frac{1}{2}\left[ \frac{L^\textbf{\textit{X}}(\tfrac{m}{2},\{\gamma_i\})^n}{L^\textbf{\textit{X}}(\tfrac{n}{2},\{\gamma_i\})^m}\right]^{\tfrac{1}{n-m}}  \; .
\end{equation}
As the Barlow lattice sums $L^\textbf{\textit{X}}(s,\gamma_0)$ are monotonically decreasing functions with increasing $s$ in the interval $s\in[1.5,\infty)$ we have the condition that $a^*_{\textrm{min}}(n,m,\gamma_0)\le 1$ for all Barlow packings. In the sticky hard-sphere limit we have $a^*_{\textrm{min}}(\gamma)= 1$. Similarly we have ${E_{\textrm{coh}}^\textbf{\textit{X}}}^*(a^*_{\textrm{min}},n,m,\gamma_0)\le 0$. The minimum value of the cohesive energy with respect to the optimized $\{\gamma_i\}$ is obtained through a Newton-Raphson procedure as detailed in Appendix \ref{app:Derivatives}.

The bulk modulus $B$ can be derived in a similar way. In reduced units we have
\begin{equation} \label{eq:BM}
B^*(V)=V^*\frac{\partial^2 {E_{\textrm{coh}}^\textbf{\textit{X}}}^*(n,m,a^*,\gamma_T)}{\partial {V^*}^2}
\end{equation}
with $V^*=\frac{\sqrt{3}}{2}\gamma_0 {a^*}^3=\frac{\sqrt{3}}{2N}\gamma_T {a^*}^3$. Here and for the following properties introduced we keep the distances between the hexagonal layers at constant $\gamma_i=\gamma_0$. This is justified as the deviations from the ideal $\gamma_0$ value are very small as discussed in detail in section \ref{sec:Results}. Using zero pressure $P=dE/dV=0$ we get the following expression for the bulk modulus in terms of lattice sums\cite{Smits2021},
\begin{equation} \label{eq:BM1}
B^*(n,m,a^*,\gamma_T)=\frac{Nnm}{9\sqrt{3}\gamma_{T}(n-m)} \; \left[ (n+3)(a^*)^{-(n+3)}L^\textbf{\textit{X}}(\tfrac{n}{2},\gamma_T)
- (m+3)(a^*)^{-(m+3)}L^\textbf{\textit{X}}(\tfrac{m}{2},\gamma_T) \right] 
\end{equation}
Using \eqref{eq:RLJnmmin} we get
\begin{equation}\label{eq:BM2}
    \begin{aligned}
        B^*(n,m,a^*_{\textrm{min}},\gamma_T)&=\frac{Nnm}{9\sqrt{3}\gamma_T(n-m)} \; \Bigg[ (n+3) L^\textbf{\textit{X}}(\tfrac{n}{2},\gamma_T)
\left( \frac{L^\textbf{\textit{X}}(\tfrac{m}{2},\gamma_T)}{L^\textbf{\textit{X}}(\tfrac{n}{2},\gamma_T)}\right)^{\tfrac{n+3}{n-m}}\\
&- (m+3) L^\textbf{\textit{X}}(\tfrac{m}{2},\gamma_T)
\left( \frac{L^\textbf{\textit{X}}(\tfrac{m}{2},\gamma_T)}{L^\textbf{\textit{X}}(\tfrac{n}{2},\gamma_T)}\right)^{\tfrac{m+3}{n-m}}
\Bigg] 
    \end{aligned}
\end{equation}
For the vibrational contribution to the entropy, discussed in the next section, we require the average Einstein frequency $\omega_E$ in the lattice sum model by taking the expression from Ref.\onlinecite{Smits2021},
\begin{equation}\label{eq:vibfreq}
    \begin{aligned}
        \omega_E(n,m,a^*,\gamma_T)&=\sqrt{\frac{\epsilon}{r_e^2M}} \sqrt{\frac{nm}{3(n-m)}}\Bigg\{ (n-1) L^\textbf{\textit{X}}(\tfrac{n+2}{2},\gamma_T) (a^*)^{-n-2}\\
&- (m-1) L^\textbf{\textit{X}}(\tfrac{m+2}{2},\gamma_T) (a^*)^{-m-2} \Bigg\}^{1/2}
    \end{aligned}
\end{equation}
where $M$ is the mass of the atom in the lattice for a homonuclear system. In order to obtain a frequency that is only dependent on lattice constants (in dimensionless units) and some prefactors, we rewrite this equation and use \eqref{eq:RLJnmmin}, 
\begin{equation}\label{eq:vibfreq2}
    \begin{aligned}
        \omega_E^{'}=\sqrt{\frac{r_e^2M}{\epsilon}}\omega_E
&= \sqrt{\frac{nm}{3(n-m)}}\Bigg\{ (n-1) L^\textbf{\textit{X}}(\tfrac{n+2}{2},\gamma_T) 
\left( \frac{L^\textbf{\textit{X}}(\tfrac{m}{2},\gamma_T)}{L^\textbf{\textit{X}}(\tfrac{n}{2},\gamma_T)}\right)^{\tfrac{n+2}{n-m}}\\
&- (m-1) L^\textbf{\textit{X}}(\tfrac{m+2}{2},\gamma_T) 
\left( \frac{L^\textbf{\textit{X}}(\tfrac{m}{2},\gamma_T)}{L^\textbf{\textit{X}}(\tfrac{n}{2},\gamma_T)}\right)^{\tfrac{m+2}{n-m}}
\Bigg\}^{1/2}
    \end{aligned}
\end{equation}
which results in a zero-point vibrational energy of
\begin{align} \label{eq:ZPVE}
E_0 = \frac{3}{2} \hbar \omega_E = \frac{3}{2} \sqrt{\frac{\epsilon}{r_e^2M}}~ \hbar \omega_E^{'}
\end{align}
The Einstein frequency describes the vibration of an atom in the field of all other atoms. However, this simplified model, although analytical in terms of lattice sums, neglects the frequency splitting between the parallel and perpendicular components of an atom in the hexagonal layer, i.e. for Barlow packing one generally has $\omega_E^{\|} \ne \omega_E^{\perp}$. Only for the fcc phase are these three frequency components perfectly degenerate because of symmetry\cite{Smits2021}. Moreover, one should in principle consider the change in geometry due to zero-point vibrations. In addition, phonon dispersion cannot be neglected anymore when small zero-point energy differences or differences in vibrational entropies between different structures are considered\cite{Schwerdtfeger-2016}. We therefore decided to include phonon dispersion calculations for both fcc and hcp by diagonalizing the $(3N,3N)$ hermitian dynamical matrix $D(\vec{k})$ ($N=1$ for fcc and $N=2$ for hcp) in $k$-space \cite{Rosciszewski2007},
\begin{align} \label{eq:phonons}
D(\vec{k})v(\vec{k}) = \omega^2(\vec{k})v(\vec{k})
\end{align}
where the matrix $D(\vec{k})$ is related to the mass-weighted force-field matrix $F$ in real space through a Fourier transformation, $v(\vec{k})$ the corresponding eigenvector and the $3N$ real solutions $\omega(\vec{k})$ at a specific $\vec{k}$-point are the phonon frequencies. In our case, $D(\vec{k})v(\vec{k})$ becomes a real and symmetric matrix because of roto-inversion symmetry for the Barlow packings. This implies that only the cosine term in the Fourier transformation $e^{i\vec{k}\cdot\vec{R}}=\cos(\vec{k}\cdot\vec{R})+i\sin(\vec{k}\cdot\vec{R})$ needs to be considered in our calculations. Hence for Barlow packings we get the following $(3N,3N)$ dynamical matrix,
\begin{align} \label{eq:dynamic matrix}
D_{\alpha\kappa,\beta\kappa'}(\vec{k}) = \frac{1}{M}\sum_b F_{\alpha\kappa,\beta\kappa'}(0,b) \cos(\vec{k}\cdot\vec{R_b})
\end{align}
where $\alpha$ and $\beta$ label the coordinate axes $x,y,z$, $\kappa$ and $\kappa'$ characterize the two coupled atoms with mass $M$ in the vibration, and $F$ is the force-field matrix in real space between an atom in the reference unit cell 0 and an atom in unit cell $b$. The harmonic phonon frequencies can be normalized such that the mass and the $\epsilon$ parameter of the $(n,m)$-LJ potentials become just multiplying scalars. Then, we have in a similar way to \eqref{eq:vibfreq2}
\begin{equation}
    \omega'(\vec{k}) = \sqrt{\frac{r_e^2M}{\epsilon}}\omega(\vec{k})
\end{equation}
This has the advantage that due to the simple mass scaling we need to determine the phonon branches from one calculation for a specific $(n,m)$-LJ potential. Diagonalizing $D$ gives $3N$ solutions ($N$ being the period) from which the frequencies can be obtained. Hence, instead of diagonalizing a $3N_A$-dimensional (mass-weighted) force field matrix which computationally is $\mathcal{O}(N_A^3)$ ($N_A$ being the number of vibrating atoms chosen in the solid), one has to diagonalize a far smaller $(3N,3N)$ dynamical matrix at many different $k$-points. Nevertheless, both methods were used for cross checking. A detailed description of lattice dynamics theory can be found in the book by Born and Huang\cite{born-1998}.

\subsection{Entropy}
\label{subsec:entropy}

The entropy for most ordered materials with a sizable electronic band gap comes mainly from the vibrational contributions, which  within the Einstein approximation is given by the expression
\begin{equation} \label{eq:PhonoEntropy}
S_{\rm vib}(T) = 3k_B\left[ {\rm -ln}\left( 1-e^{-\beta \omega_E}\right) 
- \frac{\beta \omega_E}{1-e^{\beta \omega_E}} \right]
\end{equation}
where $\beta=\hbar/k_BT$. The second term in this expression appears with opposite sign in the internal energy and drops out for the Gibbs free energy expression. For more details see Appendix \ref{app:LatticeThermodynamics}. This approximation neglects phonon dispersion and anharmonicity effects and therefore is only approximate for multi-lattices, but nevertheless useful as we shall see. Moreover, it has been shown that for the hcp lattice the error coming from such an approximate treatment is small for the Einstein frequency \cite{Smits2021}. We re-express this in terms of our dimensionless Einstein frequency for better comparison,
\begin{equation} \label{eq:PhonoEntropy1}
S_{\rm vib}(T') = 3k_B\left[ {\rm -ln}\left( 1-e^{-\beta^{'} \omega_E^{'}}\right) 
- \frac{\beta^{'} \omega_E^{'}}{1-e^{\beta^{'} \omega_E^{'}}} \right]
\end{equation}
where $\beta^{'}=1/T^{'}$ with the temperature
\begin{equation} \label{eq:Temperature}
T^{'}=\frac{k_Br_e}{\hbar} \sqrt{\frac{M}{\epsilon}}  T
\end{equation}
and we keep the entropy in the usual SI units. $T^{'}$ is a dimensionless mass-, distance- and energy-weighted temperature which means that we have shifted the mass dependency conveniently out of the Einstein frequency and can study the temperature dependence of the function $S_{\rm vib}(T')$ (alternatively we can set the mass, energy in equilibrium distance to 1 as for reduced units and reintroduce them at a later stage). To give an example for using dimensionless units, the argon triple (melting) point is at 83.81 K. Using a mass of $M=39.962$ g/mol for the 40-Ar isotope, a distance of $r_e=3.7624$ \AA~ and dissociation energy of $\epsilon=1.1885$ kJ/mol for diatomic argon\cite{PatkowskiSzalewicz2010}, we obtain $T^{'}=23.941$ for the triple point. Hence we can easily convert the dimensionless temperature $T^{'}$ into the real temperature for specific isotopes of solids such as the rare gas elements.  We note that $S_{\rm vib}\rightarrow 0$ for $T^{'}\rightarrow 0$ or $M\rightarrow 0, S_{\rm vib}\rightarrow \infty$ for $T^{'}\rightarrow \infty, M\rightarrow \infty$. Further, $S_{\rm vib}(T')$ increases monotonically from zero with increasing $T'$. Another important fact is that the high temperature limit for the vibrational entropy can also be interpreted as the high mass limit $(M\rightarrow\infty)$ or the classical limit $(\hbar\rightarrow 0)$.

Eq.\eqref{eq:PhonoEntropy} can be easily extended to include phonon dispersion,
\begin{equation} \label{eq:PhonoEntropypdisp}
S_{\rm vib}(T') = k_B \sum_{\textbf{k},\mu}\left[ {\rm -ln}\left( 1-e^{-\beta^{'} \omega^{'}_\mu(\textbf{k})}\right) 
- \frac{\beta^{'} \omega^{'}_\mu(\textbf{k})}{1-e^{\beta^{'} \omega^{'}_\mu(\textbf{k})}} \right]
\end{equation}
where the index $\mu$ runs over the $3N$ solutions ($N=1$ for fcc and $N=2$ for a bi-lattice such as hcp) for \eqref{eq:phonons} at a specific $\textbf{k}$-point. As we discuss the high temperature (or low frequency) limit in this work, we can use a simple Taylor expansion (see appendix \ref{app:LatticeThermodynamics}) of the exponential and obtain
\begin{equation} \label{eq:PhonoEntropypHT1}
S_{\rm vib}(T'\gg 0) \approx k_B \left[ 1-\sum_{\textbf{k},\mu}{\rm ln}\left( \beta^{'} \omega^{'}_\mu(\textbf{k})\right) 
\right]
\end{equation}
As $\beta^{'} \omega^{'}_\mu(\textbf{k)}\ll 1$, the entropy remains positive but approaches infinity for $T'\rightarrow\infty$.

For the following we focus on entropy difference between a specific Barlow packing and the Barlow structure with the smallest periodicity, i.e. the hcp structure, 
\begin{equation} \label{eq:PhonoEntropy3}
\Delta S_{\rm vib}^\textbf{\textit{X}}(T')=S_{\rm vib}^\textbf{\textit{X}}(T')-S_{\rm vib}^\text{hcp}(T')
\end{equation}
which, as we shall see, is very sensitive to phonon dispersion. The formalism can be extended if we take anharmonicity effects into account affecting the phonon frequencies, which we however neglect in this work as it comes at a significant computational cost.
For the high-temperature limit we simply get from \eqref{eq:PhonoEntropypHT1}
\begin{equation} \label{eq:PhonoEntropypHT3}
\Delta S_{\rm vib}^\textbf{\textit{X}}(T'\rightarrow\infty)= -k_B \left[ \sum_{\textbf{k},\mu}{\rm ln}\left( \omega^{'~{\textit{X}}}_\mu(\textbf{k})\right) -  \sum_{\textbf{k},\mu}{\rm ln}\left( \omega^{'~{\text{hcp}}}_\mu(\textbf{k})\right)\right]
\end{equation}
that is we get a finite value for $\Delta S_{\rm vib}$ as the term $\beta'$ causing the infinity drops out. In fact, in this limit it does not matter what units are used for the phonon frequencies as the conversion factor $f$ drops out for entropy differences because of the simple relation for the logarithm,  $\ln(f(T)\omega)=\ln(f(T))+\ln(\omega)$, where in the usual units we have $f(T)=\hbar/k_BT$.

\subsection{Density Functional Calculations}
\label{sec:Lithium}

In order to assess the stability for Barlow packings for more realistic systems, we carried out DFT calculations\cite{VASP1,VASP2,VASP3,VASP4,VASP5} for solid argon where we used the Perdew-Burke-Ernzerhof (PBE) exchange-correlation functional\cite{Perdew1996} with dispersion corrections by means of the DFT-D3 approach with the Becke-Johnson damping function\cite{Grimme2010,Grimme2011,Grimme2011b}. The atomic cores are described by the Projector Augmented Wave (PAW) method \cite{Blochl1994PAW}, and the electronic minimization was done using Gaussian smearing with an energy width of $0.1$ eV and a single-point evaluation of the energy using the tetrahedron method with Bl\"ochl corrections.\cite{Blochl1994BZInt} The $k$-point grid was set using the keyword KSPACING=0.10 centered at the $\Gamma$ point and an energy cut-off of $500$ eV. The crystal parameters $a$ and $\gamma_T$ where optimized using a Newton-Raphson procedure with the gradient and Hessian matrix calculated numerically applying a threshold for convergence of $10^{-5}$ kJ/mol for the electronic energy. For the isolated atom required for the cohesive energy we used a large orthorhombic unit cell of $14\times 14.001\times 14.002$ \AA. We investigated Barlow packings of periodicity $12$ since this number of layers is the least common multiple of the periodicity of important structures such as hcp, $(AB)$, fcc $(ABC)$, and dhcp $(ABAC)$. 

For comparison to our LJ entropy calculations including phonon dispersion, we decided to carry out additional DFT calculations to obtain the phonon frequencies and subsequently the entropy difference between the fcc and hcp phases of solid argon at the optimized lattice geometries. For this we used the DFT-D3 approach as described above. The phonon frequencies were obtained through a finite difference method with a $6\times6\times6$ supercell centered at the $\Gamma$ point. The temperature dependent entropy was obtained through Eq. \eqref{eq:PhonoEntropypdisp}.

\section{Results and discussions}
\label{sec:Results}

The Barlow lattice sums $L^\textbf{\textit{X}}(N,s,\{\gamma_i\})$ with equal packing fraction $\alpha^\textbf{\textit{X}}(2)$ and period $N$ differ only minutely from each other and lie approximately on a straight line with respect to $\alpha^\textbf{\textit{X}}(2)$, with a very small root mean square error of RMS = $2.361\times 10^{-7}$ for $s=\frac{n}{2}=3$ as shown in our previous paper\cite{Cooper2025}. As we shall see, this translates into a linear behavior for many other properties as well. We therefore introduce the following definition for a property $P$ of a specific Barlow packing with packing fraction $\alpha^\textbf{\textit{X}}(2)$,
\begin{equation}\label{eq:line}
\tilde{P}^\textbf{\textit{X}}(n,m,\{\gamma_i\}) = P^\text{hcp}(n,m,\{\gamma_i\}) + \alpha^\textbf{\textit{X}}(2) \left[ P^\text{fcc}(n,m,\{\gamma_i\}) - P^\text{hcp}(n,m,\{\gamma_i\})\right]
\end{equation}
with $\bar{\gamma}=\frac{1}{N}\gamma_T=\frac{1}{N}\sum_{i=1,\dots N}\gamma_i\approx \gamma_0$, and the deviation from the ideal linear behavior by
\begin{equation}\label{eq:linedev}
\Delta P^\textbf{\textit{X}}(n,m,\{\gamma_i\}) = P^\textbf{\textit{X}}(n,m,\{\gamma_i\}) - \tilde{P}^\textbf{\textit{X}}(n,m,\{\gamma_i\})
\end{equation}
Furthermore, $\Delta L^\textbf{\textit{X}}(s,\gamma_i=\gamma_0)$ values with $\gamma_0=\sqrt{\frac{2}{3}}$ lie approximately within a triangle along the packing fraction $\alpha$.\cite{Cooper2025} 

For the following discussion we neglect the * notation (e.g. $P\equiv P^*$) and work with reduced units if not otherwise stated. We further distinguish between three different cases for our optimization procedure for the different $(n,m)$-LJ potentials: 
\begin{itemize}
    \item Case 1: The Barlow packings are kept at ideal $\gamma_T=N\gamma_0=N\sqrt{\frac{2}{3}}$; 
    \item Case 2: $\gamma_T$ is optimized with the restriction that $\gamma_i=\gamma_j$, for all $i\ne j$ and $i=1,\dots,N$;
    \item Case 3: All $\gamma_i$ are optimized. 
\end{itemize}
For case 2 or 3 we refer to the optimized $\gamma_i$ values as $\gamma_\text{min}=\frac{1}{N}\gamma_T=\frac{1}{N}\sum_{i=1,N}\gamma_i\approx\sqrt{\frac{2}{3}}$.

\subsection{Lattice constants}
\label{sec:LatticConstants}

\begin{figure}[htb!]
\centering
a)\includegraphics[width=.47\columnwidth]{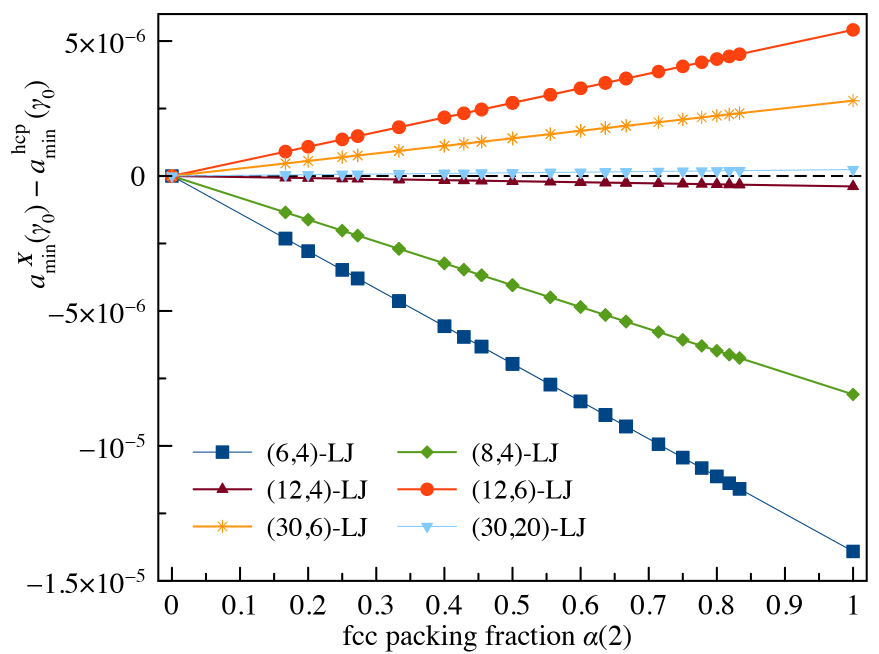}
b) \includegraphics[width=.47\columnwidth]{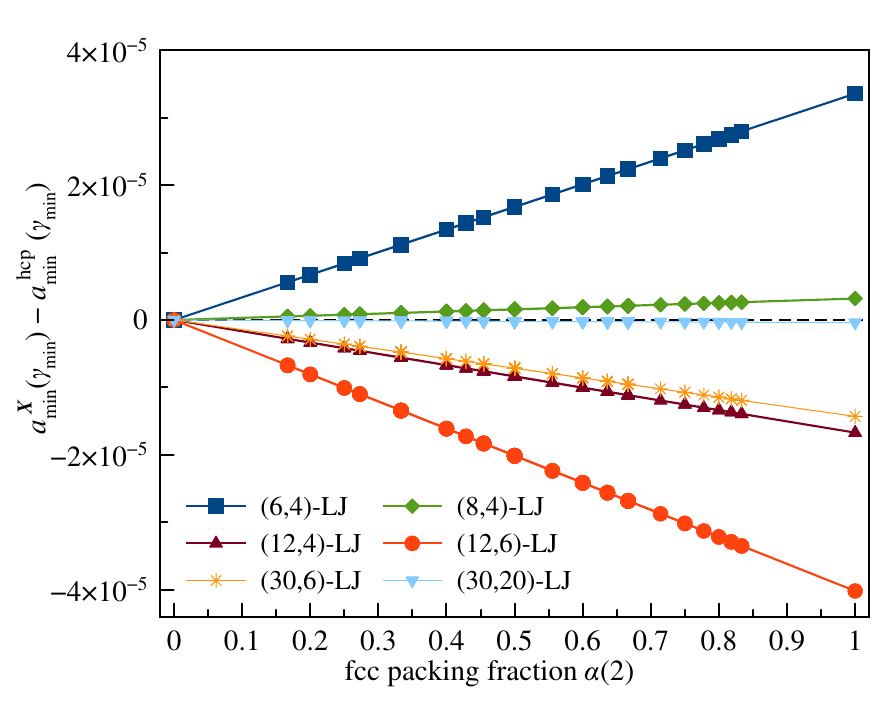}
\caption{(a) Difference in the minimum lattice constants $a_\text{min}$ (in reduced units) at fixed $\gamma_T=N\gamma_0=N\sqrt{\frac{2}{3}}$ (case 1) for Barlow packings compared to the reference hcp structure ($\alpha(2)=0$) as a function of the fcc stacking fraction $\alpha(2)$ for various $(n,m)$-LJ potentials. (b) Same as (a) but at the optimized $\gamma_T$ or $\{\gamma_i\}$ lattice parameters, i.e. case 2 or 3 (the difference between the two cases is not visible on this scale, see discussion).}
\label{fig:aminpack}
\end{figure}

\begin{figure}[htb!]
\centering
a)\includegraphics[width=.47\columnwidth]{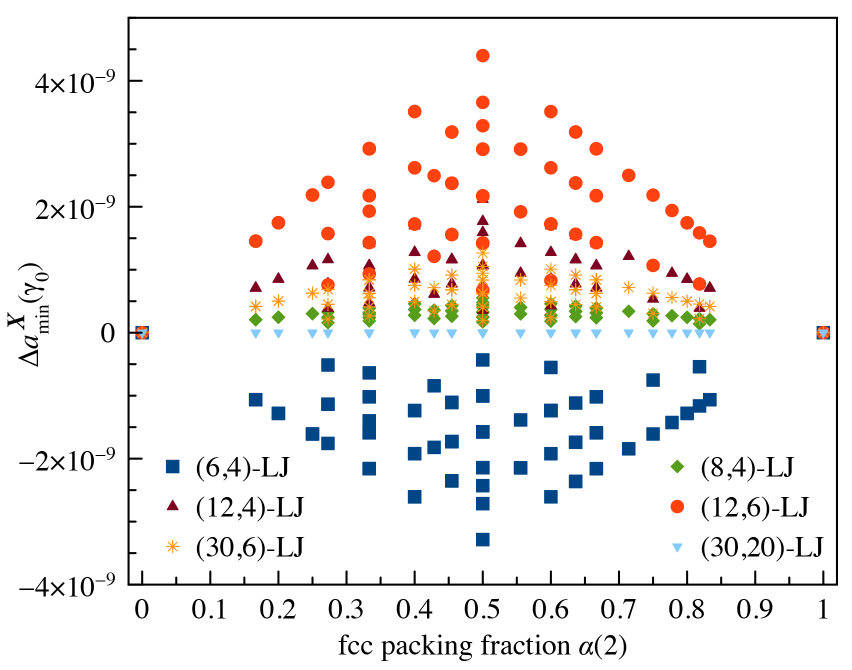}
b) \includegraphics[width=.47\columnwidth]{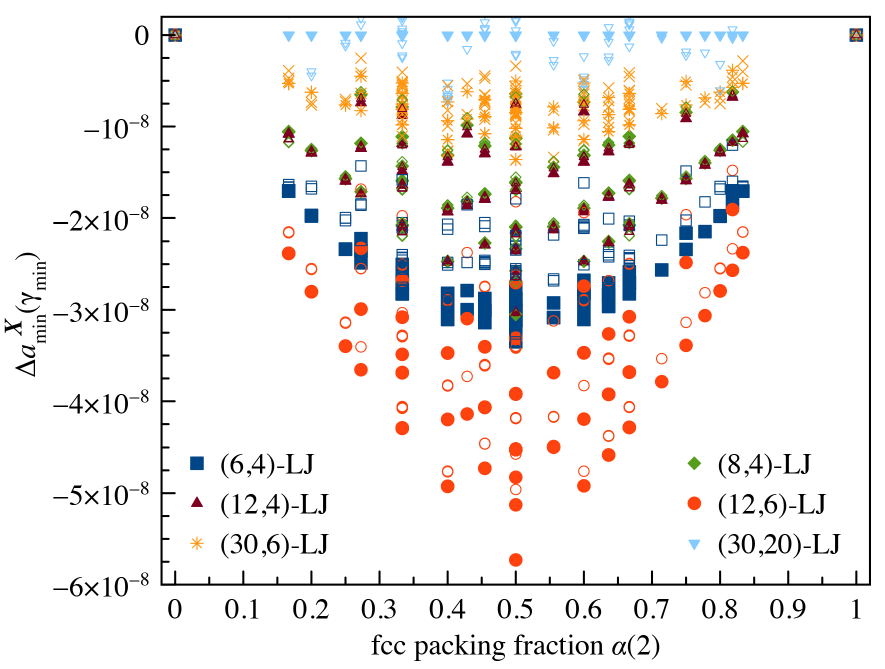}
\caption{(a) Distribution of the difference in the minimum lattice constants $\Delta a_\text{min}^\textbf{\textit{X}}(n,m,\gamma_i=\gamma_0)$ (case 1) for Barlow packings at $\gamma_0=\sqrt{2/3}$ compared to the hcp structure ($\alpha(2)=0$) as a function of the fcc stacking fraction $\alpha(2)$ for various $(n,m)$-LJ potentials according to eq.\eqref{eq:linedev}. (b) Same as (a) but at the optimized $\gamma_\text{min}$ (filled symbols for case 2 and open symbols for case 3).}
\label{fig:aminpack1}
\end{figure}

The optimized minimum lattice constants with respect to hcp, $\Delta a_\text{min}^\textbf{X}(\{\gamma_i\})=a_\text{min}^\textbf{X}(\{\gamma_i\})-a_\text{min}^\text{hcp}(\{\gamma_i\})$, for case 1 are shown in Figure \ref{fig:aminpack}a. We observe a very strong linear behavior with respect to the fcc packing fraction ($\tilde{a}_\text{min}\cong a_\text{min}$) with a tiny RMS error for the 102 different Barlow structures considered (up to $N=12$), e.g. $2.280\times 10^{-9}$  for the (12,6)-LJ potential. This can be rationalized by a simple Taylor expansion in terms of the ratio between the Barlow and reference hcp lattice sums (see appendix \ref{app:Taylor}), where the strong linear correlation with respect to the fcc packing fraction is directly related to the linear behavior seen for the lattice sums\cite{Cooper2025}. In fact, the deviation from the ideal linear behavior, $\Delta a_\text{min}^\textbf{X}$, according to eq.\eqref{eq:linedev} is four orders of magnitude smaller compared to the actual $\tilde{a}_\text{min}^\textbf{X}$ values, as shown in Figure \ref{fig:aminpack1}a. Here we observe an interesting triangle pattern with both positive and negative deviations from the ideal line depending on the choice of the exponents $(n,m)$ in the LJ potential. This originates from the similar pattern  found for the lattice sums\cite{Cooper2025}. We note that the largest deviation from the ideal linear behavior is found for the (ABAC) ($N=4$) Barlow packing with fcc packing fraction $\alpha(2)=0.5$.

Cases 2 and 3 involve the optimization of the lattice parameters $\gamma_i$ through a Newton-Raphson procedure as outlined in appendix \ref{app:Derivatives}. Because of the small differences between the $\gamma_i$ values for a specific Barlow packing, analytical first and second derivatives of the lattice sums with respect to $\gamma_i$ were required as outlined in appendix \ref{app:Bessel}. One could assume that the small $\Delta\gamma$ values (see eq.\eqref{eq:linedev} for the definition) translates into small changes in the other lattice constant such as $a_\text{min}^\textbf{X}$. While this is indeed the case, for the differences $\Delta a_\text{min}^\textbf{X}=a_\text{min}^\textbf{X}-a_\text{min}^\text{hcp}$, which are already very small, it nevertheless becomes significant as a comparison between the two figures \ref{fig:aminpack} (a) and (b) shows. Here $a_\text{min}^\text{fcc}(n,m,\{\gamma_i\}) - a_\text{min}^\text{hcp}(n,m,\{\gamma_i\})$ become very sensitive to the $\gamma_i$-values chosen which can change significantly the slope of the individual linear curves for the different $(n,m)$-LJ potentials leading even to a different sign in the slops. However, the strong linear behavior still persists with a small RMS error of $3.739\times 10^{-8}$ for case 2 and $3.368\times 10^{-8}$ for case 3 for the (12,6)-LJ potential. However, case 2 and 3 lead to a very similar linear behavior as shown in Figure \ref{fig:aminpack}b.

\begin{figure}[htb!]
\centering
a)\includegraphics[width=.47\columnwidth]{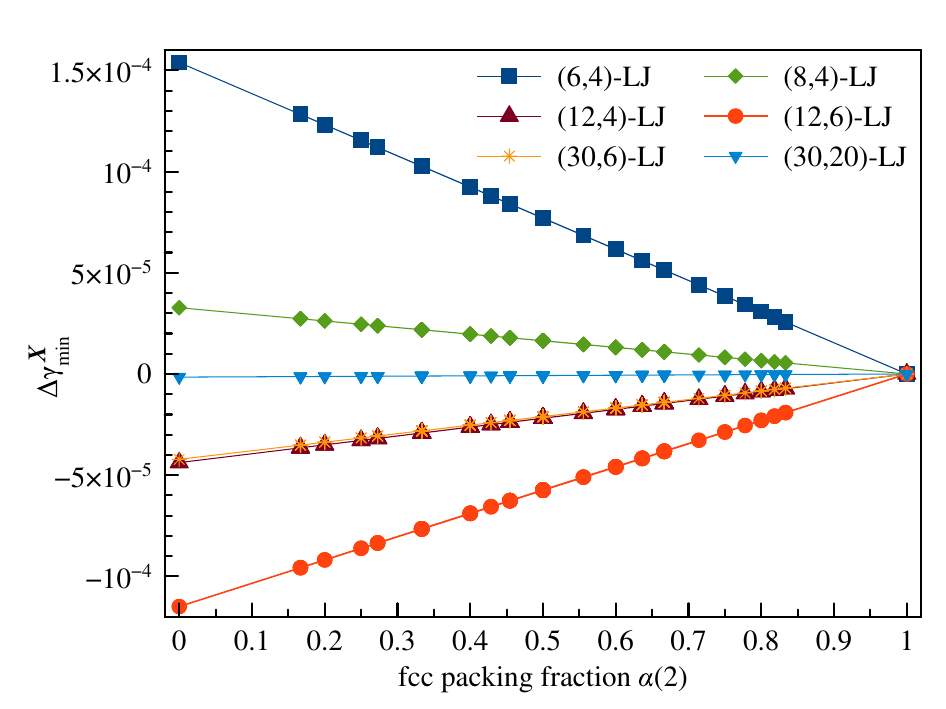}
b) \includegraphics[width=.47\columnwidth]{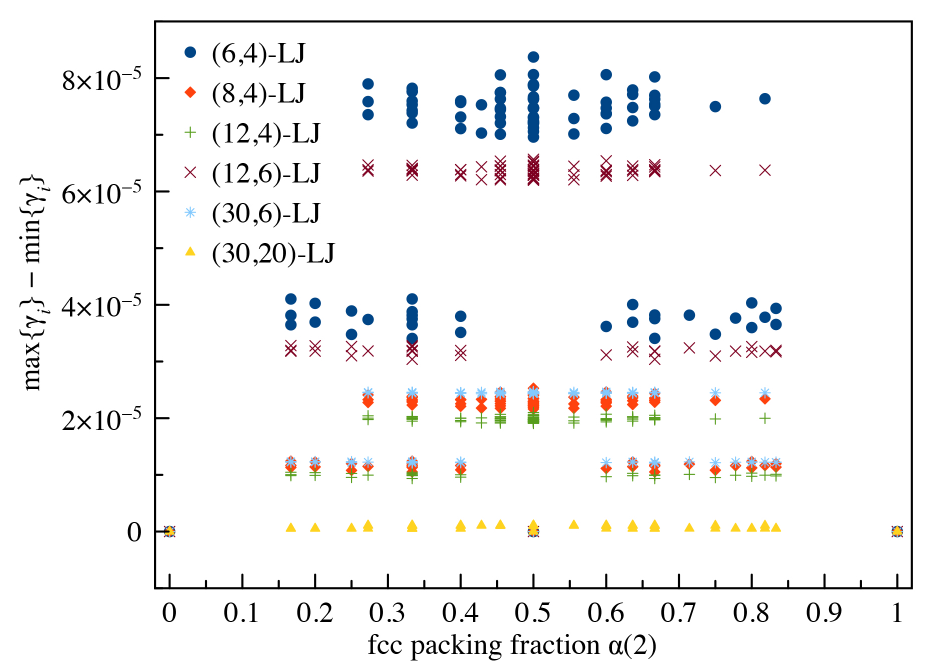}
\caption{(a) Difference $\Delta\gamma_\text{min}=\gamma_\text{min}-\gamma_0$ between the optimized $\gamma_\text{min}$ value and the ideal one of $\gamma_0=\sqrt{2/3}$ for Barlow packings as a function of the fcc stacking fraction $\alpha(2)$ for various $(n,m)$-LJ potentials for cases 2 and 3 (the difference between case 2 and 3 cannot be seen by the naked eye). (b) Spread of $\gamma_i$ values, $\text{max}(\gamma_i)-\text{min}(\gamma_i)$.}
\label{fig:gammamin}
\end{figure}

\begin{figure}[htb!]
\centering
\includegraphics[width=.47\columnwidth]{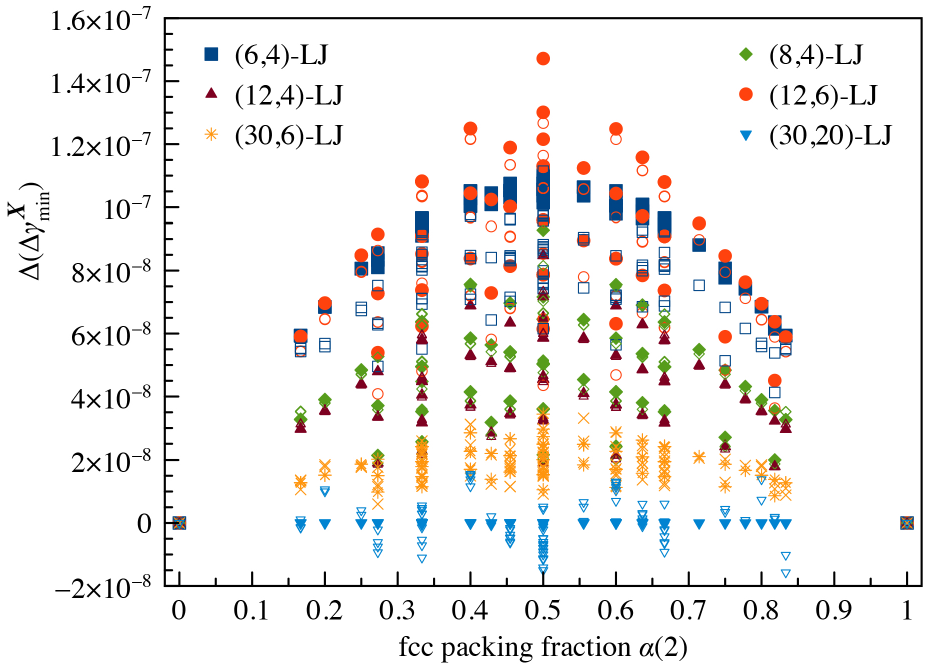}
\caption{Deviation of the $\Delta\gamma_\text{min}$ values from the ideal straight line connecting the fcc and hcp structures according to eq.\eqref{eq:linedev} using case 2 (filled symbols) and case 3 (open symbols).}
\label{fig:gammamin1}
\end{figure}

The deviations from the ideal $\gamma_0=\sqrt{\frac{2}{3}}$ value, $\Delta\gamma_\text{min}=\gamma_\text{min}-\gamma_0$, are listed in Table \ref{tab:barlowdeltagamma}, appendix \ref{app:CohesiveEnergies}, and are shown in Figure \ref{fig:gammamin}a. Howard\cite{Howard1970} already showed as early as in 1970 by direct lattice summations over 450 shells within the hcp lattice, using a $(12,6)$LJ potential, that the $\gamma_0=\sqrt{\frac{2}{3}}$ value is reduced by tiny amount of $\Delta\gamma_\text{min}=-1.15\times 10^{-4}$. Through exact lattice summations we get a more accurate value of $\Delta\gamma=-1.14928459\times 10^{-4}$, see Table \ref{tab:barlowdeltagamma} and Ref.\onlinecite{Schwerdtfeger2023}. Not surprisingly, for the fcc lattice we have exactly $\Delta\gamma_\text{min}=0$ for all $(n,m)$ combinations, while hcp shows the largest absolute deviation compared to all other Barlow structures for all the LJ potentials studied here. Even for the case where hcp and fcc are energetically degenerate, e.g. at $n=8.0$ and  $m=4.208890947988$, we get the largest absolute value for $\Delta\gamma_\text{min}$ for the hcp structure. As a result, we see again an almost perfect linear behavior for $\Delta\gamma_\text{min}(\alpha(2))$ with deviations from the ideal straight line three orders of magnitude smaller than the actual $\Delta\gamma_\text{min}$ value. In addition, we get this interesting triangle pattern for the deviations in the $\Delta\gamma_\text{min}$ values (see Figure \ref{fig:gammamin1}). The spread of the optimized $\gamma_i$ values as shown in Figure \ref{fig:gammamin}b shows larger values for the soft potentials. For the hard potentials approaching the hard-sphere limit one obtains $\text{max}(\gamma_i)-\text{min}(\gamma_i)=0$ as expected.

We briefly consider the volume per atom $\bar{V}=V/N$ using the linear relationships we found for the lattice constant. Using a Taylor expansion up to first order because of the smallness of the hcp/fcc differences in the lattice constants we get
\begin{align}
\bar{V}^\textbf{\textit{X}} = \frac{\sqrt{3}}{2} (a^\textbf{\textit{X}})^3 \gamma_T^\textbf{\textit{X}}
&= \frac{\sqrt{3}}{2} \left[ a^\text{hcp} + \alpha^\textbf{\textit{X}}(2) \left( a^\text{fcc} - a^\text{hcp}\right) \right]^3
\left[ \gamma_\textrm{min}^\text{hcp} + \alpha^\textbf{\textit{X}}(2) \left( \gamma_\textrm{min}^\text{fcc} - \gamma_\textrm{min}^\text{hcp}\right) \right]\\
\nonumber
&\approx \bar{V}^\text{hcp} \left[ 1 + 3\alpha^\textbf{\textit{X}}(2) \frac{\left( a^\text{fcc} - a^\text{hcp}\right)}{a^\text{hcp}} \right]
\left[ 1 + \alpha^\textbf{\textit{X}}(2) \frac{\left( \gamma_\textrm{min}^\text{fcc} - \gamma_\textrm{min}^\text{hcp}\right)}{\gamma_\textrm{min}^\text{hcp}} \right]
\label{eq:vol1} \\
&\approx \bar{V}^\text{hcp} + \alpha^\textbf{\textit{X}}(2) \bar{V}^\text{hcp} \left[ 3 \frac{\left( a^\text{fcc} - a^\text{hcp}\right)}{a^\text{hcp}} + \frac{\left( \gamma_\textrm{min}^\text{fcc} - \gamma_\textrm{min}^\text{hcp}\right)}{\gamma_\textrm{min}^\text{hcp}} \right]
\end{align}
As for $\alpha(2)=1$ we have $V=V^\text{fcc}$ we get the relationship
\begin{align}
\bar{V}^\text{fcc}-\bar{V}^\text{hcp}\approx \bar{V}^\text{hcp} \left[ 3 \frac{\left( a^\text{fcc} - a^\text{hcp}\right)}{a^\text{hcp}} + \frac{\left( \gamma_\textrm{min}^\text{fcc} - \gamma_\textrm{min}^\text{hcp}\right)}{\gamma_\textrm{min}^\text{hcp}} \right]
\end{align}
We thus obtain to a very good approximation a linear relationship for the volume of the Barlow packings, which will become useful for the discussion of the free energy of the system at finite pressure, see section \ref{sec:FreeEnergy}.

\subsection{Cohesive energies}
\label{sec:Cohesive energies}

\begin{figure}[htb!]
\centering
a)\includegraphics[width=.44\columnwidth]{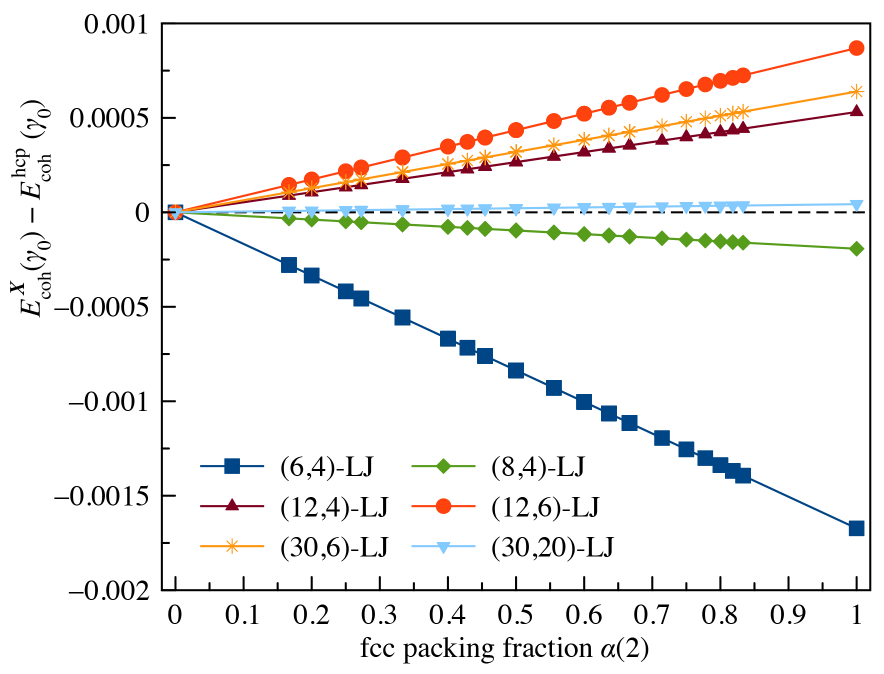}
b) \includegraphics[width=.47\columnwidth]{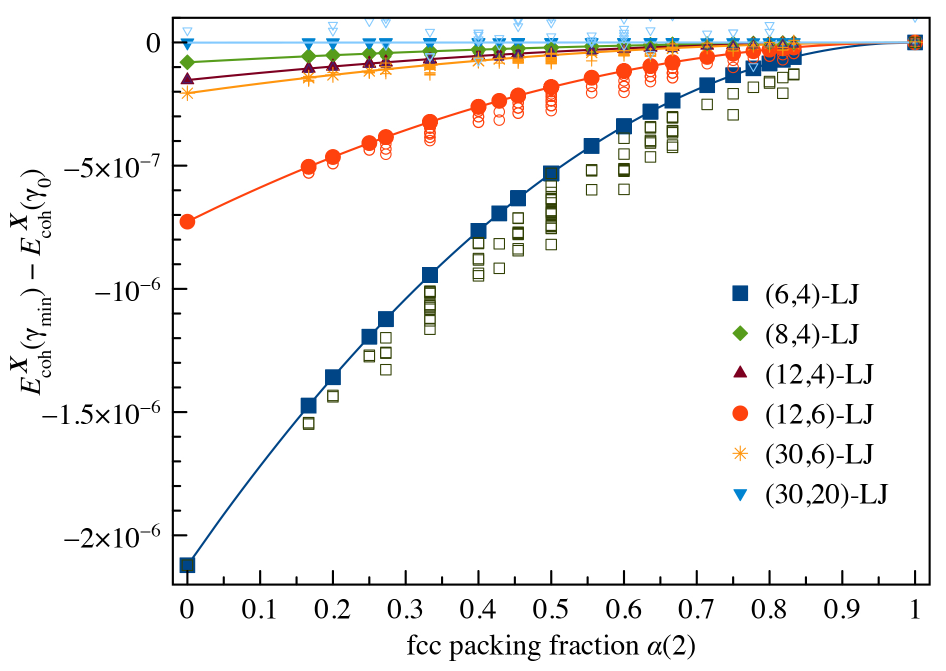}
\caption{(a) Cohesive energy difference for Barlow packings with respect to the hcp structure as a function of the fcc stacking fraction $\alpha(2)$ for various $(n,m)$-LJ potentials and $\gamma_0=\sqrt{\frac{2}{3}}$ (case 1). Data are listed in Table \ref{tab:barlowEcohenerg} and supporting information. (b) Difference in cohesive energies $\Delta E_\text{coh}(\gamma_\text{min})=E_\text{coh}(\gamma_\text{min})-E_\text{coh}(\gamma_0)$. A quadratic polynomial was used for the fit procedure for case 2 (filled symbols). Case 3 results are shown by open symbols.}
\label{fig:cohendiffpack}
\end{figure}

\begin{figure}[htb!]
\centering
a) \includegraphics[width=.44\columnwidth]{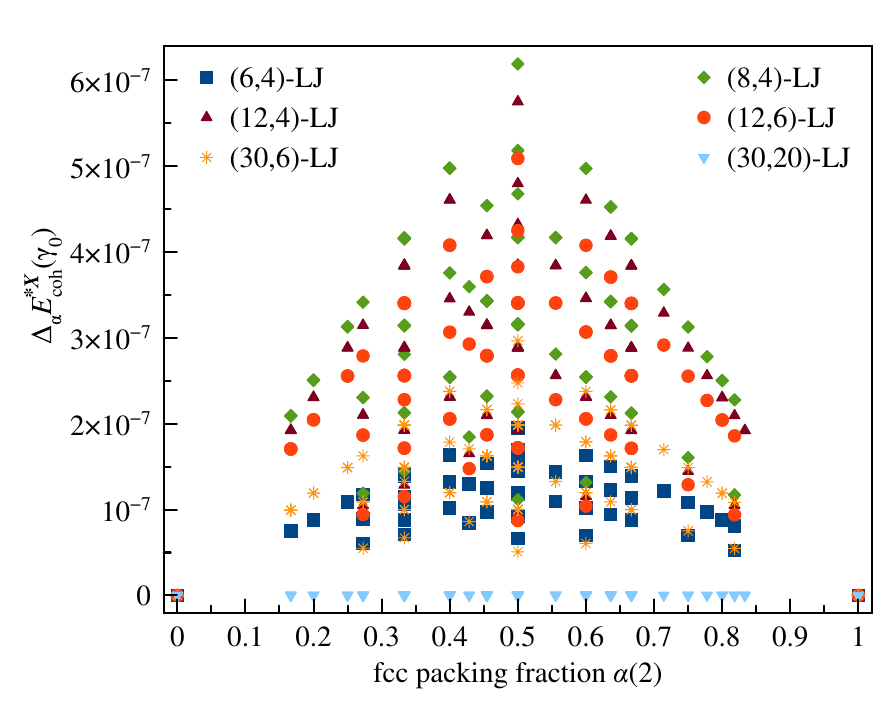}
b) \includegraphics[width=.47\columnwidth]{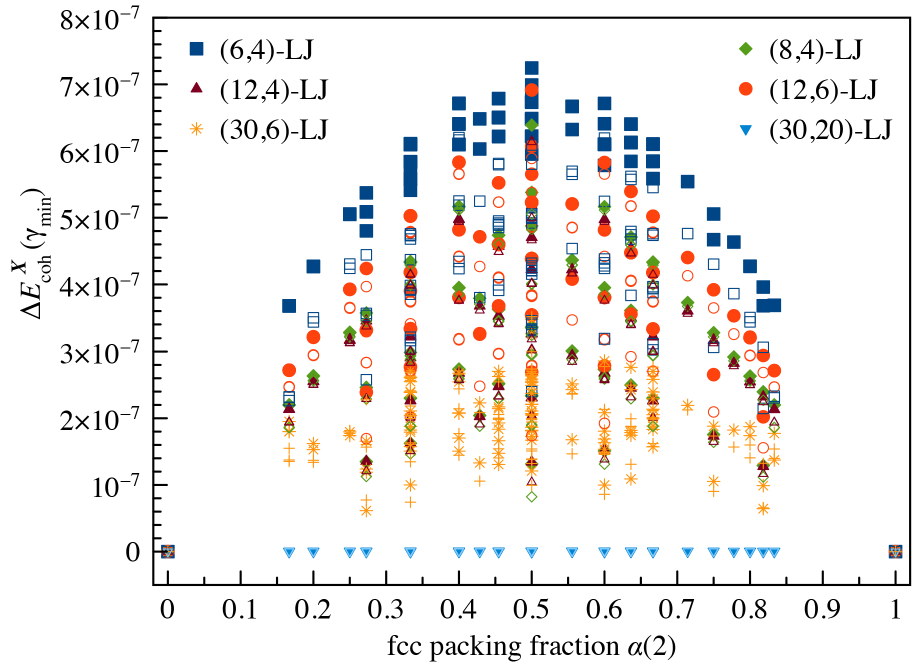}
\caption{(a) Deviation of cohesive energies of Barlow packings from the line connecting the fcc and hcp cohesive energy values according to eq.\eqref{eq:linedev} for various $(n,m)$-LJ potentials and for $\gamma_0=\sqrt{\frac{2}{3}}$. (b) Deviation of the  $\Delta E_\text{coh}(\gamma_\text{min})$ values from the line of Barlow packings  connecting the fcc and hcp values for various $(n,m)$-LJ potentials. Filled symbols for case 2 and open symbols for case 3.}
\label{fig:Egammachange}
\end{figure}

In order to compare the thermodynamic stabilities of the Barlow packings we need to analyze their cohesive energies first. Figure \ref{fig:cohendiffpack}a shows the difference in cohesive energies of the Barlow packings as a function of the fcc stacking fraction $\alpha(2)$ for various $(n,m)$-LJ potentials. We see again a nice linear behavior as a function of $\alpha(2)$ originating from the linear behavior in the lattice sums\cite{Cooper2025}, as demonstrated by a Taylor expansion in Appendix \ref{app:Taylor}. As for the lattice constants, the deviation from linearity for the cohesive energies is rather small with an RMS error of $2.673\times 10^{-7}$ for case 1 using the (12,6)-LJ potential. Figure \ref{fig:Egammachange}a shows a triangular distributions pattern for these deviations with the largest absolute value for the $ABAC$ Barlow packing ($\alpha(2)=0.5$). This implies that the LJ cohesive energy for a Barlow packing can be estimated to relatively high accuracy from the fcc stacking fraction and the cohesive energies for fcc and hcp. The cohesive energy for the sticky hard-sphere (SHS) limit is exactly $-6$ and $\Delta {E^\textbf{\textit{X}}_{\text{SHS}}}=0$, which is nearly satisfied by the (30,20)-LJ potential as shown in Table \ref{tab:barlowEcohenerg} and Figure \ref{fig:cohendiffpack}a. This linear relationship will be important for the discussion of the natural appearance of Barlow packings as we shall see when we discuss the free energy of the system. We note that for very soft potentials the fcc phase has a lower cohesive energy compared to hcp as already discussed by us before.\cite{Smits2021,Schwerdtfeger2023,Cooper2024}

Considering the effect of changing $\gamma_0\rightarrow\gamma_\text{min}$ upon optimization, case 2 and 3 leads to almost identical linear curves compared to case 1 not distinguishable by the naked eye, Figure \ref{fig:cohendiffpack}a. Figure \ref{fig:cohendiffpack}b shows that the cohesive energy difference between the two $\gamma$ values (difference between case 1 and 2) changes quadratically with the fcc stacking fraction. This change is however about three orders of magnitude smaller compared to the fcc/hcp cohesive energy difference (compare Figures \ref{fig:cohendiffpack}a and \ref{fig:cohendiffpack}b) and therefore has only a minor effect on the linear behavior. Case 3 where all $\gamma_i$-values are optimized leads interestingly to a spread in the  $\gamma_i$-values for a specific fcc stacking fraction $\alpha(2)$, see Figure \ref{fig:cohendiffpack}b. We also see a slight change in the distribution for the deviation from the linear behavior of the cohesive energy difference, i.e. compares Figures \ref{fig:Egammachange} a and b. Nevertheless, the important result here is that the cohesive energy difference for the Barlow structures with the respect to hcp behaves linearly with the fcc stacking fraction and optimizing the lattice parameters $\gamma_i$-values does not significantly change this result. Moreover, these deviations are all positive for the LJ potentials considered, that is the Barlow differences in cohesive energies are all above the line.

The results obtained using Lennard-Jones (LJ) potentials show that the very small differences in lattice constants and cohesive energies require exact lattice summations to accurately capture both the linear dependence and the subtle deviations from it. Although the LJ potential provides a useful toy model and is particularly relevant for the rare gases at low pressures, many-body effects may also play an important role in determining the linear dependence of the cohesive energy on $\alpha(2)$.\cite{maltby2026} Assessing these effects, however, poses a significant challenge for density functional theory (DFT), as the energy differences involved demand exceptionally high numerical accuracy. 

As an initial assessment, we performed DFT calculations for solid argon considering Barlow packings with periodicities of 2, 3, and 4. However, to ensure an equivalent comparison among these structures, all calculations were carried out using unit cells containing the same number of hexagonal layers. This is essential because the cohesive energy differences between the different structures are extremely small and comparable numerical precision must be maintained across all stackings. We therefore selected a unit cell containing 12 hexagonal layers, corresponding to the least common multiple of the considered periodicities, which yields a total of 43 distinct Barlow packings for $N=2,3,4$ and 12. 

\begin{figure}[ht!]
    \centering 
    \includegraphics[width=.5\columnwidth]{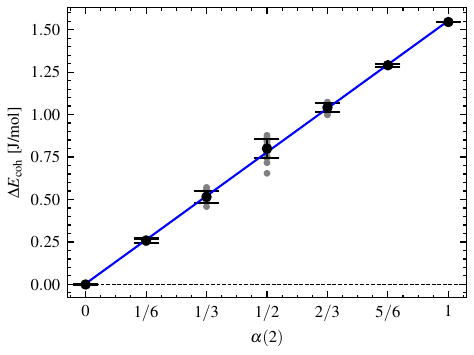}
    \caption{Cohesive energy difference for Barlow packings with periodicity $N=12$ with respect to the hcp structure as a function of the fcc stacking fraction $\alpha(2)$ for solid argon at the PBE-D3 level of theory. The mean cohesive energy for each $\alpha(2)$ is shown in black dots with error bars representing the standard deviation at that fcc stacking fraction and its linear fit in blue. The gray dots are the actual data.}
    \label{fig:EcohDFT}
\end{figure}

The cohesive energies as a function of the fcc stacking fraction, $\alpha(2)$, are shown in Figure \ref{fig:EcohDFT}. Although the energy differences are extremely small, the data are distributed around the straight line connecting the hcp and fcc structures, with slight but systematic deviations. Fitting the mean cohesive energy at each value of $\alpha(2)$ yields
\begin{equation}
    \Delta E_{\text{coh}} = 1.5473\alpha(2) + 0.0048
\end{equation}
with an $R^{2}=0.99967$. The excellent linear correlation indicates that the cohesive energy remains primarily governed by the local stacking environment, while the larger deviations from linearity compared to the LJ potentials reflect the many-body contributions captured by the density functional theory (DFT) calculations.
\cite{Cencek-2013,Klein1967}

Solid argon is known to exhibit predominantly two-body interactions, with dispersion forces providing the dominant contribution to its cohesion. Nevertheless, previous studies have shown that many-body interactions become important when discriminating between nearly degenerate crystal structures.\cite{Cencek-2013,Klein1967} The small deviations observed here are therefore consistent with the expected magnitude of these many-body effects for a rare gas element.

\subsection{Bulk moduli}

\begin{figure}[htb!]
\centering
a)\includegraphics[width=.47\columnwidth]{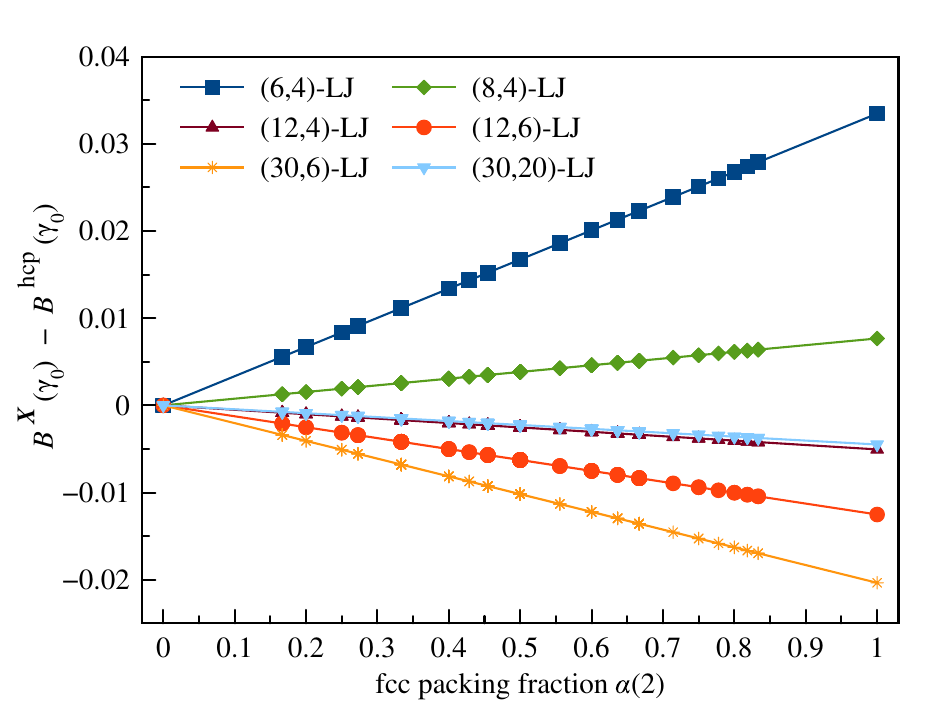}
b)\includegraphics[width=.47\columnwidth]{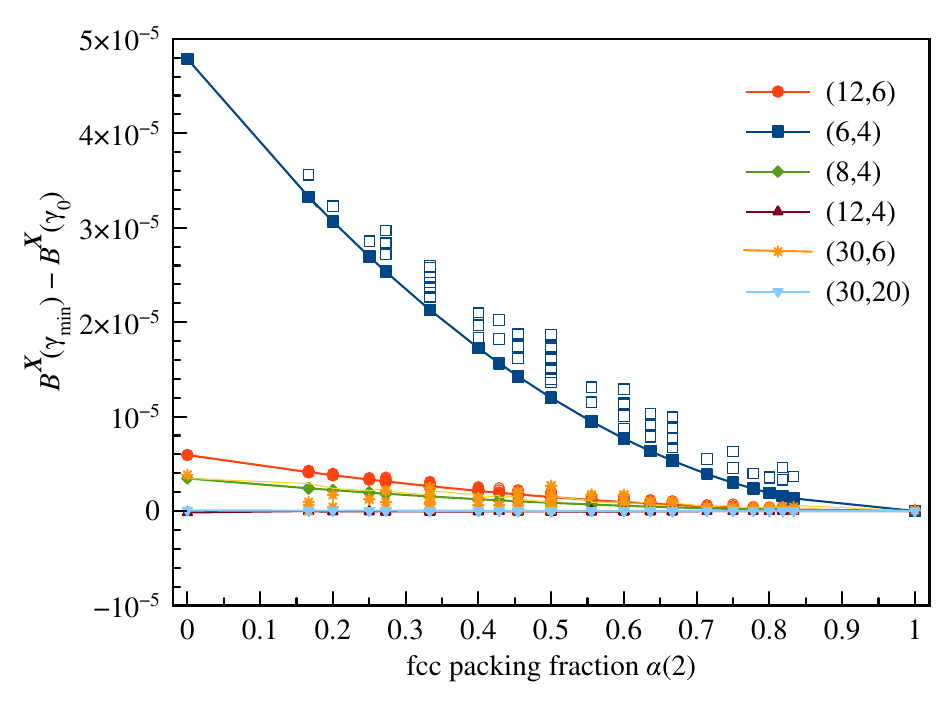}
\caption{(a) Differences in bulk moduli for Barlow packings compared to the hcp structure ($\alpha(2)=0$) as a function of the fcc stacking fraction $\alpha(2)$ for various $(n,m)$-LJ potentials. $\gamma_0$ is set to $\sqrt{\frac{2}{3}}$. Data are listed in the supplementary material. (b) Difference in bulk moduli $\Delta B(\gamma_\text{min})=B(\gamma_\text{min})-B(\gamma_0)$. Case 2 results are shown by (filled symbols) and case 3  open symbols.}
\label{fig:BMdiffpack}
\end{figure}

\begin{figure}[htb!]
\centering
a)\includegraphics[width=.47\columnwidth]{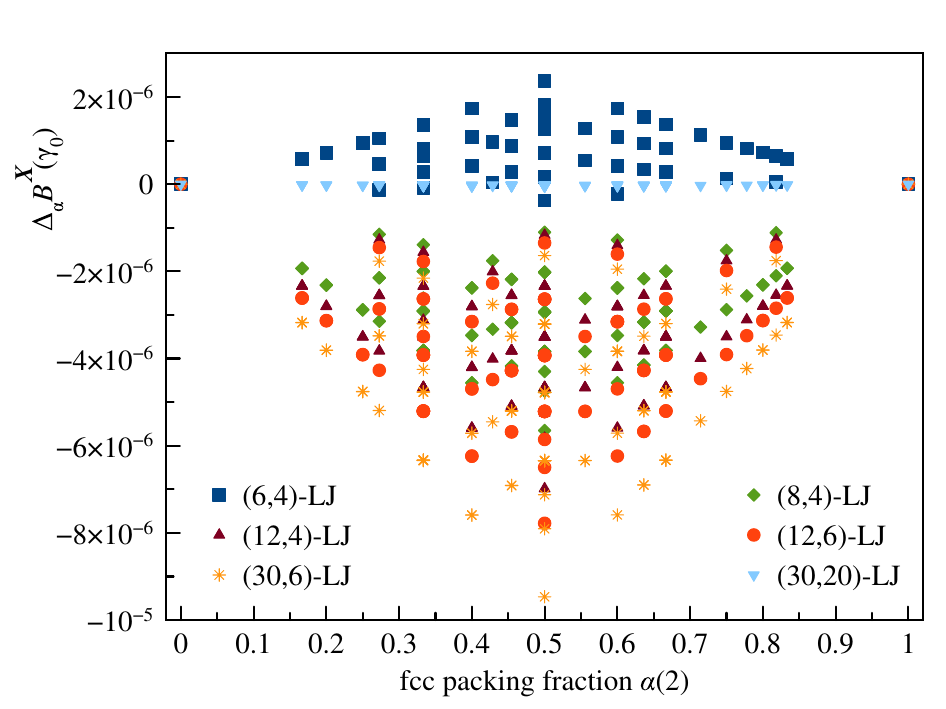}
b)\includegraphics[width=.47\columnwidth]{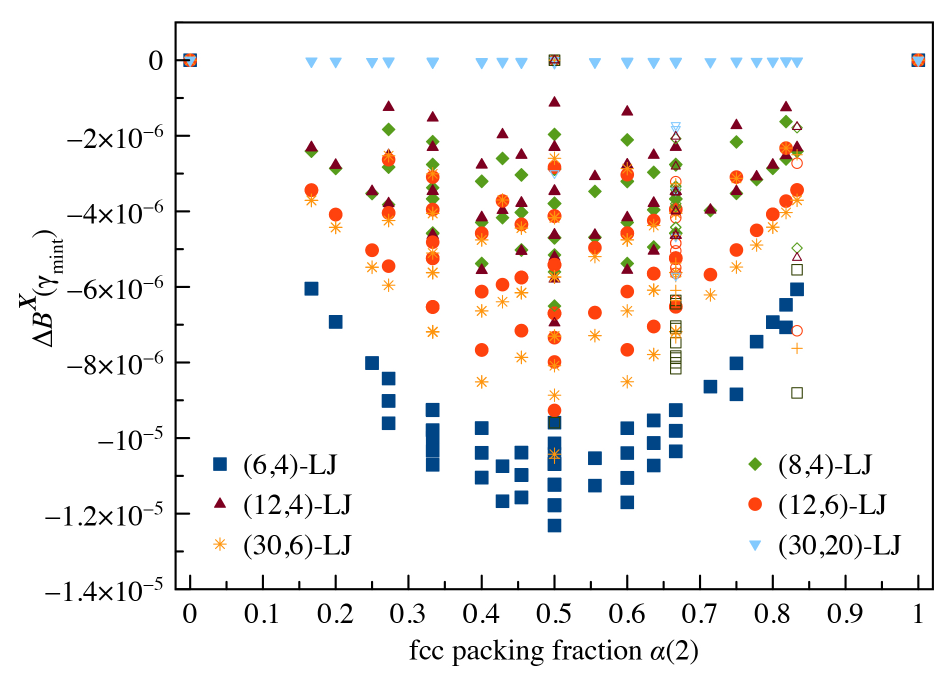}
\caption{(a) Deviation of bulk moduli of Barlow packings from the line connecting the fcc and hcp values for various $(n,m)$-LJ potentials (case 1). (b) Same as for (a) but for case 2 (closed symbols) and case 3 (open symbols).}
\label{fig:BMdiffpack2}
\end{figure}

Figure \ref{fig:BMdiffpack}a shows the bulk moduli for various $(n,m)$-LJ potentials as a function of the fcc stacking fraction $\alpha(2)$, again showing this strong linear behavior. One cannot distinguish the differences in the bulk moduli for the different cases 1-3 from this graph. Figure \ref{fig:BMdiffpack}b therefore shows the differences between the bulk moduli with optimized $\gamma$-values with respect to case 1 ($\gamma_i=\gamma_0=\sqrt{\frac{2}{3}}$). These differences are 8 orders of magnitude smaller than the actual bulk modulus values. We therefore see already some numerical noise for the LJ potentials with large repulsive walls ($n=30$). The RMS error for the deviation from the linear line in Figure \ref{fig:BMdiffpack}a is again very small, i.e RMS = $4.09\times 10^{-6}$ for the (12,6)-LJ potential, and the distribution of the differences from the linear behavior as shown in Figure \ref{fig:BMdiffpack2} exhibits the same features as discussed for the other properties. Note the different behavior in the deviations between the case 1 and 2, i.e. compare Figures \ref{fig:BMdiffpack2} a and b.

As the (12,6)-LJ potential is rather close to the exact potential curve for solid argon in the long-range\cite{Smits2021}, we may compare the calculated bulk modulus to more accurate calculations\cite{Schwerdtfeger-2016} and to experimental results \cite{Dorner-1971,Batchelder-1979}. For the fcc lattice we have $B^{*}=106.327846$ (in reduced units) using the (12,6)-LJ potential, which translates into a value of $B=B^{*}\epsilon/r_e^3=3.947$ GPa using the equilibrium distance for Ar$_2$ of $r_e=3.7624$  \AA~ and the dissociation energy of $\epsilon=1188.5$ J/mol from accurate relativistic coupled-cluster calculations of Jaeger et al.\cite{Jaeger2009}. This compares well with the more accurate two-body potential result of 3.732 GPa.\cite{Schwerdtfeger-2016} The lower temperature experimental values are 2.64 GPa\cite{Batchelder-1979} by Batchelder et al. and 2.38(1.7) GPa by Dorner et al.\cite{Dorner-1971} Thess significantly lower values are a manifestation of many-body and vibrational effects.\cite{Schwerdtfeger-2016} For example, using a many-body expansion including anharmonic vibrational effects results in a bulk modulus for fcc argon of 2.811 GPa and a difference to the hcp value of only -0.0017 GPa\cite{Schwerdtfeger-2016} in qualitative agreement with the (12,6)-LJ result. Such small differences are perhaps beyond current experimental limits. 

\subsection{Vibrational frequencies}

The Einstein (or Debye) model for vibrational frequencies generally provides a reasonable approximation for estimating zero-point vibrational energies, specific heats, and entropic contributions to the free energy\cite{Mermin}. The corresponding results for the different LJ potentials are listed in Table \ref{tab:freq}. For example, using the (12,6)-LJ potential with $r_e=3.7618~\AA$, $\epsilon=1.1900$ kJ/mol,\cite{Jaeger2009} and $M=39.948$ g/mol, Eq.~\eqref{eq:ZPVE} yields a harmonic zero-point vibrational energy of 79.6 cm$^{-1}$ for fcc argon. This value is in reasonable agreement with the more accurate two-body result of 68.0 cm$^{-1}$ obtained in our previous work.\cite{Schwerdtfeger-2016} However, as we shall see, for zero-point vibrational energy or entropy differences between the two quasi-degenerate phases fcc and hcp, this simple approximation breaks down requiring a more accurate treatment of phonon dispersion. We therefore only discuss case 1 for the Einstein frequencies and introduce phonon dispersion for the two extreme Barlow structures fcc and hcp.

\begin{figure}[htb!]
\centering
a)\includegraphics[width=.47\columnwidth]{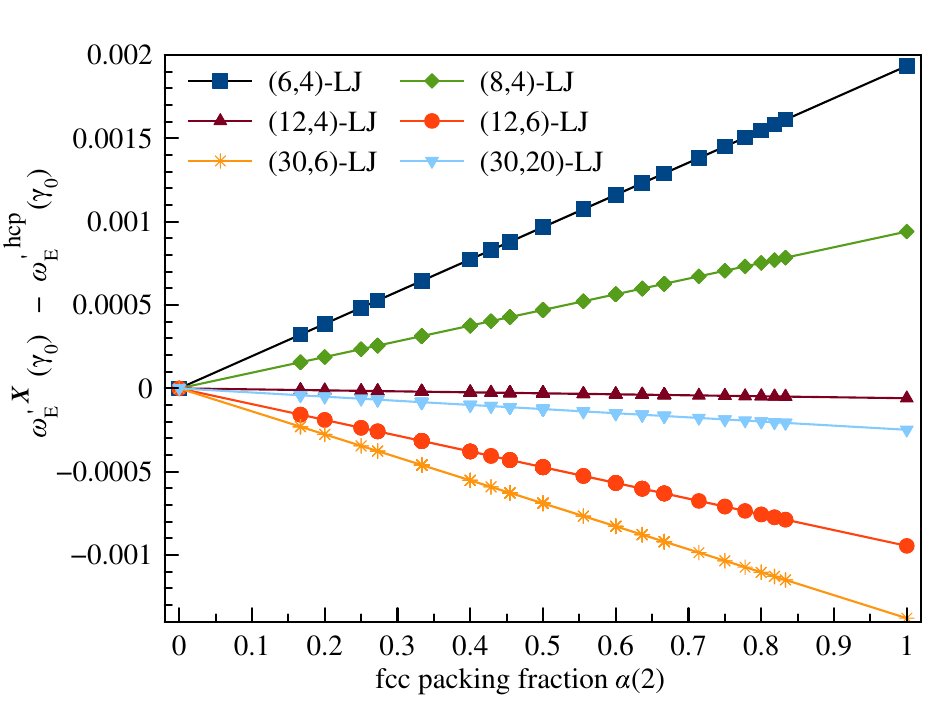}
b)\includegraphics[width=.47\columnwidth]{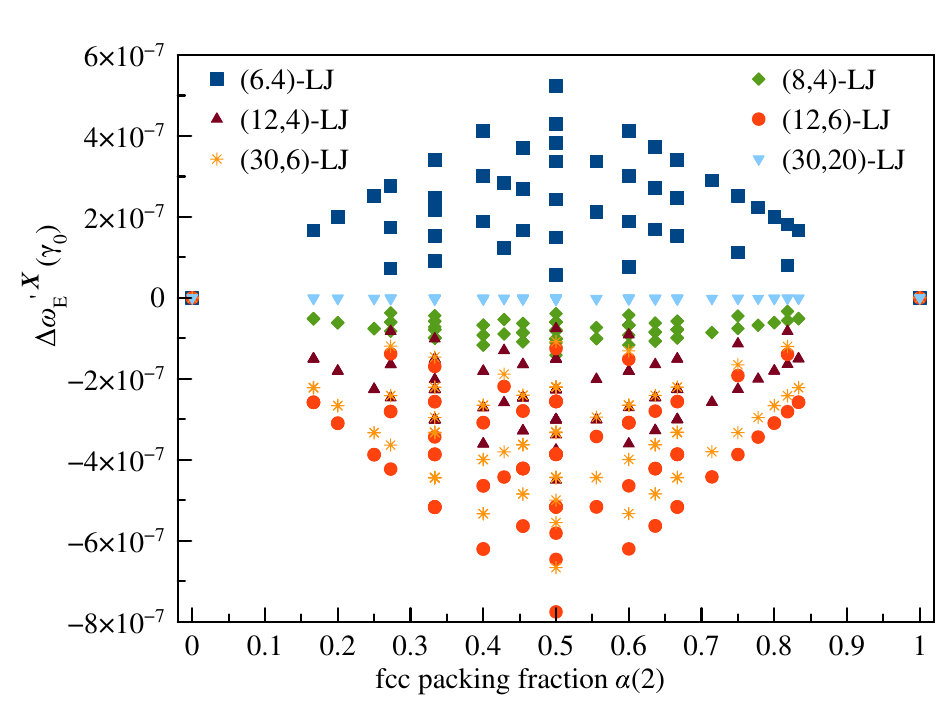}
\caption{(a) Differences in Einstein frequencies (in dimensionless units) for Barlow packings compared to the hcp structure ($\alpha(2)=0$) as a function of the fcc stacking fraction $\alpha(2)$ for various $(n,m)$-LJ potentials. $\gamma_0=\sqrt{\frac{2}{3}}$ is used (case 1). (b) Deviation of Einstein frequencies of Barlow packings from the line connecting the fcc and hcp values for various $(n,m)$-LJ potentials.}
\label{fig:Einsteinpack}
\end{figure}

The calculated Einstein frequencies are shown in Figure \ref{fig:Einsteinpack}a. They exhibit the same nearly perfect linear behavior observed for the lattice energies, with an RMS error of only $4.04\times10^{-7}$ for the (12,6)-LJ potential. The corresponding distribution of the residuals from the fitted line is shown in Figure~\ref{fig:Einsteinpack}b. Consequently, the zero-point vibrational energy, $E_0=\frac{3}{2}\omega_E'$ (in dimensionless units), also follows an almost perfectly linear dependence on the stacking parameter. We briefly note that replacing $\gamma_0$ by $\gamma_{\mathrm{min}}$ produces only negligible changes in the Einstein frequencies. These changes are far too small to affect the essentially perfect linear behavior and merely alter the distribution of the residuals. 
\begin{center}
\begin{table}[htb!] 
\setlength{\tabcolsep}{4pt}
    \fontsize{7.5}{12}\selectfont
\begin{tabular}{|c|c|c|c|c|c|c|c|c|} 
 \hline
 LJ & \multicolumn{3}{c}\textit{Einstein} && \multicolumn{3}{c} \textit{Phonon dispersion} &\\
  \hline
$(n,m)$ & $\omega^{'}_\text{hcp}$ & $\omega^{'}_\text{fcc}$ &
$\Delta\omega^{'}$ & $\Delta S$ & $\omega^{'}_\text{hcp}$ & $\omega^{'}_\text{fcc}$ &
$\Delta\omega^{'}$ & $\Delta S$\\
\hline
 $(6,4)$  & 38.5165422 & 38.5184768 & 1.9346$\times 10^{-3}$  &-1.5068$\times 10^{-4}$& $7.0560514$ & $6.9453406$ & $-0.11071$ & $0.118253$ \\
 $(8,4)$  & 30.7329725 & 30.7339135 & 9.4102$\times 10^{-4}$  &-9.1856$\times 10^{-5}$& $8.4748053$ & $8.3468950$ & $-0.12391$ & $0.112104$ \\
 $(12,4)$ & 28.4954081 & 28.4953490 & -5.9129$\times 10^{-5}$ &6.2254$\times 10^{-6}$& $11.0551343$ & $10.9009752$ & $-0.15416$ & $0.102934$ \\
 $(12,6)$ & 21.7963095 & 21.7953645 & -9.4504$\times 10^{-4}$ &1.3008$\times 10^{-4}$& $13.9503325$ & $13.7628320$ & $-0.18750$ & $0.099128$ \\
 $(30,6)$ & 30.7370491 & 30.7356701 & -1.3790$\times 10^{-3}$ &1.3460$\times 10^{-4}$& $26.9804930$ & $26.6397369$ & $-0.34076$ & $0.092906$ \\
 $(30,20)$& 49.0421487 & 49.0419002 & -2.4847$\times 10^{-4}$ &1.5200$\times 10^{-5}$& $55.5361793$ & $54.8483352$ & $-0.68784$ & $0.091216$ \\
\hline
\end{tabular}
\caption{Harmonic frequencies $\omega^{'}$ for the Einstein model and for the phonon dispersion (averaged over all frequencies) for the fcc and hcp lattices and differences $\Delta\omega^{'}=\omega^{'}_\text{fcc}-\omega^{'}_\text{hcp}$ for various $(n,m)$-LJ potentials (in dimensionless units) and entropy differences $\Delta S=S_\text{fcc}-S_\text{hcp}$ in the high temperature limit $T^{'}\rightarrow\infty$ in units of $k_B$ according to eq.\eqref{eq:PhonoEntropypHT3}.}
\label{tab:freq}
\end{table}
\end{center}

\begin{figure}[htb!]
\centering
a)\includegraphics[width=.47\columnwidth]{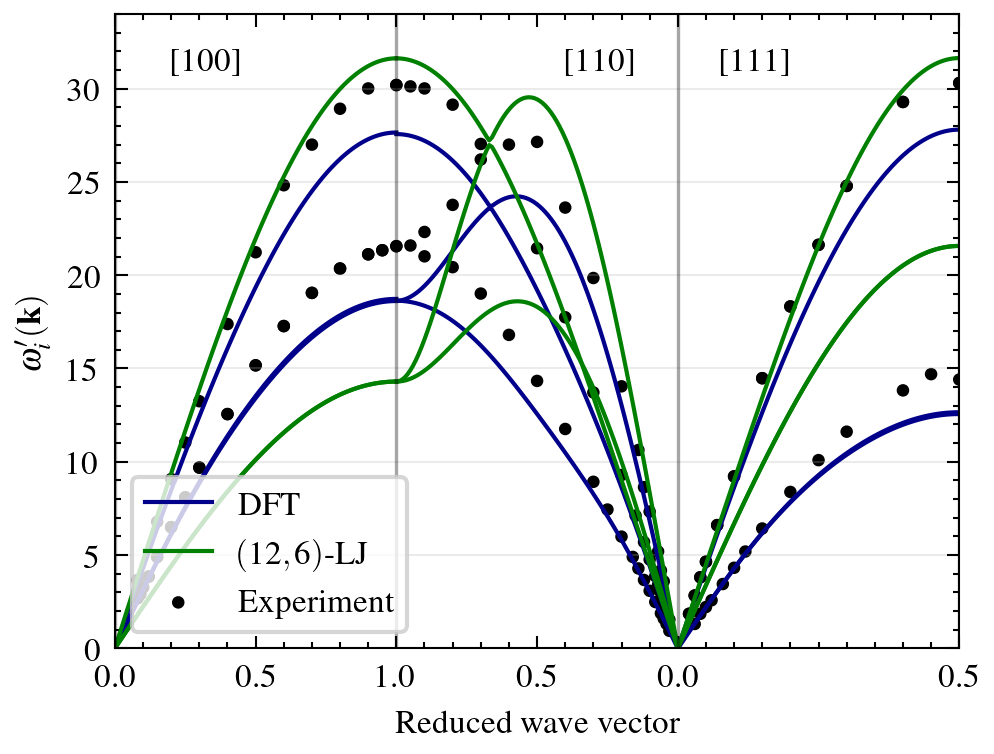}
b)\includegraphics[width=.47\columnwidth]{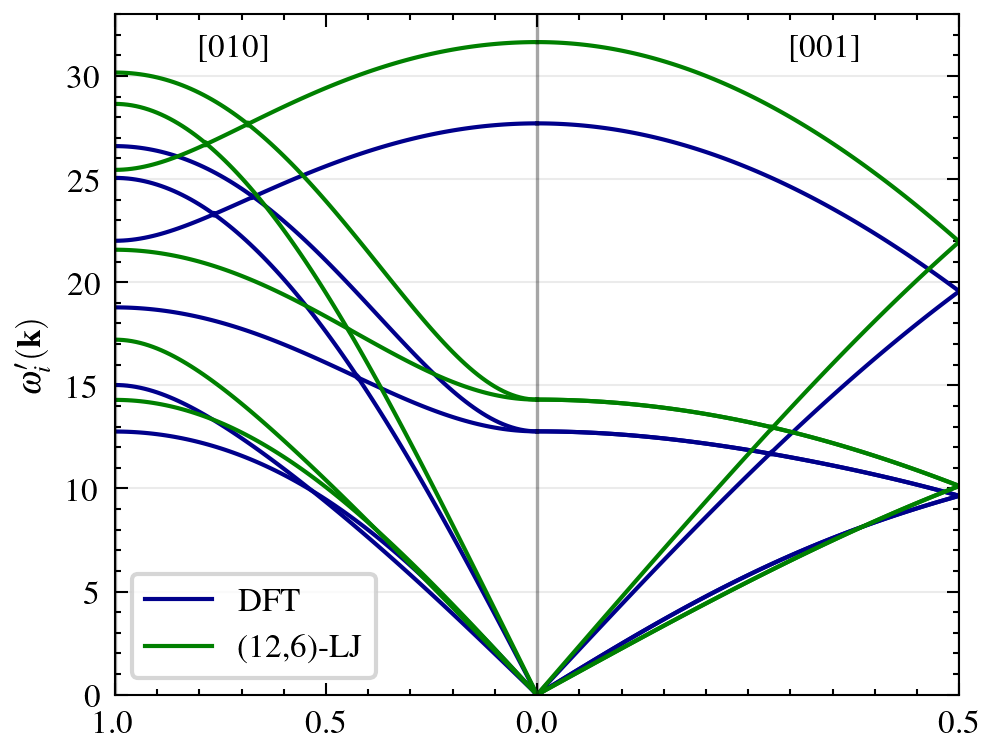}
\caption{Phonon dispersion curves for (a) the fcc and (b) the hcp crystals using the (12,6)-LJ potential. The solid lines are the LJ results, dashed lines are PBE-D3 phonon dispersion and the dots on the fcc plot are the experimental data from Ref. \onlinecite{Fujii1974}.}
\label{fig:phonondispersion}
\end{figure}

The harmonic phonon dispersion curves for the fcc and hcp structures are shown in Figure~\ref{fig:phonondispersion} for the (12,6)-LJ potential. In general, a Barlow packing with $N$ atoms in the primitive unit cell has up to $3N$ phonon branches; thus hcp exhibits six branches. A smaller number of branches may occur because of symmetry. For example, fcc has only three branches since its primitive unit cell contains a single atom.

The Brillouin-zone-averaged phonon frequencies for hcp and fcc are 13.9503325 and 13.7628320, respectively, compared with the corresponding Einstein frequencies of 21.796309541 and 21.795364501 obtained from the exact lattice summations. The difference between the average phonon frequencies, $|\Delta\omega'|$, is 0.1875005, more than two orders of magnitude larger than the corresponding difference between the Einstein frequencies (9.4504$\times10^{-4}$). Furthermore, whereas the Einstein approximation predicts a sign change in $\Delta\omega'$ for sufficiently soft LJ potentials, no such sign change is observed when phonon dispersion is included. This demonstrates that the Einstein approximation must be used with caution when comparing the subtle differences in vibrational properties between nearly degenerate Barlow packings. Nevertheless, it is reasonable to expect that the nearly perfect linear dependence on the stacking parameter observed for the Einstein frequencies in Figure~\ref{fig:Einsteinpack}a will also be retained for the Brillouin-zone-averaged phonon frequencies. A computer program for calculating phonon dispersion in general Barlow packings is currently under development in our group.

\subsection{Entropy}

From the Shannon information entropy,\cite{Chakrabarti2005} one might naively expect that quasi-degenerate Barlow packings with random stacking sequences ${X_1,X_2,\ldots}$, where $X_i\in{A,B,C}$ and $X_i\ne X_{i+1}$, should be entropically favored. However, the information entropy associated with a sequence of stacking letters is fundamentally different from the thermodynamic entropy of a real quantum solid. The relationship between Gibbs and Shannon entropies in the context of a vibrating solid has recently been discussed by Huang and Widom.\cite{Huang2022}

Furthermore, Eq.~\eqref{eq:PhonoEntropy1} shows that $\Delta S_{\rm vib}^{\textbf{\textit{X}}}(T)\rightarrow 0$ as $T\rightarrow 0$ K for a perfect crystal. This is consistent with the third law of thermodynamics, according to which the entropy of a crystalline material with a nondegenerate ground state approaches zero in the zero-temperature limit. Physically, this follows from the quantum-mechanical suppression of thermal excitations of vibrational modes as their energies become large compared with $k_BT$.\cite{Huang2022,HirschfelderCurtissBird1964} Consequently, there is no residual vibrational entropy difference between ideal periodic Barlow packings at 0 K. Entropy differences between different periodic, defect-free Barlow structures therefore arise only at finite temperatures ($T>0$), through their different vibrational spectra.

\begin{figure}[htb!]
\centering
a)\includegraphics[width=.47\columnwidth]{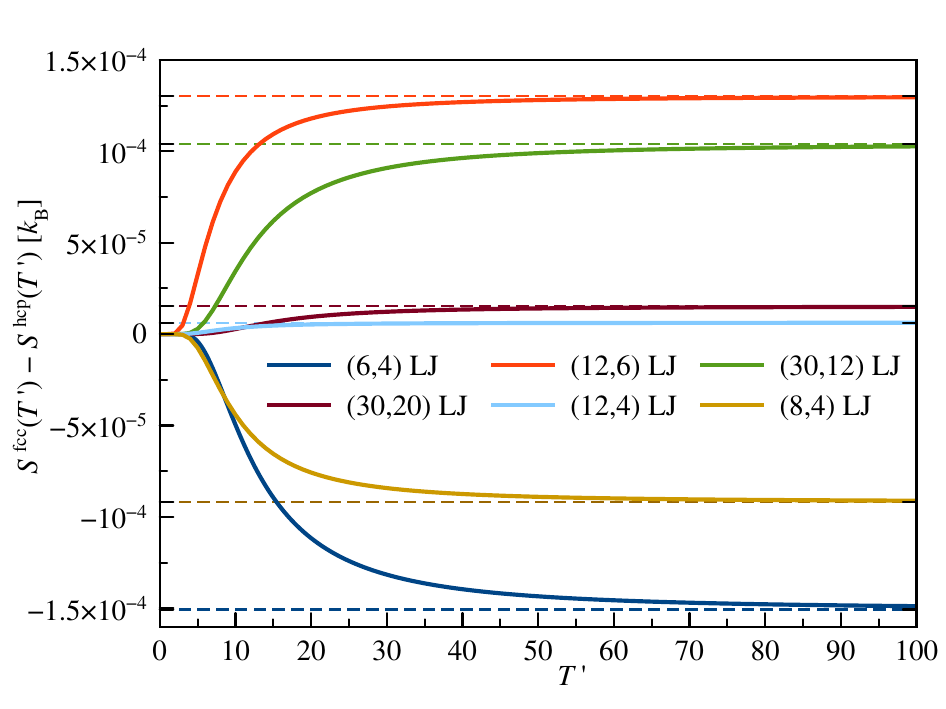}
b) \includegraphics[width=.47\columnwidth]{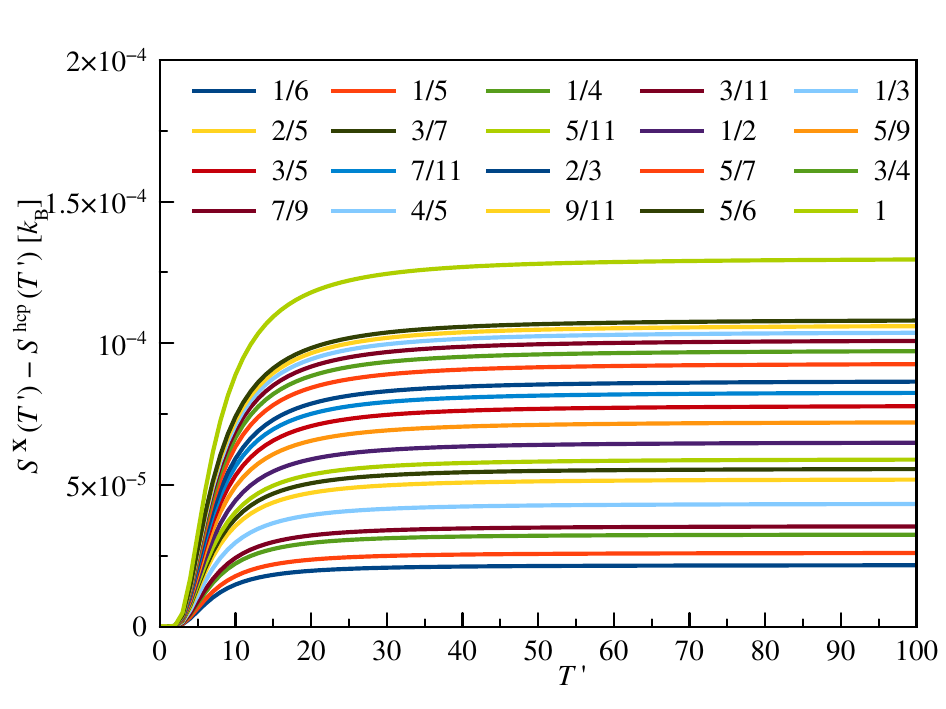}
\caption{(a) Entropy difference $\Delta S(T')= S^\text{fcc}(T') - S^\text{hcp}(T')$ for different temperatures $T'$ and $(n,m)$-LJ potentials. The dashed lines represent the high temperature limits $T'\rightarrow\infty$. (b) Entropy difference $\Delta S(T')$ for the $(12,6)$-LJ potential and various Barlow packings with different fcc stacking fractions $\alpha(2)$.}
\label{fig:vibentropy1}
\end{figure}

\begin{figure}[htb!]
\centering
\includegraphics[width=.47\columnwidth]{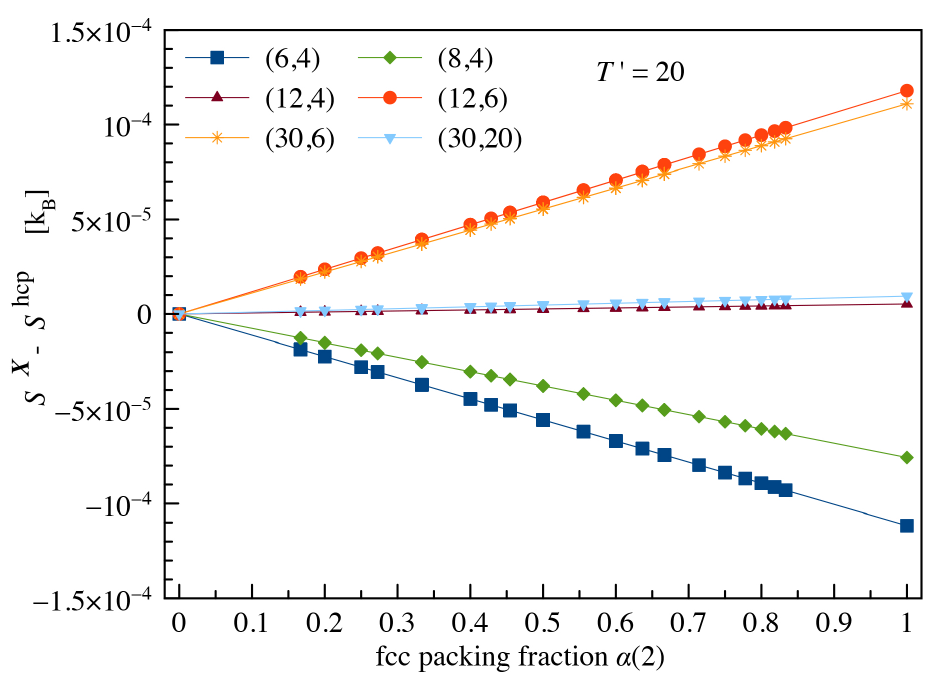}
\caption{Vibrational entropy difference at $T'=20$ with respect to the fcc stacking fraction $\alpha(2)$ for various LJ potentials.}
\label{fig:vibentropy2}
\end{figure}

The entropy difference, $\Delta S(T')= S^\text{fcc}(T') - S^\text{hcp}(T')$, as a function of the dimensionless temperature $T'$ for various $(n,m)$-LJ potentials within the Einstein vibrational model is shown in Figure~\ref{fig:vibentropy1}. The corresponding entropy differences at a fixed temperature of $T'=20$ are presented in Figure~\ref{fig:vibentropy2}. This temperature is close to the melting (triple) point, $T'=23.941$ (see Section~\ref{subsec:entropy}), with $\Delta S(T')$ close to the high-temperature limit for the (12,6)-LJ potential.

\begin{figure}
    \centering
a)    \includegraphics[width=.45\columnwidth]{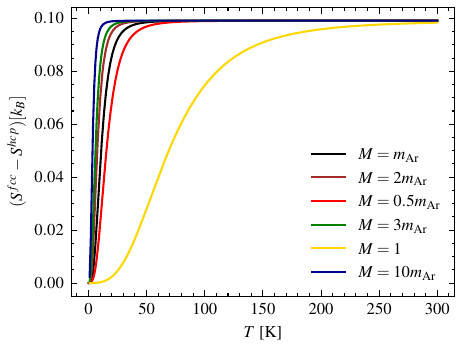} b)    
\includegraphics[width=.45\columnwidth]{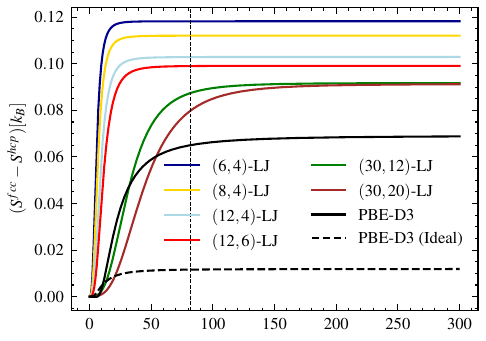}
    \caption{(a) Mass ($M$) and temperature $T$ dependency of the entropy difference between fcc and hcp due to harmonic phonon contribution using the $(12,6)$-LJ potential. The parameters of argon are used in the LJ potential and the mass is in terms of the argon mass $m_{\text{Ar}}$. (b) Variation of the entropy difference between fcc and hcp due to harmonic phonon contribution with temperature $T$ for different $(n,m)$-LJ potentials and PBE-D3 for solid argon. PBE-D3 (ideal) describes the entropy fcc/hcp difference with respect the ideal hcp structure ($\gamma_T=\sqrt{\tfrac{8}{3}}$). The dashed vertical line is at the melting point of argon.}
    \label{fig:PhononEntropy}
\end{figure}

Once again, the entropy differences exhibit an almost perfectly linear dependence on the stacking parameter across the Barlow packings. Moreover, $|\Delta S(T')|$ increases monotonically with temperature and reaches its maximum in the high-temperature limit, $T'\rightarrow\infty$. Figure~\ref{fig:PhononEntropy}a shows the $\Delta S(T')$ curves for different atomic masses, demonstrating the usefulness of our definition of a mass-weighted temperature. In the high-temperature limit (Eq.~\eqref{eq:PhonoEntropypHT3}), however, the Einstein model predicts entropy differences that are at least one order of magnitude smaller than those obtained from previous Monte Carlo simulations.\cite{Mau1999,Travesset2014} As discussed in more detail below, this discrepancy can be attributed to the neglect of phonon dispersion. In addition, Figure~\ref{fig:vibentropy2} suggests that, within the Einstein approximation, vibrational entropy favors the hcp structure for sufficiently small values of the attractive LJ exponent $m$. To understand this apparent failure of the Einstein model, we next examine the role of phonon dispersion in the high-temperature limit in more detail.


Within the Einstein model, the high-temperature limit is given by 
\begin{equation} \label{eq:PhonoEntropy4}
\Delta S_{\rm vib}^\textbf{\textit{X}}(T'\rightarrow\infty)= 3 {\rm ln}\left( \frac{\omega_E^{'\text{hcp}}}{\omega_E^{'\textbf{\textit{X}}}}\right) = 3 {\rm ln}\left(  1+\cfrac{\omega_E^{'\textbf{\textit{X}}}-\omega_E^{'\text{hcp}}}{\omega_E^{'\text{hcp}}}\right)^{-1}\approx -3\frac{\omega_E^{'\textbf{\textit{X}}}-\omega_E^{'\text{hcp}}}{\omega_E^{'\text{hcp}}}
\end{equation}
where the final expression follows from a first-order Taylor expansion of the logarithm. This implies that $\Delta S_{\rm vib}^\textbf{\textit{X}}(T'\rightarrow\infty)=0$ only if ${\omega_E^{'\text{hcp}}}={\omega_E^{'\textbf{\textit{X}}}}$. Consequently, the entropy difference between hcp and any other Barlow packing becomes extremely sensitive to even minute differences in the corresponding Einstein frequencies. 

As shown in the previous section, although the Einstein frequency itself provides a reasonable estimate of the zero-point vibrational energy, the differences between the Einstein frequencies of nearly degenerate structures are not described with sufficient accuracy. This limitation is directly reflected in the entropy differences. Indeed, comparison of the high-temperature fcc-hcp entropy differences obtained from the Einstein model with those including full phonon dispersion (Table \ref{tab:freq}) shows that the Einstein approximation consistently underestimates the entropy difference by approximately three orders of magnitude. Moreover, once phonon dispersion is included, the high-temperature entropy differences become remarkably similar for the different $(n,m)$-LJ potentials. 

To further assess the validity of the entropy calculations based on LJ potentials, we carried out more computationally demanding DFT calculations for solid argon to obtain phonon frequencies and vibrational entropies for the fcc and hcp phases. It is well established that the (12,6)-LJ potential provides a reasonable description of solid argon under ambient conditions.\cite{Dobbs1957} For example, the phonon dispersion curves of optimized argon obtained from DFT, shown in Figure \ref{fig:phonondispersion}, closely resemble those calculated with the (12,6)-LJ potential. However, when considering entropy differences, more significant deviations from the simple LJ model become apparent, as illustrated in Figure~\ref{fig:PhononEntropy}. In particular, the convergence to the high-temperature limit is considerably slower at the PBE-D3 level of theory, and the overall fcc-hcp entropy difference is reduced compared with the (12,6)-LJ potential. For example, the high-temperature entropy difference predicted by PBE-D3 is 30.6\% smaller than the corresponding (12,6)-LJ value.

Because of this, it is important to consider two potential reasons for discrepancies when comparing different calculations of the entropy difference. First of all, as can be seen from Figure \ref{fig:phonondispersion}, at the $\Gamma$ point all phonon frequencies tend to zero, but for the low-frequency modes, we can approximate the terms in Equation \eqref{eq:PhonoEntropypdisp} $1-e^{-x}\approx x$ and $x/(e^{x}-1)\approx 1-(x/2)$ such that each individual contribution is
\begin{equation}
    S_{\text{vib}}(T',\mathbf{k}) \approx k_{B} \sum_{\mu}\left[1-\ln(\beta'\omega'_{\mu}(\mathbf{k}))\right]
\end{equation}
Hence, when $\omega'\to 0$, we have $S_{\text{vib}}\to\infty$ diverging logarithmically. Therefore, for very low positive frequencies the entropy becomes extremely large and sensitive to small numerical errors in the calculated phonon frequencies. Thus, the soft phonons can contribute significantly to differences in vibrational entropy. Consequently, even a relatively small number of soft modes can make a substantial contribution to calculated free-energy differences. Note that in the Debye model, the phonon density of states is proportional to the square of the frequency,\cite{Kantorovich2004quantum} and as a consequence the soft-phonon contribution to the entropy is integrable and does not lead to an infinite vibrational entropy. Furthermore, we can see that for each term in the sum, $S_{\text{vib}}$, the variation of the entropy with respect to a change in the frequency gives
\begin{equation}
    \frac{\partial S_{\text{vib}}}{\partial \omega'} \approx \frac{-k_{B}}{\omega'}
\end{equation}
This means that even small differences in the soft phonons may lead to a non-negligible change in the entropy difference due to its high sensitivity to the frequencies.

Second, slight distortions in the optimized structures may produce phonons with a small difference compared to the ideal lattices, hence producing a another source of variation for the entropy. For example, the entropy difference calculated with the PBE-D3 optimized structures produces a high-temperature limit $\Delta S/k_{B}=0.068812$, whereas the difference between the ideal fcc and hcp (only optimizing lattice parameter $a$) is only $\Delta S/k_{B}=0.011848$. This is shown in Figure \ref{fig:PhononEntropy}b.

It is also instructive to compare the fcc/hcp entropy differences predicted by our (optimized) DFT approximation with those obtained from the (12,6)-LJ potential for solid argon. For several representative temperatures, the entropy difference, $\Delta S = S^\mathrm{fcc} - S^\mathrm{hcp}$ (in units of $k_B$), is found to be: $0.003793$ ($0.04362693$) at $T=10$ K, $0.02821$ ($0.08562311$) at $T=20$ K, $0.054018$ ($0.0977023$) at $T=40$ K, $0.061801$ ($0.09881417$) at $T=60$ K, and $0.064872$ ($0.09902527$) at $T=80$ K, where the corresponding (12,6)-LJ values are given in parentheses. 
Concerning previous work on entropy differences, Mau and Huse showed from Monte-Carlo simulations of hard-spheres close to the maximum packing density that the entropy of the fcc lattice is greater than that of hcp, i.e. $\Delta S=1.156\times 10^{-3}k_B $ per sphere.\cite{Mau1999,Elser2014} Using their parametrization one can in fact show that the entropy of a Barlow packing is correlated to the fcc stacking fraction\cite{Cooper2025} and lying in-between that of fcc and hcp\cite{Mau1999}. Travesset\cite{Travesset2014} used classical dynamic lattice theory (CDLT) for a (12,6)-LJ potential and obtained very similar results compared to Mau and Huse, i.e. $S^\text{fcc}-S^\text{hcp}=1.4754 \times 10^{-3}k_B$ in the phonon hard-sphere limit. He argued that the higher entropy of fcc is the reason why it is the equilibrium phase at sufficiently high temperatures, albeit bcc is often preferred close to the melting line usually explained by Landau theory\cite{Alexander1978}. For a solid, most of the entropy contribution comes from thermal fluctuations described by the vibrating atoms, see appendix \ref{app:LatticeThermodynamics}. This is especially the case for insulators at low temperatures where electronic entropy contributions can be neglected. We therefore focused our attention solely on the vibrational contribution to the entropy for Barlow packings described by Lennard-Jones interactions. We note, however, that our simplified (static) model neglects dynamic and quantum effects\cite{Wiebe2020}, the former becoming important at higher temperatures close to the melting point, which requires a larger supercell treatment. Most notably, for the (30,20)-LJ potential which is close to the hard-sphere limit we have $\Delta S_{\rm vib}^\textbf{\textit{X}}(T'\rightarrow\infty)=9.9128\times 10^{-2}k_B$ from phonon dispersion calculations in the high temperature limit as compared to the much lower value of Mau and Huse.\cite{Mau1999} This demonstrates the fcc/hcp entropy difference is very sensitive to the theoretical model applied, and because of its small value, remains a challenge to quantum theoretical methods.

\subsection{Free energy}
\label{sec:FreeEnergy}

We can now turn to the question why fcc and hcp are the preferred crystal packings over the infinitely many Barlow structures even so they have equal packing density and are energetically quasi-degenerate. For this we consider the free energy $G$ (per atom), 
\begin{equation} \label{eq:FreeEnergy}
G = U + PV - TS
\end{equation}
with the internal energy of the system
\begin{equation} \label{eq:InternalEnergy}
U = E_\text{coh} + E_0  + E_\text{th}
\end{equation}
In our simplified static model used in this work we neglect dynamic effects, such as temperature effects on lattice constants which changes the internal energy, anharmonicity effects and other smaller contributions. Nevertheless, we may assume a near linear behavior of the internal energy including thermal contributions ($E_\text{th}$), volume and (to a certain approximation)\cite{Mau1999,Cooper2025} the entropy as well.
We therefore have
\begin{equation} \label{eq:FreeEnergylinear}
G^\textbf{\textit{X}}(s,\gamma,P,T) \approx G^\text{hcp}(s,\gamma,P,T) + \alpha^\textbf{\textit{X}}(2) \left[ G^\text{fcc}(s,\gamma,P,T) - G^\text{hcp}(s,\gamma,P,T)\right]
\end{equation}
We note that for the free energy the internal energy change associated with thermal phonon occupation is balanced by the entropy term. Assuming \eqref{eq:FreeEnergylinear} to be valid (which requires testing for more complex systems such as metals), we have the Barlow packings lying in-between the two limiting phases of either fcc or hcp at a given temperature $T$, i.e.
\begin{equation} \label{eq:FreeEnergyinequ}
G^\text{Ph1}(s,\gamma,P,T) \le G^\textbf{\textit{X}}(s,\gamma,P,T) \le G^\text{Ph2}(s,\gamma,P,T)
\end{equation}
where (Ph1,Ph2) either (fcc,hcp) or (hcp,fcc). Exceptions are for the near degeneracy case, i.e. when $G^\text{Ph1}(s,\gamma,T) = G^\text{Ph2}(s,\gamma,T)$ (co-existence line) or very close to, at certain temperatures and pressures. This implies that the free energy will prefer either the fcc or the hcp phase over any of the other Barlow packings in most cases. However, the difference in free energies between the Barlow packings is predicted to be very small, e.g. for solid argon the fcc phase dominates over the hcp phase by only a few J/mol\cite{Schwerdtfeger-2016}. Here we mention that experiments by solid argon a phase transition from fcc to hcp starts at 4.2 GPa with both phases coexisting in a very wide pressure domain up to >100 GPa\cite{Dewaele2021}. It is assumed that in this co-existence region of the two phases the bulk is dominated by stacking faults of the two-dimensional close-packed hexagonal layers, i.e. by different Barlow packings. In this respect kinetic effects at finite temperatures and pressures become important as well giving access to metastable Barlow structures apart from the extremal cases fcc and hcp.

\section{Conclusions}

We have shown, using exact lattice summations for Lennard-Jones (LJ) potentials, that a remarkably linear relationship with respect to the fcc stacking fraction $\alpha(2)$ exists for a number of important solid-state properties of Barlow packings. The use of rapidly convergent Bessel-function expansions for the lattice sums made it possible to resolve the extremely small energy differences between the competing bulk phases with high numerical accuracy. Based on these findings, we hypothesize that the free energy likewise exhibits an almost perfectly linear dependence on $\alpha(2)$ at low temperatures and pressures, with fcc and hcp representing the two limiting cases. Consequently, there is little thermodynamic driving force to stabilize intermediate stacking sequences unless the free energies of fcc and hcp become nearly degenerate. Furthermore, recent studies\cite{Schwerdtfeger-2026,Robles-Navarro2026,maltby2026} showed that, along one of the possible transformation pathways -- the Burgers-Bain path -- the activation barrier for the fcc-hcp phase transition in solid argon is below 1 kJ mol$^{-1}$. We expect similarly low barriers for the other rare-gas solids, with the various Barlow packings corresponding to local minima on the Gibbs free-energy hypersurface. The combination of an almost linear free-energy dependence on $\alpha(2)$ and the low transition barriers suggests that specific Barlow packings are only weakly stabilized relative to one another. Consequently, it is likely to be experimentally challenging to freeze in or isolate individual Barlow stacking sequences in such systems.

Although Eq.~\eqref{eq:FreeEnergylinear} is expected to provide a good approximation for weakly interacting systems such as argon, it remains unclear how many-body interactions -- known to play a crucial role in metallic and open-shell materials -- affect the energetics of Barlow stacking sequences. Testing this hypothesis for real materials, such as the transition metals, therefore represents a major challenge, notwithstanding the well-known limitations of DFT.\cite{DFTExchange2022} Even within DFT, accurately resolving the extremely small free-energy differences between competing stackings is computationally demanding. Furthermore, a meaningful numerical comparison of periodic Barlow packings requires a common supercell that can accommodate all stacking periods under consideration, i.e., a supercell whose periodicity is given by the least common multiple of the individual periods, $N_i$.

We also demonstrated that accurately determining the small vibrational entropy differences between the fcc and hcp structures is considerably more demanding than predicting their vibrational energies, especially if these differences are very small as for the different Barlow packings. For more accurate entropy differences one requires to minimize lattice constant for the Gibbs free energy at constant pressure and temperature including phonon dispersion which is a challenging computational task. For further improvements one has to switch to self-consistent phonon theory, which renormalizes the phonon frequencies through anharmonic interactions,\cite{Ehsan2021} or to more elaborate path-integral molecular dynamics methods.\cite{Marx1996} Nevertheless, our calculations confirm the established result that the vibrational entropy of fcc is slightly larger than that of hcp. The present treatment neglects phonon dispersion for Barlow packings other than fcc and hcp, finite-pressure and finite-temperature effects on the phonon spectrum, anharmonic contributions, and many-body interactions and dynamic effects through minimizing the Gibbs free energy. Extending the present formalism to include such effects will be the subject of future work. Whether kinetic effects, defects, impurities, or even external fields can selectively stabilize particular Barlow stackings remains an interesting question for future experimental and theoretical investigations.

\section*{Supplementary Material}

The supplementary material is available upon request and consists of a more detailed derivation of lattice sum derivatives in SupplMaterial1.pdf, and an Excel spreadsheet file called SupplMaterial2.xlsx containing data for the lattice constants, cohesive energies, bulk moduli, Einstein frequencies and entropies for various Barlow packings and various Lennard-Jones potentials.

\section*{Acknowledgment}
We thank the Sir Neil Waters Trust at Massey University for financial support. We are grateful to Odile Smits (University of Queensland, Brisbane) for valuable discussions.

\appendix
\setcounter{table}{0}
\renewcommand{\thetable}{A\arabic{table}}


\section{Bessel function expansions of lattice sums and their derivatives}
\label{app:Bessel}

Formulae for the expansions of the Barlow lattice sums \eqref{eq:LA}, \eqref{eq:LAA} and \eqref{eq:LAB} in terms of Bessel functions have already been provided in our previous paper\cite{Cooper2025}. For the corresponding lattice sum derivatives with respect to the lattice constants $\{\gamma_i\}$ required for the Newton-Raphson procedure we rewrite the Bessel function expansions in a more convenient short-hand notation and set $\gamma\equiv\gamma_T=\sum_{i=1,...N}\gamma_i$,
\begin{align}\label{eq:latsumshort1}
 L^{A}(s,r,\gamma) &= A_1(s) + \gamma^{2-2s} A_2(s)  + 2\gamma^{1-s} A_3(s) \sum_{k,n\in\mathbb{N}} t_{kn}(s) 
K_{s-1}\left(\gamma p_{kn} \right)\\
\label{eq:latsumshort12}
 L^{AA}(s,r,\gamma) &= \gamma^{2-2s} A_2(s,r)  + \gamma^{1-s} A_3(s) \sum_{k\in\mathbb{Z}} \sum_{n\in\mathbb{N}} t_{kn}(s,r) 
K_{s-1}\left(\gamma p_{kn}(r) \right)\\
\label{eq:latsumshort13}
L^{AB}(s,r,\gamma)&= \gamma^{2-2s} A_2(s,r) 
 + \gamma^{1-s}  A_3(s) \sum_{k\in\mathbb{Z}} \sum_{n\in\mathbb{N}} u_{kn}(s,r) K_{s-1}\left(  \gamma p_{kn}(r) \right).
\end{align}
with the coefficients,
\begin{align}\label{eq:latsumshort1coeff}
\nonumber
A_1(s)&=6\zeta(s) \,L_{-3}(s) \;\;\;,\;\;\; A_2(s,r)=\frac{2\pi}{\sqrt{3}(s-1)}  \left(\zeta(2s-2,r) + \zeta(2s-2,1-r)\right)\\
A_3(s)&=\frac{2}{\Gamma(s)} \left(\frac{2\pi}{\sqrt{3}}\right)^s \;\;\;,\;\;\; t_{kn}(s,r)=u_2(n)n^{(s-1)/2}\left| k+r \right|^{1-s} \\
\nonumber
u_{kn}(s,r)&=u_2(n)\left(\frac{n}  {3(k+r)^2}\right)^{(s-1)/2}\cos{\left(\frac{2\pi n}{3}\right)}
\;\;\;,\;\;\;
p_{kn}(r)=\frac{4\pi}{3}\sqrt{n}\left| k+r \right|
\end{align}
where $A_2(s)=A_2(s,r=0)$, $t_{kn}(s)=t_{kn}(s,r=0)$, $u_{kn}(s)=u_{kn}(s,r=0)$, $p_{kn}=p_{kn}(r=0)$ and $r=(k_2-k_1)/N\in\mathbb{Q}, 0<r< 1$, if we neglect symmetry breaking effects along the $c$-axis, and $r\equiv r_{k_1k_2}=\gamma^{-1}d_{k_1,k_2}$ with $d_{k_1k_2}=\sum_{i=k_1,...,k_2-1} \gamma_i$ otherwise. Here, $u_2(n)$ is the number of representations of the positive integer $n$ by the form $i^2+ij+j^2$. The gamma $\Gamma(s) (s>0)$ and modified Bessel functions $K_s(x)$ may be defined by\cite{andrews1999,abramowitzstegun2008}
\begin{equation}
\label{gamma1}
\Gamma(s) = \int_{[0,\infty)}  t^{s-1}\, e^{-t}\,\ud t \; .
\end{equation}
and
\begin{equation}
\label{bessel1}
\int_{[0,\infty)} x^{s-1}e^{-ax-b/x} \ud x = 2\left(\frac{b}{a}\right)^{s/2} K_s(2\sqrt{ab}) \; .
\end{equation}
with the properties
\begin{equation}
\label{bessel1.5}
K_{s}(z) = K_{-s}(z) \; .
\end{equation}
\begin{equation}
\label{bessel1.6}
K_{s}(z) = K_{s+2}(z)- \frac{2s+2}{z}K_{s+1}(z) \; ,
\end{equation}
and the asymptotic formula,
\begin{equation}
\label{bessel3}
K_s(z) \sim \sqrt{\frac{\pi}{2z}} \, e^{-z} \quad \text{as} \quad z\rightarrow \infty, \quad (\,|\text{arg} \,z| < 3\pi/2) \; .
\end{equation}
The $L_{-3}$ function is defined by
\begin{align}
L_{-3}(s) &= 3^{-s}  \left[ \zeta(s,\tfrac{1}{3}) -  \zeta(s,\tfrac{2}{3})\right]. 
\label{zeta3}
\end{align}
where $h(s,x)$ is the Hurwitz zeta function, which for $x=1$ reduces to the Riemann zeta function, i.e. $\zeta(s,1)=\zeta(s)$.

We also introduced a second set of Bessel function expansions in our previous paper\cite{Cooper2025},
 \begin{align}\label{eq:latsumshort2}
L^{AA}(s,r,\gamma)&=\gamma^{-2s}B_1(s,r) + \gamma^{-1} B_2(s) +\gamma^{-(s+\tfrac{1}{2})} B_3(s) {\sum_{k,n\in\mathbb{N}}} v_{kn}(s,r) K_{s-\frac{1}{2}} \left(\gamma^{-1} f_{kn}   \right) \\
L^{AB}(s,r,\gamma)&= \gamma^{-1}B_4(s)
		+ \gamma^{-(s+\tfrac{1}{2})} B_5(s) {\sum_{k,n\in\mathbb{N}}} w_{kn}(s,r) K_{s-\frac{1}{2}}\left(\gamma^{-1} g_{kn} \right)
\end{align}
with the coefficients
\begin{align}\label{eq:latsumshort2coeff}
\nonumber
B_1(s,r)&=\left(\zeta(2s,r) + \zeta(2s,1-r)\right) \;\;\;,\;\;\; B_3(s)=\frac{4\pi^s}{\Gamma(s)}=\frac{1}{2}B_5(s)\\ 
B_2(s)&=\frac{2\sqrt{\pi}}{3^{s-\frac32}} \frac{\Gamma(s-\frac12)}{\Gamma(s)} \; 
\zeta(s-\textstyle\frac12)\, \left(\zeta(s-\textstyle\frac12,\textstyle{\frac13}) - \zeta(s-\textstyle\frac12,\textstyle{\frac23})\right)\\
\nonumber
B_4(s)&=3\sqrt{\pi}\,(1 - 3^{-(s-1/2)})\frac{\Gamma(s-\frac12)}{\Gamma(s)}    \, 
\zeta\left(s-\tfrac{1}{2}\right) \left( \zeta(s-\tfrac12,\tfrac13) - \zeta(s-\tfrac12,\tfrac23) \right)\\
\nonumber
v_{kn}(s,r)&= \frac{k^{s-\frac12}}{n^{(2s-1)/4}} u_2(n) \cos(2\pi kr) \;\;\;,\;\;\; w_{kn}(s,r)=k^{s-\frac12}\cos(2\pi kr) \frac{u_2(3n-2)}{(n-\tfrac{2}{3})^{(2s-1)/4}}\\
\nonumber
f_{kn}&=2\pi k\sqrt{n} \;\;\;,\;\;\; g_{kn}= 2\pi k \sqrt{n-\tfrac{2}{3}}
\end{align}
For the lattice sum derivatives we require the derivatives of the Bessel functions,
\begin{align}\label{eq:derivBes1}
\partial_x K_{s}(x)&=-K_{s+1}(x)+\frac{s}{x}K_{s}(x) \\
\label{eq:derivBes2}
\partial_x K_{s}(x)&=-K_{s-1}(x)-\frac{s}{x}K_{s}(x) \,.
\end{align}

If we neglect symmetry breaking effects along the $c$-axis we have to consider the derivatives with respect to the single parameter $\gamma\equiv\gamma_T$ only. For this, the lattice sum derivatives are easily derived. For the first set we get,
\begin{align}\label{eq:latsumderiv1a}
\partial_\gamma L^{A}(s,\gamma)&=2(1-s)A_2(s)\gamma^{1-2s} -2\gamma^{1-s}A_3(s)\sum_{k,n\in\mathbb{N}} t_{kn}(s) p_{kn}
K_s\left(\gamma p_{kn} \right)\\
\partial_\gamma L^{AA}(s,\gamma)&=2(1-s)A_2(s,r)\gamma^{1-2s} -\gamma^{1-s}A_3(s)\sum_{k\in\mathbb{Z}} \sum_{n\in\mathbb{N}} t_{kn}(s,r) p_{kn}(r)
K_s\left(\gamma p_{kn}(r) \right)\\
\partial_\gamma L^{AB}(s,\gamma)&=2(1-s)A_2(s,r)\gamma^{1-2s} -\gamma^{1-s}A_3(s)\sum_{k\in\mathbb{Z}} \sum_{n\in\mathbb{N}} u_{kn}(s,r) p_{kn}(r)
K_s\left(\gamma p_{kn}(r) \right)
\end{align}
and
\begin{align}\label{eq:latsumderiv2a}
\partial^2_\gamma L^{A}(s,r,\gamma)&=(1-2s)\gamma^{-1} \partial_\gamma L^{A}(s,r,\gamma)
+2\gamma^{1-s}A_3(s) \sum_{k,n\in\mathbb{N}} t_{kn}(s) p^2_{kn}
K_{s-1}\left(\gamma p_{kn} \right)\\
\partial^2_\gamma L^{AA}(s,r,\gamma)&=(1-2s)\gamma^{-1} \partial_\gamma L^{AA}(s,r,\gamma)
+\gamma^{1-s}A_3(s)\sum_{k\in\mathbb{Z}} \sum_{n\in\mathbb{N}} t_{kn}(s,r) p^2_{kn}(r)
K_{s-1}\left(\gamma p_{kn}(r) \right)\\
\partial^2_\gamma L^{AB}(s,r,\gamma)&=(1-2s)\gamma^{-1} \partial_\gamma L^{AB}(s,r,\gamma)
+\gamma^{1-s}A_3(s)\sum_{k\in\mathbb{Z}} \sum_{n\in\mathbb{N}} u_{kn}(s,r) p^2_{kn}(r)
K_{s-1}\left(\gamma p_{kn}(r) \right)
\end{align}
For the second set we have,
\begin{align}\label{eq:latsumderiv1b}
\partial_\gamma L^{AA}(\gamma,s)&=
 -\gamma^{-1}L^{AA}(s,r,\gamma) -(2s-1)\gamma^{-2s-1}B_1(s,r)\\
 \nonumber
 &+ \gamma^{-(s+\tfrac{5}{2})} B_3(s) {\sum_{k,n\in\mathbb{N}}} v_{kn}(s,r) f_{kn} K_{s-\frac{3}{2}}(\gamma^{-1} f_{kn} ) \\
\partial_\gamma L^{AB}(\gamma,s)&
=-\gamma^{-1}L^{AB}(s,r,\gamma) + \gamma^{-(s+\tfrac{5}{2})} B_5(s) {\sum_{k,n\in\mathbb{N}}} w_{kn}(s,r) g_{kn} K_{s-\frac{3}{2}}(\gamma^{-1} g_{kn} )
\end{align}
and
\begin{align}\label{eq:latsumderiv2}
\partial^2_\gamma L^{AA}(s,r,\gamma)
&= -2s\gamma^{-2} L^{AA}(\gamma,s)  -2(s+1)\gamma^{-1} \partial_\gamma L^{AA} \\
\nonumber
&+\gamma^{-(s+\tfrac{9}{2})} B_3(s) {\sum_{k,n\in\mathbb{N}}} v_{kn}(s,r) f^2_{kn} K_{s-\frac{1}{2}}(\gamma^{-1} f_{kn} ) \\
\partial^2_\gamma L^{AB}(s,r,\gamma)
&= -2s\gamma^{-2} L^{AB}(\gamma,s)  -2(s+1)\gamma^{-1} \partial_\gamma L^{AB} \\
\nonumber
&+\gamma^{-(s+\tfrac{9}{2})} B_5(s) {\sum_{k,n\in\mathbb{N}}} w_{kn}(s,r) g^2_{kn} K_{s-\frac{1}{2}}(\gamma^{-1} g_{kn} )
\end{align}

We now extend the lattice sum derivatives to the case where we include symmetry breaking in the location of the hexagonal layers along the main $c$-axis. For the first set of derivatives we use eqs.\eqref{eq:latsumshort1}-\eqref{eq:latsumshort13} and remember that $\gamma\equiv\gamma_T$ and $r\rightarrow r_{k_1k_2}=\gamma^{-1}d_{k_1,k_2}$
\begin{align}\label{eq:latsumderiv1sym}
\partial_{\gamma_j} L^\text{A}&(s,\{\gamma_i\})=
2(1-s)\gamma^{1-2s} A_2(s) -2\gamma^{1-s}A_3(s)\sum_{k,n\in\mathbb{N}} t_{kn}(s) p_{kn}
K_s\left(\gamma p_{kn} \right)\\
\partial_{\gamma_j} L^\text{AA}&(s,\{\gamma_i\})= \frac{4\pi}{\sqrt{3}}\gamma^{1-2s}\left\{ H_-(2s-1,r_{k_1k_2})\left( r_{k_1k_2} - \delta_{j,[k_1k_2-1]}\right)-H_+(2s-2,r_{k_1k_2})\right\}\\
\nonumber
&-\gamma^{1-s}A_3(s) \sum_{k\in\mathbb{Z}} \sum_{n\in\mathbb{N}} t_{kn}(s,r_{k_1k_2}) p_{kn}(r_{k_1k_2})\left[ 1-(k+r_{k_1k_2})^{-1} \left( r_{k_1k_2} - \delta_{j,[k_1k_2-1]} \right)\right] K_s(\gamma^{-1}p_{kn}(r_{k_1k_2})) \\
\partial_{\gamma_j} L^\text{AB}&(s,\{\gamma_i\})=
\frac{4\pi}{\sqrt{3}}\gamma^{1-2s}\left\{ H_-(2s-1,r_{k_1k_2})\left( r_{k_1k_2} - \delta_{j,[k_1k_2-1]}\right)-H_+(2s-2,r_{k_1k_2})\right\}\\
\nonumber
&-\gamma^{1-s}A_3(s) \sum_{k\in\mathbb{Z}} \sum_{n\in\mathbb{N}} u_{kn}(s,r_{k_1k_2}) p_{kn}(r_{k_1k_2})\left[ 1-(k+r_{k_1k_2})^{-1} \left( r_{k_1k_2} - \delta_{j,[k_1k_2-1]} \right)\right] K_s(\gamma^{-1}p_{kn}(r_{k_1k_2})) 
\end{align}
and
\begin{align}\label{eq:latsumderiv2asym}
\nonumber
\partial_{\gamma_i} \partial_{\gamma_j} &L^\text{A}(s,\{\gamma_i\})=\partial^2_{\gamma_j}L^{A}(s,\{\gamma_i\}=
(2s-1)(2s-2)\gamma^{-2s} A_2(s)  \\
&+ 2(2s-1)\gamma^{-s} A_3(s) \sum_{k,n\in\mathbb{N}} t_{kn}(s) p_{kn} K_{s}\left(\gamma p_{kn} \right) 
+ 2\gamma^{1-s} A_3(s) \sum_{k,n\in\mathbb{N}} t_{kn}(s) p^2_{kn}
K_{s-1}\left(\gamma p_{kn} \right)\\
\nonumber
\partial_{\gamma_i} \partial_{\gamma_j} &L^{AA}(s,\{\gamma_i\})=
\frac{4\pi}{\sqrt{3}}\gamma^{-2s} (2s-1) \bigg\{ 
H_+(2s-2,r_{k_1k_2}) + H_+(2s,r_{k_1k_2})
(r_{k_1k_2}-\delta_{i,[k_1k_2-1]})(r_{k_1k_2}-\delta_{j,[k_1k_2-1]}) \\
&-H_-(2s-1,r_{k_1k_2}) \left[ 2r_{k_1k_2}- \delta_{i,[k_1k_2-1]} - \delta_{j,[k_1k_2-1]}\right] \bigg\}\\
\nonumber
&+ \gamma^{-s} A_3(s) \sum_{k\in\mathbb{Z}} \sum_{n\in\mathbb{N}}  t_{kn}(s,r_{k_1k_2})  p_{kn}(r_{k_1k_2}) 
 \left\{ p_{kn}(r_{k_1k_2}) K_{s-1}\left(\gamma p_{kn}(r_{k_1k_2}) \right) +(2s-1) \gamma K_{s}\left(\gamma p_{kn}(r_{k_1k_2}) \right)\right\} \\
\nonumber
&\quad\times \{ 1 - (k+r_{k_1k_2})^{-1}\left( r_{k_1k_2} - \delta_{i,[k_1,k_2-1]} \right) \} \left\{ 1 - (k+r_{k_1k_2})^{-1}\left( r_{k_1k_2} - \delta_{j,[k_1,k_2-1]} \right) \right\}\\
\nonumber
\partial_{\gamma_i} \partial_{\gamma_j} &L^\text{AB}(s,\{\gamma_i\})=
\frac{4\pi}{\sqrt{3}}\gamma^{-2s} (2s-1) \bigg\{ 
H_+(2s-2,r_{k_1k_2}) + H_+(2s,r_{k_1k_2})
(r_{k_1k_2}-\delta_{i,[k_1k_2-1]})(r_{k_1k_2}-\delta_{j,[k_1k_2-1]}) \\
&-H_-(2s-1,r_{k_1k_2}) \left[ 2r_{k_1k_2}- \delta_{i,[k_1k_2-1]} - \delta_{j,[k_1k_2-1]}\right] \bigg\}\\
\nonumber
&+ \gamma^{-s} A_3(s) \sum_{k\in\mathbb{Z}} \sum_{n\in\mathbb{N}}  u_{kn}(s,r_{k_1k_2})  p_{kn}(r_{k_1k_2}) 
 \left\{ p_{kn}(r_{k_1k_2}) K_{s-1}\left(\gamma p_{kn}(r_{k_1k_2}) \right) +(2s-1) \gamma K_{s}\left(\gamma p_{kn}(r_{k_1k_2}) \right)\right\} \\
\nonumber
&\quad\times \{ 1 - (k+r_{k_1k_2})^{-1}\left( r_{k_1k_2} - \delta_{i,[k_1,k_2-1]} \right) \} \left\{ 1 - (k+r_{k_1k_2})^{-1}\left( r_{k_1k_2} - \delta_{j,[k_1,k_2-1]} \right) \right\}
\end{align}
Here we used the extension of the Kronecker symbol $\delta_{j,[k_1,k_2-1]}=1$ if $j\in [k_1,k_2-1]$ and 0 otherwise, and the definition for the Hurwitz sums $H_\pm(s,r_{k_1,k_2})=\zeta(s,r_{k_1,k_2})\pm\zeta(s,1-r_{k_1,k_2})$. We see that the first and second derivatives of $L^A(s,\{\gamma_i\})$ with respect to $\gamma_i$ and $\gamma_j$ are only dependent of $\gamma$.

For the second set we have,
\begin{align}\label{eq:latsumderiv1b2}
\nonumber
\partial_{\gamma_j} &L^\text{AA}(s,\{\gamma_i\})=
-2s \gamma^{-2s-1}\left[ H_+(2s,r_{k_1k_2}) - H_-(2s+1,r_{k_1k_2}) \left( r_{k_1k_2} - \delta_{j,[k_1,k_2-1]} \right) \right] - \gamma^{-2} B_2(s) \\
\nonumber
&+\gamma^{-(s+\tfrac{3}{2})} B_3(s) {\sum_{k,n\in\mathbb{N}}} v_{kn}(s,r_{k_1k_2}) \left\{(2\pi k) \tan(2\pi kr_{k_1k_2}) 
\left( r_{k_1k_2} - \delta_{j,[k_1,k_2-1]} \right) - 2s\right\} K_{s-\frac{1}{2}} \left(\gamma^{-1} f_{kn}   \right) \\
&+\gamma^{-(s+\tfrac{5}{2})} B_3(s) {\sum_{k,n\in\mathbb{N}}} v_{kn}(s,r_{k_1k_2}) f_{kn} K_{s+\frac{1}{2}} \left(\gamma^{-1} f_{kn} \right) \\
\partial_{\gamma_j} &L^\text{AB}(s,\{\gamma_i\})= -\gamma^{-2}B_4(s)
+\gamma^{-(s+\tfrac{5}{2})} B_5(s) {\sum_{k,n\in\mathbb{N}}} w_{kn}(s,r_{k_1k_2}) g_{kn} K_{s+\frac{1}{2}} \left(\gamma^{-1} g_{kn} \right)  \\
\nonumber
&+\gamma^{-(s+\tfrac{3}{2})} B_5(s) {\sum_{k,n\in\mathbb{N}}} w_{kn}(s,r_{k_1k_2}) \left\{(2\pi k) \tan(2\pi kr_{k_1k_2}) 
\left( r_{k_1k_2} - \delta_{j,[k_1,k_2-1]} \right) - 2s\right\} K_{s-\frac{1}{2}} \left(\gamma^{-1} g_{kn}   \right) \\
\nonumber
&\text{and}\\
\nonumber
\partial_{\gamma_i} \partial_{\gamma_j} &L^\text{AA}(s,\{\gamma_i\})=
2s(2s+1) \gamma^{-2(s+1)}\bigg\{ H_+(2s,r_{k_1k_2}) +H_+(2s+2,r_{k_1k_2}) \left( r_{k_1k_2} - \delta_{i,[k_1,k_2-1]} \right)\left( r_{k_1k_2} - \delta_{j,[k_1,k_2-1]} \right) \\
& - H_-(2s+1,r_{k_1k_2}) \left( 2r_{k_1k_2} - \delta_{i,[k_1,k_2-1]} - \delta_{j,[k_1,k_2-1]} \right) \bigg\} +2 \gamma^{-3} B_2(s) \\
\nonumber
&+\gamma^{-(s+\tfrac{5}{2})} B_3(s) {\sum_{k,n\in\mathbb{N}}}  v_{kn}(s,r_{k_1k_2}) C_{kn}(s,k_1,k_2,\gamma) K_{s-\frac{1}{2}} \left(\gamma^{-1} f_{kn}   \right) \\
\nonumber
&+ \gamma^{-(s+\tfrac{7}{2})} B_3(s) {\sum_{k,n\in\mathbb{N}}} v_{kn}(s,r_{k_1k_2}) f_{kn} D_{kn}(s,k_1,k_2,\gamma)  K_{s+\frac{1}{2}} \left(\gamma^{-1} f_{kn}   \right) \\
\nonumber
\partial_{\gamma_i} \partial_{\gamma_j} &L^\text{AB}(s,\{\gamma_i\})=2\gamma^{-3}B_4(s)
+\gamma^{-(s+\tfrac{5}{2})} B_5(s) {\sum_{k,n\in\mathbb{N}}}  w_{kn}(s,r_{k_1k_2}) E_{kn}(s,k_1,k_2,\gamma) K_{s-\frac{1}{2}} \left(\gamma^{-1} g_{kn}   \right) \\
&+ \gamma^{-(s+\tfrac{7}{2})} B_5(s) {\sum_{k,n\in\mathbb{N}}} w_{kn}(s,r_{k_1k_2}) g_{kn} D_{kn}(s,k_1,k_2,\gamma)  K_{s+\frac{1}{2}} \left(\gamma^{-1} g_{kn}   \right) 
\end{align}
Here the coefficients are defined as
\begin{align}\label{eq:2ndLAAi}
\nonumber
C_{kn}(s,k_1,k_2,\gamma)&= 2s(2s+1) + \gamma^{-2} f^2_{kn} -(2\pi k)^2 \left( r_{k_1k_2} - \delta_{i,[k_1,k_2-1]} \right) \left( r_{k_1k_2} - \delta_{j,[k_1,k_2-1]} \right)  \\
&-2(2s+1)(2\pi k)\tan(2\pi k  r_{k_1k_2}) \left( 2r_{k_1k_2} - \delta_{i,[k_1,k_2-1]}  - \delta_{j,[k_1,k_2-1]}\right)  \\
D_{kn}(s,k_1,k_2,\gamma)&=- 2(2s+1) + (2\pi k) \tan(2\pi kr_{k_1k_2}) 
\left( 2r_{k_1k_2} - \delta_{i,[k_1,k_2-1]} - \delta_{j,[k_1,k_2-1]} \right)  \\
E_{kn}(s,k_1,k_2,\gamma)&= 2s(2s+1) + \gamma^{-2} g^2_{kn} -(2\pi k)^2 \left( r_{k_1k_2} - \delta_{i,[k_1,k_2-1]} \right) \left( r_{k_1k_2} - \delta_{j,[k_1,k_2-1]} \right)  \\
\nonumber
&-2(s+1)(2\pi k)\tan(2\pi k  r_{k_1k_2}) \left( 2r_{k_1k_2} - \delta_{i,[k_1,k_2-1]}  - \delta_{j,[k_1,k_2-1]}\right) 
\end{align}
A more detailed derivation is given in the supplementary material. 

We see that because of the simple linear dependence of $\gamma$ on the $\gamma_i$ for the lattice sums $L^A(s,\{\gamma_i\})$ the first and second derivatives lead to the same functional form compared to eqs.\eqref{eq:latsumderiv1a} and \eqref{eq:latsumderiv1b}. However, the derivative for $L^{AA}(s,\{\gamma_i\})$ and $L^{AB}(s,\{\gamma_i\})$ become more complicated. The advantage of the second set of equations is that the Bessel functions are not dependent on $k_1$ and $k_2$ which saves computer time.

\section{Derivatives from Direct Summations}
\label{app:DirSum}
The derivatives of the lattice sums $L^{A}$, $L^{AA}$ and $L^{AB}$ using directly formulas eqs.\eqref{eq:LA}, \eqref{eq:LAA} and \eqref{eq:LAB}, respectively, are shown in this section. They become useful for large exponents in the Lennard-Jones potential, i.e. $n\ge 10$. Furthermore, to speed up computer time we can use permutation symmetry between indices when applies. We only consider the more general case here where symmetry breaking for the location of the hexagonal layers along the $c$-axis  is considered. We remember that $\gamma\equiv\gamma_T$. The first lattice sum derivative with respect to to $\gamma_{j}$ according to \eqref{eq:LA} becomes,
\begin{equation}
    \begin{aligned}
        \partial_{\gamma_{j}}L^{A}(s,\gamma) &= -2s\gamma{\sum_{\vec{i}\in\mathbb{Z}^{3}}}' i_{3}^{2} S^{A}(\vec{i},\gamma)^{-(s+1)}
    \end{aligned}
\end{equation}
and for the second derivative we get
\begin{equation}
    \partial_{\gamma_{j}}^{2}L^{A}(s,\gamma) = 2s{\sum_{\vec{i}\in\mathbb{Z}^{3}}}' \left[2(s+1)i_{3}^{4}\gamma^{2}S^{A}(\vec{i},\gamma)^{-(s+2)} - i_{3}^{2}S^{A}(\vec{i},\gamma)^{-(s+1)}\right]
\end{equation}
Note that since $\partial_{\gamma_{j}}\gamma=1$, the mixed second derivatives are equal to $\partial_{\gamma_{j}}^{2}L^{A}$. Turning to the lattice sums \eqref{eq:LAA} and \eqref{eq:LAB} and its derivatives, we need the following relations (remember that $r_{k_{1}k_{2}}=\gamma^{-1}d_{k_{1}k_{2}}$),
\begin{equation}
    \partial_{\gamma_{j}}r_{k_{1}k_{2}} = \gamma^{-1}\left(\chi_{j}^{[k_{1},k_{2}-1]} - r_{k_{1}k_{2}}\right)
\end{equation}
\begin{equation}
    \partial_{\gamma_{j}}S^{AA}(\vec{i},k_1,k_2,\{\gamma_i\}) = 2\gamma\left(i_{3}+r_{k_{1}k_{2}}\right)\left[i_{3} + \chi_{j}^{[k_{1},k_{2}-1]}\right]
\end{equation}
and
\begin{equation}
    \partial_{\gamma_{i}\gamma_{j}}^{2}S^{AA}(\vec{i},k_1,k_2,\{\gamma_i\}) = 2\left(i_{3} + \chi_{i}^{[k_{1},k_{2}-1]}\right)^{2}
\end{equation}
where 
\begin{equation}
    \chi_{j}^{[k_{1},k_{2}-1]} = 
    \begin{cases}
        0 & \text{if }j\notin\left[k_{1},k_{2}-1\right] \\
        1 & \text{if }j\in\left[k_{1},k_{2}-1\right]
    \end{cases}
\end{equation}
where $1\leq k_{1}<k_{2}$ are the labels of two different layers of the Barlow sequences. The first derivative of $L^{AA}$ becomes,
\begin{align}
    \partial_{\gamma_{j}}L^{AA} &= -2\gamma\sum_{\vec{i}\in\mathbb{Z}^{3}} s\left(i_{3}+r_{k_{1}k_{2}}\right)S^{AA}(\vec{i},k_1,k_2,\{\gamma_i\})^{-(s+1)}\left[i_{3} + \chi_{j}^{[k_{1},k_{2}-1]}\right]
\end{align}
and the second derivatives are
\begin{align}
    \partial_{\gamma_{j}}^{2}L^{AA} &= \sum_{\vec{i}\in\mathbb{Z}^{3}} 2s(s+1)\gamma\left(i_{3}+r_{k_{1}k_{2}}\right)S^{AA}(\vec{i},k_1,k_2,\{\gamma_i\})^{-(s+2)}\left[i_{3}+\chi_{j}^{[k_{1},k_{2}-1]}\right] \\
    \nonumber
    &- 2sS^{AA}(\vec{i},k_1,k_2,\{\gamma_i\})^{-(s+1)}\left[i_{3}+\chi_{j}^{k_{1},k_{2}-1}\right]^{2}
\end{align}
\begin{align}
    \partial_{\gamma_{i}\gamma_{j}}^{2}L^{AA} &= \sum_{\vec{i}\in\mathbb{Z}^{3}} 2s(s+1)\gamma\left(i_{3}+r_{k_{1}k_{2}}\right)S^{AA}(\vec{i},k_1,k_2,\{\gamma_i\})^{-(s+2)}\left[i_{3}+\chi_{j}^{[k_{1},k_{2}-1]}\right] \\
    \nonumber
    &- 2sS^{AA}(\vec{i},k_1,k_2,\{\gamma_i\})^{-(s+1)}\left[i_{3}+\chi_{i}^{k_{1},k_{2}-1}\right]^{2}
\end{align}
For the last lattice sum $L^{AB}$ we get similar expressions,
\begin{equation}
    \partial_{\gamma_{j}}L^{AB} = -2\gamma\sum_{\vec{i}\in\mathbb{Z}^{3}} s\left(i_{3}+r_{k_{1}k_{2}}\right)S^{AB}(\vec{i},k_1,k_2,\{\gamma_i\})^{-(s+1)}\left[i_{3} + \chi_{j}^{[k_{1},k_{2}-1]}\right]
\end{equation}
\begin{align}
    \partial_{\gamma_{j}}^{2}L^{AB} &= \sum_{\vec{i}\in\mathbb{Z}^{3}} 2s(s+1)\gamma\left(i_{3}+r_{k_{1}k_{2}}\right)S^{AB}(\vec{i},k_1,k_2,\{\gamma_i\})^{-(s+2)}\left[i_{3}+\chi_{j}^{[k_{1},k_{2}-1]}\right] \\
    \nonumber
    &- 2sS^{AB}(\vec{i},k_1,k_2,\{\gamma_i\})^{-(s+1)}\left[i_{3}+\chi_{j}^{k_{1},k_{2}-1}\right]^{2}
\end{align}
\begin{align}
    \partial_{\gamma_{i}\gamma_{j}}^{2}L^{AB} &= \sum_{\vec{i}\in\mathbb{Z}^{3}} 2s(s+1)\gamma\left(i_{3}+r_{k_{1}k_{2}}\right)S^{AB}(\vec{i},k_1,k_2,\{\gamma_i\})^{-(s+2)}\left[i_{3}+\chi_{j}^{[k_{1},k_{2}-1]}\right] \\
    \nonumber
    &- 2sS^{AB}(\vec{i},k_1,k_2,\{\gamma_i\})^{-(s+1)}\left[i_{3}+\chi_{i}^{k_{1},k_{2}-1}\right]^{2}
\end{align}

\section{Taylor expansions involving Barlow lattice sums}
\label{app:Taylor}

In order to discuss the linear trends observed for the properties with respect to the fcc stacking fraction, it is convenient to introduce a Taylor expansion for the lattice sum differences. Consider the variable
\begin{equation}
x = \frac{L^\textit{\textbf{X}}(s,\gamma)}{L^\text{hcp}(s,\gamma)} - 1 = \frac{\Delta L^\textit{\textbf{X}}(s,\gamma)}{L^\text{hcp}(s,\gamma)} 
\end{equation}
where $x>0$ is very small in the range $s\in (1.5,\infty)$ for $\gamma_0=\sqrt{\frac{2}{3}}$ as pointed out in our first paper\cite{Cooper2025}. 
We assume a near linear behavior in the lattice sums, that is
\begin{equation}\label{eq:linearbehave}
\Delta L^\textit{\textbf{X}}(s,\gamma) \approx \alpha^\textit{\textbf{X}}(2) \left( L^\text{fcc}(s,\gamma) -  L^\text{hcp}(s,\gamma)\right)
\end{equation}
with the fcc stacking fraction for a specific Barlow packing in the range of $0\le\alpha^\textbf{X}(2)\le 1$. We now introduce the Taylor expansion around $x=0$ for
\begin{equation}
(1+x)^q = 1+qx+\frac{1}{2}q(q-1)x^2 + ...
\end{equation}
with $q\in \mathbb{R}$. We apply this first to the minimum lattice constant \eqref{eq:RLJnmmin} and keep only the first order terms,
\begin{align} \label{eq:Tayloramindiff}
 \Delta a^{*\textit{\textbf{X}}}_{\textrm{min}}(n,m,\gamma)&=\left( \frac{L^\textbf{\textit{X}}(\frac{n}{2},\gamma)}{L^\textbf{\textit{X}}(\frac{m}{2},\gamma)}\right)^{\tfrac{1}{n-m}} 
 -\left( \frac{L^\text{hcp}(\frac{n}{2},\gamma)}{L^\text{hcp}(\frac{m}{2},\gamma)}\right)^{\tfrac{1}{n-m}}\\
 \nonumber
 &= \left( \frac{L^\text{hcp}(\frac{n}{2},\gamma)}{L^\text{hcp}(\frac{m}{2},\gamma)}\right)^{\tfrac{1}{n-m}}
 \left[ \left( \frac{\Delta L^\textbf{\textit{X}}(\frac{n}{2},\gamma)}{L^\text{hcp}(\frac{n}{2},\gamma)} \right)^{\tfrac{1}{n-m}} \left( \frac{\Delta L^\textbf{\textit{X}}(\frac{m}{2},\gamma)}{L^\text{hcp}(\frac{m}{2},\gamma)} \right)^{-\tfrac{1}{n-m}} -1 \right] \\
 \nonumber
 &\approx \frac{1}{n-m}\left( \frac{L^\text{hcp}(\frac{n}{2},\gamma)}{L^\text{hcp}(\frac{m}{2},\gamma)}\right)^{\tfrac{1}{n-m}}
 \left[ \frac{\Delta L^\textit{\textbf{X}}(\frac{n}{2},\gamma)}{L^\text{hcp}(\frac{n}{2},\gamma)} - \frac{\Delta L^\textit{\textbf{X}}(\frac{m}{2},\gamma)}{L^\text{hcp}(\frac{m}{2},\gamma)}\right]
 \,,
\end{align}
Inserting \eqref{eq:linearbehave} finally gives
\begin{align} \label{eq:Tayloramindiff1}
\Delta a^{*\textit{\textbf{X}}}_{\textrm{min}}(n,m,\gamma) \approx 
\frac{\alpha^\textit{\textbf{X}}(2)}{n-m}
\left( \frac{L^\text{hcp}(\frac{n}{2},\gamma)}{L^\text{hcp}(\frac{m}{2},\gamma)}\right)^{\tfrac{1}{n-m}}
 \left[ \frac{L^\text{fcc}(\frac{n}{2},\gamma)}{L^\text{hcp}(\frac{n}{2},\gamma)} - \frac{L^\text{fcc}(\frac{m}{2},\gamma)}{L^\text{hcp}(\frac{m}{2},\gamma)}\right]\\
 \nonumber
 = \frac{\alpha^\textit{\textbf{X}}(2)}{n-m} 
 \left[ \frac{L^\text{fcc}(\frac{n}{2},\gamma)}{L^\text{hcp}(\frac{n}{2},\gamma)} - \frac{L^\text{fcc}(\frac{m}{2},\gamma)}{L^\text{hcp}(\frac{m}{2},\gamma)}\right] a^{*\text{hcp}}_{\textrm{min}}(n,m,\gamma)
 \,,
\end{align}
Deviations between the two equations \eqref{eq:Tayloramindiff} and \eqref{eq:Tayloramindiff1} are smaller than $5\times 10^{-9}$. It is therefore clear that a near linear behavior in $\Delta L^\textbf{X}(s,\gamma)$ with respect to the fcc stacking fraction $\alpha(2)$ translates into a near linear behavior in $\Delta a^{*\textbf{X}}_{\textrm{min}}(n,m,\gamma)$ because of the small differences between the lattice sums of Barlow packings. We notice that the term $L^\text{fcc}(\frac{m}{2},\gamma)/L^\text{hcp}(\frac{m}{2},\gamma)$ is close to 1.0 and has a minimum around $m=7.05$ (or $s=3.53$). This implies that the term in the square bracket of \eqref{eq:Tayloramindiff1} can be positive or negative depending on the $(n,m)$ combination for the LJ potential.

It is straightforward to derive the Taylor expansions to first order for the other properties as well, e.g. for the cohesive energy
\begin{align} \label{eq:ELJnmmindiff}
\Delta {E_{\textrm{coh}}^\textbf{\textit{X}}}^*(a^*_{\textrm{min}},n,m,\gamma)&=
-\frac{1}{2}\left[ \frac{L^\textbf{\textit{X}}(\frac{m}{2},\gamma)^n}{L^\textbf{\textit{X}}(\frac{n}{2},\gamma)^m}\right]^{\tfrac{1}{n-m}} 
+\frac{1}{2}\left[ \frac{L^\text{hcp}(\frac{m}{2},\gamma)^n}{L^\text{hcp}(\frac{n}{2},\gamma)^m}\right]^{\tfrac{1}{n-m}}\\
\nonumber
&=\frac{1}{2}\left[ \frac{L^\text{hcp}(\frac{m}{2},\gamma)^n}{L^\text{hcp}(\frac{n}{2},\gamma)^m}\right]^{\tfrac{1}{n-m}}
\left[ 1 - \left( 1 + \frac{\Delta L^\textbf{\textit{X}}(\frac{m}{2},\gamma)}{L^\text{hcp}(\frac{m}{2},\gamma)}\right)^{\tfrac{n}{n-m}}
\left( 1 + \frac{\Delta L^\textbf{\textit{X}}(\frac{n}{2},\gamma)}{L^\text{hcp}(\frac{n}{2},\gamma)}\right)^{-\tfrac{m}{n-m}} \right]\\
\nonumber
&\approx -\frac{1}{2(n-m)}\left[ \frac{L^\text{hcp}(\frac{m}{2},\gamma)^n}{L^\text{hcp}(\frac{n}{2},\gamma)^m}\right]^{\tfrac{1}{n-m}}
\left[  \frac{n\Delta L^\textbf{\textit{X}}(\frac{m}{2},\gamma)}{L^\text{hcp}(\frac{m}{2},\gamma)}
-  \frac{m\Delta L^\textbf{\textit{X}}(\frac{n}{2},\gamma)}{L^\text{hcp}(\frac{n}{2},\gamma)} \right] \\
\nonumber
&=-\frac{\alpha^\textit{\textbf{X}}(2)}{2(n-m)}\left[ \frac{L^\text{hcp}(\frac{m}{2},\gamma)^n}{L^\text{hcp}(\frac{n}{2},\gamma)^m}\right]^{\tfrac{1}{n-m}}
\left[ \frac{nL^\text{fcc}(\frac{m}{2},\gamma)}{L^\text{hcp}(\frac{m}{2},\gamma)} - \frac{mL^\text{fcc}(\frac{n}{2},\gamma)}{L^\text{hcp}(\frac{n}{2},\gamma)} - (n-m)\right]\\
\nonumber
&=\alpha^\textit{\textbf{X}}(2)
\left[ \frac{n}{(n-m)}\frac{L^\text{fcc}(\frac{m}{2},\gamma)}{L^\text{hcp}(\frac{m}{2},\gamma)} - \frac{m}{(n-m)} \frac{L^\text{fcc}(\frac{n}{2},\gamma)}{L^\text{hcp}(\frac{n}{2},\gamma)} -1 \right] {E_{\textrm{coh}}^\text{hcp}}^*(a^*_{\textrm{min}},n,m,\gamma)
\; .
\end{align}
A similar analysis can be done for the bulk modulus and the Einstein frequency.

\section{Entropy for Crystal Lattices from Statistical Thermodynamics}
\label{app:LatticeThermodynamics}

The equations for the thermodynamic functions of a crystal lattice with atoms interacting through a potential such as the $(n,m)$-LJ potentials can be obtained from the thermodynamic partition function, $Z$, in the canonical ensemble. Starting from the Schr\"odinger equation of a lattice, the first approximation to introduce is the Born-Oppenheimer approximation, enabling us to separate the vibrational, electronic and nuclear degrees of freedom. Then, for a crystal, the total energy is $E = E_{\text{el}} + E_{\text{vib}} + E_{\text{nuc}}$. Since under standard conditions of temperature and pressure the nuclei are in their nuclear ground state, the nuclear energy becomes just an additive constant that drops out when taking energy differences between two crystals of the same composition (here we neglect nuclear spin effects). Hence, we only need to consider the electronic energy defining the potential energy hypersurface $V(\mathbf{R})$ in which the nuclei move, and the energy due to their relative motion by solving the vibrational Schr\"odinger equation
\begin{equation}
    \left[\sum_{I}\frac{\nabla_{\mathbf{R}_{I}}^{2}}{2M_{I}} + V(\mathbf{R}) - E_{\text{vib}}(\mathbf{R})\right]\chi(\mathbf{R}) = 0
\end{equation}
with the nuclear coordinates $\mathbf{R}$ and $\chi(\mathbf{R})$ being the wavefunction for the nuclear motion. $V(\mathbf{R})$ can be obtained from first-principles electronic structure methods fit to an empirical potential such as the $(m,n)$-LJ potential.

All thermodynamic functions for a lattice at finite $T$ can be derived from the partition function\cite{greiner2012thermodynamics} 
\begin{equation}
    Z=\text{Tr}\left(e^{-\beta\hat{H}}\right) = \sum_{n}g_{n}e^{-\beta E_{n}}
\end{equation}
where the Hamiltonian is written in the eigenstate basis, $g_{n}$ is the degeneracy of the $n$-th energy level and $\beta=1/k_BT$. Since the energy is separated into vibrational and electronic contributions (neglecting nuclear contributions), the partition function is written as the product of $Z=Z_{\text{vib}}Z_{\text{el}}$, with
\begin{equation}
    Z_{\text{el}} = \sum_{i}g_{n}e^{-\beta E_{i}} \approx g_{1} + g_{2}e^{-\beta \Delta E_{12}}
\end{equation}
for well spaced energy levels at lower temperatures, where $\Delta E_{12}$ is the energy difference to the $2$nd energy level (first excited state) with respect to the ground state which is arbitrarily set to zero. $\Delta E_{12}$ is typically in the range of a few eV and the electronic term in the partition function therefore vanishes for relatively low temperatures even for metallic systems where the electronic band gap is zero\cite{Jerabek2022}. Hence, only the first vibrational term is important for us, although we will keep the second term here for completeness.

For the vibrational contribution, let $\phi(r)$ be a two-body potential, such as LJ, describing the energy hypersurface in which the nuclei move. Then, a Taylor expansion around the equilibrium positions of the atoms in a lattice gives
\begin{equation}\label{eq:harmonic-approx}
    V = V_{\text{eq}} + \frac{1}{2}\sum_{\boldsymbol{\ell},\kappa}\sum_{\boldsymbol{\ell}',\kappa'}\left(\frac{\partial^{2}\phi_{\kappa\kappa'}(\boldsymbol{\ell},\boldsymbol{\ell}')}{\partial\mathbf{u}_{\kappa}(\boldsymbol{\ell})\partial\mathbf{u}_{\kappa'}(\boldsymbol{\ell}')}\right)_{\mathbf{u}=0}\mathbf{u}_{\kappa}(\boldsymbol{\ell})\mathbf{u}_{\kappa'}(\boldsymbol{\ell}') + \dots
\end{equation}
where $\mathbf{u}_{\kappa}(\boldsymbol{\ell})$ is the displacement from equilibrium of the $\kappa$-th atom in the $\boldsymbol{\ell}$-th unit cell. All terms explicitly shown in Eq.\eqref{eq:harmonic-approx} constitute the harmonic approximation. Higher-order anharmonic corrections become more important at higher temperatures or for weakly interacting systems. In this respect, the atom confinement in a solid makes the vibrations more harmonic compared to a diatomic molecule. From the second derivatives is it possible to find the dynamic matrix and then calculate the (mass-dependent) phonon frequencies by using Eq. \eqref{eq:phonons}.

The quantum levels (energy eigenvalues) of a crystal within the harmonic approximation are given by
\begin{equation}
    E_{\{n_{\mathbf{k}j}\}} = E_{\text{eq}} + \sum_{\mathbf{k},j}\hbar\omega_{j}(\mathbf{k})\left(n_{\mathbf{k}j}+\frac{1}{2}\right)
\end{equation}
with the first term being a static contribution to the partition function such as the cohesive energy, i.e. the energy of the lattice in the equilibrium configuration, and the second term being the dynamic contribution. The dynamic part contains all the $3N$ phonon branches at the $\mathbf{k}$ points within the first Brillouin zone for our Barlow packings. Therefore, the vibrational partition function is
\begin{equation}
    Z_{\text{vib}} = e^{-\beta E_{\text{eq}}}\prod_{\mathbf{k},j}\frac{e^{-\beta\hbar\omega_{j}(\mathbf{k})/2}}{1-e^{-\beta\hbar\omega_{j}(\mathbf{k})}}
\end{equation}

Consequently, all thermodynamic functions can be written separately into vibrational and electronic contributions. The internal energy of the crystal is given by
\begin{equation}
    U = E_{\text{eq}} + \sum_{\mathbf{k},j}\left[\frac{\hbar\omega_{j}(\mathbf{k})}{2} - \frac{\hbar\omega_{j}(\mathbf{k})}{1-e^{\beta\hbar\omega_{j}(\mathbf{k})}}\right] - \frac{1}{\beta}\left[\frac{1}{\beta}\ln{Z_{\text{el}}} + \frac{g_{2}\Delta E_{12}e^{-\beta\Delta E_{12}}}{Z_{\text{el}}}\right],
\end{equation}
the entropy by
\begin{equation}\label{eq:vibel}
    S = -k_{B}\left\{\sum_{\mathbf{k},j}\left[\frac{\beta\hbar\omega_{j}(\mathbf{k})}{(1-e^{\beta\hbar\omega_{j}(\mathbf{k})})}  + \ln{\left(1-e^{-\beta\hbar\omega_{j}(\mathbf{k})}\right)}\right] + T\ln{Z_{\text{el}}} - \frac{g_{2}\Delta E_{12}e^{-\beta\Delta E_{12}}}{Z_{\text{el}}}\right\},
\end{equation}
and the Helmholtz free energy $F=U-TS$ by
\begin{equation}
    F = E_{\text{eq}} + \sum_{\mathbf{k},j}\left[\frac{\hbar\omega_{j}(\mathbf{k})}{2} - \frac{1}{\beta}\ln{\left(1-e^{-\beta\hbar\omega_{j}(\mathbf{k})}\right)}\right] - \frac{1}{\beta}\ln{Z_{\text{el}}}
\end{equation}
The vibrational part of eq.\eqref{eq:vibel} was used in all our LJ phonon dispersion calculations using the code introduced in Ref.\onlinecite{Rosciszewski2007} and in our DFT calculations for argon using Phonopy \cite{Togo_2023}.

In CDLT one treats the vibrations in a lattice classically. Within the harmonic approximation, this leads to the following expression for the vibrational entropy (see Ref.\onlinecite{Travesset2014} for a detailed derivation)
\begin{equation}\label{Eq:vibrational_hoodlt}
    \frac{S_{vib}}{N_A k_B} = \frac{3}{2}\ln\left(\frac{2\pi e^2L^2}{\beta\epsilon\lambda_T^2}\right)-\frac{1}{2 N} \sum_{j,{\bm k}}\ln\left(\frac{ML^2}{\epsilon} \omega^2_{j}({\bm k})\right) \ .
\end{equation}
where $\lambda_T=h/p_T=h/\sqrt{2\pi k_BT}$ the thermal wavelength ($p_T$ the classical thermal momentum coming from the kinetic energy term), $N_A$ the number of atoms, $e$ is Euler's number ($\ln e=1$), and $\epsilon$ and $L$ are an arbitrary energy (e.g. the dissociation energy for a LJ potential) and length (e.g. the equilibrium length $r_e$ or hard-sphere radius $\sigma$ for a LJ potential) used to render the dynamic matrix dimensionless. The relation to the the quantum formula for the vibrational entropy of a mode is given by the Taylor expansion
\begin{equation}\label{Eq:quantum_mode}
    \frac{S_{vib}}{k_B}=\left(\frac{x}{e^{x}-1}-\ln(1-e^{-x})\right) \approx 1 -\ln\left(x\right) = \ln \left(e/x\right) 
\end{equation}
with $x=\beta\hbar\omega$, where the rhs is the classical result. This relates exactly to the high temperature limit we discussed in the main part of the paper. Therefore, the classical vibrational entropy for very large particle numbers $N_A$ is  
\begin{eqnarray}\label{Eq:vibrational_quantum}
    \frac{S_{vib}}{N k_B} &=& 3 \ln\left(\frac{2\pi e}{h\beta} \right)-\frac{1}{ 2N} \sum_{\lambda,{\bm k}} \ln( \omega^2_{\lambda}({\bm k})) 
    \nonumber \\
    &=& \frac{3}{2}\ln\left(\frac{2\pi e^2 L^2}{\beta\lambda\epsilon_T^2}\right)-\frac{1}{2 N} \sum_{\lambda,{\bm k}} \ln\left(\frac{L^2}{\epsilon} M \omega^2_{\lambda}({\bm k})\right) 
\end{eqnarray}
and the entropy computed by Eq.~\eqref{Eq:quantum_mode} becomes identical to the one computed by classical statistical mechanics Eq.~\eqref{Eq:vibrational_hoodlt}. The first term (kinetic term) drops out for entropy differences $\Delta S_{vib}$ between different phases at constant temperature and the $\Delta S_{vib}$ becomes independent of the temperature $T$.

The entropy of particles interacting with an inverse power law potential,
\begin{equation}
    V(r)=\varepsilon\left(\frac{\sigma}{r}\right)^p
\end{equation}
can be easily derived within CDLT and is given by\cite{Travesset2014}
\begin{equation}
    \frac{S_{vib}}{N k_B}=\ln\left(\frac{V}{N}\frac{e}{\lambda_T^3}\right)+3-\frac{p}{2}\ln(\gamma_p)-b(p) \ ,
\end{equation}
where $\gamma_p=\frac{N}{V}\sigma^3\left(\beta\epsilon\right)^{3/p}$, $\frac{N}{V}=\frac{c_L}{R_{NN}^2}$, $R_{NN}$ is the nearest neighbor particle distance. $c_L$ depends on the particular lattice (fcc or hcp in our case). Using these relations, we get
\begin{equation}
     \frac{S_{vib}}{N k_B} = \ln\left(e^3\left(\frac{\sigma}{\lambda_T^3}\right)^3\left(\beta\epsilon\right)^{3/2}\right)+1+\ln\left(\left(c_L\left(\frac{\sigma}{R_{NN}}\right)^3\right)^{-\frac{p}{2}+1}  \right)-b(p)
\end{equation}
If we choose $L=\sigma$ then Eq.~\eqref{Eq:vibrational_hoodlt} (or Eq.~\eqref{Eq:vibrational_quantum}) becomes
\begin{equation}\label{eq:SvibLJ}
    \frac{S_{vib}}{N k_B}=\ln\left(e^3 \left(\frac{\sigma}{\lambda_T}\right)^3\left(2\pi\beta\epsilon\right)^{3/2}\right)-\frac{1}{2N}\sum_{\lambda,{\bm k}}\ln\left(\frac{M \sigma^2  \omega^2_{\lambda}({\bm k})}{\varepsilon} \right)
\end{equation}
For entropy differences at constant $T$ only the last two terms in \eqref{eq:SvibLJ} survive. Comparing both formulas we have
\begin{equation}
    b(p)=1+\ln\left((2\pi)^{-3/2}\left(c_L\left(\frac{\sigma}{R_{NN}}\right)^3\right)^{-\frac{p}{2}+1}\right)+\frac{1}{2N}\sum_{\lambda,{\bm k}}\ln\left(\frac{M \sigma^2  \omega^2_{\lambda}({\bm k})}{\varepsilon} \right)
\end{equation}
This function is evaluated by the program HOODLT\cite{Travesset2014} for inverse power potentials, and provides the values quoted in Table II of Ref.~\onlinecite{Travesset2014}. The approximate value chosen in Ref.\onlinecite{Travesset2014} for the nearest neighbor distance was $R_{NN}=\sigma$. A comparison shows that for the (12,6)-LJ potential we have $\sigma=2^{-1/6}r_e=0.8909r_e$ and for the fcc lattice we have $R_{NN}=0.9713 r_e$ evaluated through exact lattice summations. Further, $R_{NN}$ changes due to anharmonicity effects in the zero-point vibrational term and for finite temperatures.

Using HOODLT we get the following entropy differences between the fcc and hcp phases:
$\Delta S^{(6,4)} = 4.7131\times10^{-3}\,k_{B}$,
$\Delta S^{(12,6)} = 2.8532\times10^{-3}\,k_{B}$,
$\Delta S^{(30,12)} = 1.8267\times10^{-3}\,k_{B}$,
$\Delta S^{(30,20)} = 1.7436\times10^{-3}\,k_{B}$,
and $\Delta S^{(100,25)} = 1.5412\times10^{-3}\,k_{B}$.
Our calculated value for the (100,25)-LJ potential is close to the hard-sphere limit of $\Delta S = 1.4754\times10^{-3} k_B$ reported in Ref.\onlinecite{Travesset2014}. Beside the difference in magnitude compared to our DFT or LJ results including phonon dispersion, we qualitatively agree with these results, that is fcc is entropically preferred compared to hcp. Although both the CDLT formalism and Equation \eqref{eq:vibel} produce equivalent formulas at the high-temperature limit, the slight differences in the phonon frequencies due to soft phonons are structural distortions are responsible for the differences seen when comparing the entropy difference for soft potentials, while the hard-sphere limits are in better numerical agreement.

\section{The iterative Newton–Raphson algorithm} 
\label{app:Derivatives}

The iterative Newton–Raphson algorithm \cite{hammerlin2012numerical} was used to minimize the cohesive energy with respect to a set of lattice parameters $\{\gamma_i\}$  ($\vec{\gamma}=(\gamma_1,...,\gamma_N)$), 
\begin{equation}\label{eq:mincond1}
\vec{\gamma}^{(I+1)}=\vec{\gamma}^{(I)} -  \lambda \left[ (\partial_{\gamma_i}\partial_{\gamma_j}) E_{nm}^*(\{\gamma_i^{(I)}\})\right]^{-1}\vec{\nabla} E_{nm}^*(\{\gamma_i^{(I)}\})   \,,
\end{equation}

For the Hessian and gradient of the total cohesive energy $E_{nm}^\textbf{\textit{X}}(\{\gamma_i\})$ for a specific periodic Barlow packing with sequence $\textbf{\textit{X}}$ we require the following derivatives
\begin{align}\label{eq:derivativessym}
\nonumber
\partial_{\gamma_j} &E_{nm}^\textbf{\textit{X}}(\{\gamma_i\}) =  \frac{1}{2(n-m)} \bigg\{ m \left( \frac{L^\textbf{\textit{X}}(\frac{m}{2},\{\gamma_i\})}{L^\textbf{\textit{X}}(\frac{n}{2},\{\gamma_i\})} \right)^{\frac{n}{n-m}} \partial_{\gamma_j} L^\textbf{\textit{X}}(\tfrac{n}{2},\{\gamma_i\}) \\
- &n  \left( \frac{L^\textbf{\textit{X}}(\frac{m}{2},\{\gamma_i\})}{L^\textbf{\textit{X}}(\frac{n}{2},\{\gamma_i\})} \right)^{\frac{m}{n-m}} \partial_{\gamma_j} L^\textbf{\textit{X}}(\tfrac{m}{2},\{\gamma_i\}) \bigg\}\\
\end{align}
\begin{align}\label{eq:derivativessym2}
\partial_{\gamma_j} \partial_{\gamma_k}&E_{nm}^\textbf{\textit{X}}(\{\gamma_i\}) =
\frac{nm}{2(n-m)^2} \left[ \frac{\partial_{\gamma_k}L^\textbf{\textit{X}}(\tfrac{m}{2},\{\gamma_i\})}{L^\textbf{\textit{X}}(\tfrac{m}{2},\{\gamma_i\})}
- \frac{\partial_{\gamma_k}L^\textbf{\textit{X}}(\tfrac{n}{2},\{\gamma_i\})}{L^\textbf{\textit{X}}(\tfrac{n}{2},\{\gamma_i\})} \right]\\
\nonumber
&\times
\left[ \left( \frac{L^\textbf{\textit{X}}(\frac{m}{2},\{\gamma_i\})}{L^\textbf{\textit{X}}(\frac{n}{2},\{\gamma_i\})} \right)^{\frac{n}{n-m}}
\partial_{\gamma_j}L^\textbf{\textit{X}}(\tfrac{n}{2},\{\gamma_i\})
- \left( \frac{L^\textbf{\textit{X}}(\frac{m}{2},\{\gamma_i\})}{L^\textbf{\textit{X}}(\frac{n}{2},\{\gamma_i\})} \right)^{\frac{m}{n-m}}
\partial_{\gamma_j}L^\textbf{\textit{X}}(\tfrac{m}{2},\{\gamma_i\})\right]\\
&+ \frac{1}{2(n-m)} \bigg\{ m \left( \frac{L^\textbf{\textit{X}}(\frac{m}{2},\{\gamma_i\})}{L^\textbf{\textit{X}}(\frac{n}{2},\{\gamma_i\})} \right)^\frac{n}{n-m} 
 \partial_{\gamma_k} \partial_{\gamma_j} L^\textbf{\textit{X}}(\tfrac{n}{2},\{\gamma_i\}) \\
 \nonumber
 &- n \left( \frac{L^\textbf{\textit{X}}(\frac{m}{2},\{\gamma_i\})}{L^\textbf{\textit{X}}(\frac{n}{2},\{\gamma_i\})} \right)^\frac{m}{n-m} 
 \partial_{\gamma_k} \partial_{\gamma_j} L^\textbf{\textit{X}}(\tfrac{m}{2},\{\gamma_i\})
 \bigg\}
\,.
\end{align}
and in general for the Barlow lattice sum derivatives,
\begin{equation}
\partial^{n_1}_{\gamma_k} \partial^{n_2}_{\gamma_j} L^\textbf{\textit{X}}(s,\{\gamma_i\})=\left\{   \partial^{n_1}_{\gamma_k} \partial^{n_2}_{\gamma_j} L^A(s,\{\gamma_i\}) 
+ \frac{2}{N} \sum_{k_1<k_2}^N \;\; \partial^{n_1}_{\gamma_k} \partial^{n_2}_{\gamma_j} L^{X_{k_1}X_{k_2}}(s,\{\gamma_i\}) \right\}
\end{equation}
with $n_1,n_2\in\{0,1,2\}, n_1+n_2\le 2$. This leads to the simple condition for the minimum through the first derivative
\begin{align}\label{eq:derivatives2}
n L(\tfrac{n}{2},\{\gamma_i\}) \partial_{\gamma_j} L(\tfrac{m}{2},\{\gamma_i\}) - m L(\tfrac{m}{2},\{\gamma_i\}) \partial_{\gamma_j} L(\tfrac{n}{2},\{\gamma_i\}) =  0 \,.
\end{align}
Here we define $\partial_x f(x)=\partial f(x)/\partial x$. In the case where we do not consider symmetry breaking effects between the hexagonal layers we only consider the derivatives with respect to $\gamma_T$. The gradient and Hessian is obtained analytically as discussed in the next section.

Eq.\eqref{eq:mincond1} is converging very fast. Only few iterations are required to achieve convergence to computer accuracy. For example, we start with $\gamma_i^{(1)}=\sqrt{\frac{2}{3}}$ for all $\gamma_i$ of the Barlow structures, and for the hcp structure with $\gamma_1=\gamma_2$ we obtain after the first iteration $\gamma_1^{(2)}=0.816381646735$  very close to the converged result after 2 iterations with $\gamma_1^{(3)}=0.81638165246885$.

\clearpage
\section{Tables for cohesive energies and optimized parameters}
\label{app:CohesiveEnergies}

\begin{table}[htp]
    \centering
        \setlength{\tabcolsep}{1pt}
    \fontsize{7.0}{10}\selectfont
\begin{tabular}{ |r|l|r|r|r|r|r|r|r|r| } 
 \hline
 No. & Stacking			& (6,4) 				     & (8,4) 			   & (12,4) 			   & (12,6) 				& (30,6) 			& (30,20) \\ 
 \hline
1 & AB 	 			&\textbf{-38.9325305486602} &\textbf{-25.0752959432072}& \textbf{-18.3098535057780}&\textbf{-8.6110697321365} &\textbf{-7.5716717703337}& \textbf{-6.00928320286815}\\
2 & ABC 	                  	& -1.672643684e-3        & -1.927682096e-4        & 5.312237310e-4        &  8.695754328e-4        &  6.391322107e-4        &  4.328621270e-5\\
3 & ABAC                          & -8.361267034e-4        & -9.576506098e-5        & 2.661869698e-4        &  4.352966307e-4        &  3.198626830e-4        &  2.164359436e-5\\
4 & ABABC                 	& -1.003484319e-3        & -1.154063871e-4        & 3.189655319e-4        &  5.219509746e-4        &  3.835991612e-4        &  2.597193729e-5\\
5 & ABABAC                 	& -5.574080672e-4        & -6.384034753e-5        & 1.774586756e-4        &  2.901987369e-4        &  2.132423385e-4        &  1.442906866e-5\\
6 & ABACBC                 	& -1.114955978e-3        & -1.280964162e-4        & 3.545332527e-4        &  5.800572110e-4        &  4.262864076e-4        &  2.885780622e-5\\
7 & ABABABC                   & -7.167628667e-4        & -8.243041752e-5        & 2.278325291e-4        &  3.728228336e-4        &  2.739997973e-4        &  1.855139134e-5\\
8 &  ABABCBC                  & -7.167176386e-4        & -8.225535352e-5        & 2.279974317e-4        &  3.729679322e-4        &  2.740843632e-4        &  1.855152345e-5\\
9 &  ABACABC                  & -1.194623895e-3        & -1.373351599e-4        & 3.797744533e-4        &  6.214168958e-4        &  4.566929165e-4        &  3.091900854e-5\\
10&  ABABABAC               & -4.180527461e-4        & -4.787949084e-5        & 1.330940083e-4        &  2.176492541e-4        &  1.599318664e-4        &  1.082180364e-5\\
11&  ABABACAC               & -4.180520536e-4        & -4.787876063e-5        & 1.330945268e-4        &  2.176495887e-4        &  1.599320542e-4        &  1.082180366e-5\\
12&  ABABACBC               & -8.362030818e-4        & -9.606850383e-5        & 2.659004636e-4        &  4.350440465e-4        &  3.197154429e-4        &  2.164336327e-5\\
13&  ABABCBAC               & -8.361645464e-4        & -9.591641730e-5        & 2.660439760e-4        &  4.351705059e-4        &  3.197891564e-4        &  2.164347883e-5\\
14&  ABABCABC               & -1.254413141e-3        & -1.444156842e-4        & 3.985623595e-4        &  6.523105080e-4        &  4.794242568e-4        &  3.246479443e-5\\
15 &  ABACBABC              & -1.254374610e-3        & -1.442635974e-4        & 3.987058720e-4        &  6.524369655e-4        &  4.794979699e-4        &  3.246490999e-5\\
16&  ABABABABC             & -5.574771992e-4        & -6.411137373e-5        & 1.772030807e-4        &  2.899736212e-4         &  2.131111247e-4        &  1.442886320e-5\\
17&  ABABABCBC             & -5.574420178e-4        & -6.397521224e-5        & 1.773313386e-4        &  2.900864763e-4        &  2.131768984e-4        &  1.442896595e-5\\
18&  ABABCBCAC     	 & -5.574071464e-4        & -6.383937598e-5        & 1.774593657e-4        &  2.901991824e-4        &  2.132425885e-4        &  1.442906869e-5\\
19&  ABABACABC             & -9.291373243e-4        & -1.068123280e-4        & 2.953805980e-4        &  4.833254002e-4        &  3.552062442e-4        &  2.404812685e-5\\
20&  ABABCBABC             & -9.291027654e-4        & -1.066768168e-4        & 2.955083939e-4        &  4.834379566e-4        &  3.552718505e-4        &  2.404822958e-5\\
21&  ABABCACBC             & -9.291024568e-4        & -1.066764916e-4        & 2.955086248e-4        &  4.834381056e-4        &  3.552719341e-4        &  2.404822959e-5\\
22&  ABACBCABC             & -1.300847973e-3        & -1.496527692e-4        & 4.134300799e-4        &  6.765635216e-4        &  4.972351413e-4        &  3.366727759e-5\\
23&  ABABABABAC           & -3.344406108e-4        & -3.830322317e-5        & 1.064752074e-4        &  1.741194999e-4        &  1.279455471e-4        &    8.657443938e-6\\
24&  ABABABACAC           & -3.344400568e-4        & -3.830263900e-5        & 1.064756222e-4        &  1.741197676e-4        &  1.279456974e-4        &    8.657443958e-6\\
25&  ABABABACBC           & -6.689561241e-4        & -7.685332484e-5        & 2.127203740e-4        &  3.480356232e-4        &  2.557725701e-4        &  1.731469472e-5\\
26&  ABABABCBAC           & -6.689252934e-4        & -7.673165551e-5        & 2.128351843e-4        &  3.481367913e-4        &  2.558315411e-4        &  1.731478718e-5\\
27&  ABABACACBC     	  & -6.689247409e-4        & -7.673107215e-5        & 2.128355983e-4        &  3.481370585e-4        &  2.558316912e-4        &  1.731478720e-5\\
28&  ABABACABAC           & -6.688944643e-4        & -7.660998656e-5        & 2.129499938e-4        &  3.482379589e-4        &  2.558905119e-4        &  1.731487963e-5\\
29&  ABABACBCBC           & -6.689555716e-4        & -7.685274192e-5        & 2.127207881e-4        &  3.480358905e-4        &  2.557727202e-4        & 1.731469475e-5\\
30&  ABABCBABAC           & -6.688939118e-4        & -7.660940363e-5        & 2.129504078e-4        &  3.482382262e-4        &  2.558906619e-4        &  1.731487964e-5\\
31&  ABABABCABC           & -1.003516248e-3        & -1.155292204e-4        & 3.188498951e-4        &  5.218492734e-4        &  3.835398908e-4        &  2.597184480e-5\\
32&  ABABACBABC           & -1.003484870e-3        & -1.154069688e-4        & 3.189651187e-4        &  5.219507077e-4        &  3.835990114e-4        &   2.597193727e-5\\
33&  ABABCBCABC     	  & -1.003484593e-3        & -1.154066767e-4 	    & 3.189653261e-4        &  5.219508415e-4        &  3.835990866e-4        &  2.597193728e-5\\
34&  ABABCBACBC           & -1.003453765e-3        & -1.152850068e-4        & 3.190801364e-4        &  5.220520089e-4        &  3.836580574e-4        &  2.597202973e-5\\
35&  ABABCACABC           & -1.003484317e-3        & -1.154063859e-4        & 3.189655327e-4        &  5.219509750e-4        &  3.835991615e-4        &  2.597193729e-5\\
36&  ABACABACBC           & -1.003423213e-3        & -1.151636277e-4        & 3.191947401e-4        &  5.221530429e-4        &  3.837169533e-4        &  2.597212217e-5\\
37&  ABACABCABC           & -1.338027118e-3        & -1.539644410e-4        & 4.252092379e-4        &  6.958646215e-4        &  5.114247970e-4        &  3.462917155e-5\\
38&  ABACBACABC           & -1.338026842e-3        & -1.539641501e-4        & 4.252094446e-4        &  6.958647550e-4        &  5.114248719e-4        &  3.462917156e-5\\
39&  ABABABABABC         & -4.561150873e-4        & -5.245414948e-5        & 1.449843400e-4        &  2.372513040e-4        & 1.743637366e-4         &  1.180543522e-5\\
40&  ABABABABABAC       & -2.786996279e-4        & -3.191914736e-5        & 8.872933992e-5        &  1.450996370e-4        &  1.066213193e-4        &  7.214537187e-6\\ 
\hline
\end{tabular}
  \caption{Difference in cohesive energies relative to the hcp structure for different Barlow structures and various $(n,m)$-LJ potentials at fixed $\gamma_i=\sqrt{\frac{2}{3}}$. Notice that $s=n/2$. For the hcp structure the cohesive energy is shown in bold (in reduced units). For the corresponding energy differences at the optimized $\gamma_i$ values see supporting information.}
\label{tab:barlowEcohenerg}
\end{table}

\begin{table}[htp]
    \centering
        \setlength{\tabcolsep}{2pt}
    \fontsize{7.5}{12}\selectfont
\begin{tabular}{|r|l|r|r|r|r|r|r|r|r|r|r|r|r|} 
 \hline
  & 	   & \multicolumn{2}{c}{(6,4)} & \multicolumn{2}{c}{(8,4)}  & \multicolumn{2}{c}{(12,4)} & \multicolumn{2}{c}{(12,6)} & \multicolumn{2}{c}{(30,6)} & \multicolumn{2}{c}{(30,20)} \\ \hline
No. & Stacking & $\Delta \gamma$ & $\Delta E_\text{coh}$ & $\Delta \gamma$ & $\Delta E_\text{coh}$& $\Delta \gamma$ & $\Delta E_\text{coh}$& $\Delta \gamma$ & $\Delta E_\text{coh}$& $\Delta \gamma$ & $\Delta E_\text{coh}$ & $\Delta \gamma$ & $\Delta E_\text{coh}$\\

 \hline
1	 & AB 	        & 1.53893	& -2.12134	& 0.32733	& -0.08054	& -0.43794	& -0.15266	& -1.14928	& -0.72723	& -0.42157	& -0.20603	& -0.01541	& -0.00071 \\
2	 & ABC 	        & 0.00000	& 0.00000	& 0.00000	& 0.00000	& 0.00000	& 0.00000	& 0.00000	& 0.00000	& 0.00000	& 0.00000	& 0.00000	& 0.00000 \\
3	 & ABAC		    & 0.77058	& -0.53124	& 0.16460	& -0.02036	& -0.21812	& -0.03788	& -0.57317	& -0.18101	& -0.21044	& -0.05135	& -0.00771	& -0.00018 \\
4	 & ABABC		& 0.61658	& -0.34004	& 0.13135	& -0.01296	& -0.17480	& -0.02433	& -0.45888	& -0.11604	& -0.16846	& -0.03291	& -0.00617	& -0.00011 \\
5	 & ABABAC		& 1.02692	& -0.94384	& 0.21886	& -0.03600	& -0.29138	& -0.06759	& -0.76511	& -0.32247	& -0.28080	& -0.09143	& -0.01028	& -0.00032 \\
6	 & ABACBC		& 0.51394	& -0.23622	& 0.10975	& -0.00905	& -0.14540	& -0.01683	& -0.38201	& -0.08043	& -0.14028	& -0.02282	& -0.00514	& -0.00008 \\
7	 & ABABABC	    & 0.88040	& -0.69356	& 0.18737	& -0.02638	& -0.24997	& -0.04975	& -0.65601	& -0.23709	& -0.24076	& -0.06721	& -0.00881	& -0.00023 \\
8	 & ABABCBC	    & 0.88043	& -0.69362	& 0.18761	& -0.02645	& -0.24974	& -0.04966	& -0.65571	& -0.23688	& -0.24068	& -0.06717	& -0.00881	& -0.00023 \\
9	 & ABACABC	    & 0.44058	& -0.17357	& 0.09407	& -0.00665	& -0.12463	& -0.01237	& -0.32742	& -0.05909	& -0.12024	& -0.01677	& -0.00440	& -0.00006 \\
10	 & ABABABAC	    & 1.15500	& -1.19421	& 0.24598	& -0.04547	& -0.32801	& -0.08565	& -0.86112	& -0.40842	& -0.31599	& -0.11577	& -0.01156	& -0.00040 \\
11	 & ABABACAC	    & 1.15500	& -1.19421	& 0.24599	& -0.04547	& -0.32801	& -0.08565	& -0.86111	& -0.40842	& -0.31599	& -0.11577	& -0.01156	& -0.00040 \\
12	 & ABABACBC	    & 0.77052	& -0.53116	& 0.16417	& -0.02025	& -0.21851	& -0.03802	& -0.57368	& -0.18134	& -0.21058	& -0.05142	& -0.00771	& -0.00018 \\
13	 & ABABCBAC	    & 0.77055	& -0.53120	& 0.16438	& -0.02030	& -0.21832	& -0.03795	& -0.57343	& -0.18118	& -0.21051	& -0.05139	& -0.00771	& -0.00018 \\
14	 & ABABCABC	    & 0.38551	& -0.13288	& 0.08210	& -0.00506	& -0.10924	& -0.00950	& -0.28673	& -0.04532	& -0.10528	& -0.01285	& -0.00385	& -0.00005 \\
15	 & ABACBABC	    & 0.38554	& -0.13290	& 0.08232	& -0.00509	& -0.10905	& -0.00947	& -0.28647	& -0.04524	& -0.10521	& -0.01284	& -0.00385	& -0.00005 \\
16	 & ABABABABC	& 1.02686	& -0.94374	& 0.21848	& -0.03587	& -0.29173	& -0.06775	& -0.76557	& -0.32285	& -0.28093	& -0.09151	& -0.01028	& -0.00032 \\
17	 & ABABABCBC	& 1.02689	& -0.94379	& 0.21867	& -0.03593	& -0.29155	& -0.06767	& -0.76534	& -0.32266	& -0.28087	& -0.09147	& -0.01028	& -0.00032 \\
18	 & ABABCBCAC	& 1.02692	& -0.94384	& 0.21886	& -0.03600	& -0.29138	& -0.06759	& -0.76511	& -0.32246	& -0.28080	& -0.09143	& -0.01028	& -0.00032 \\
19	 & ABABACABC	& 0.68500	& -0.41974	& 0.14594	& -0.01600	& -0.19423	& -0.03004	& -0.50990	& -0.14327	& -0.18718	& -0.04063	& -0.00685	& -0.00014 \\
20	 & ABABCBABC 	& 0.68503	& -0.41978	& 0.14613	& -0.01604	& -0.19405	& -0.02998	& -0.50967	& -0.14314	& -0.18712	& -0.04060	& -0.00685	& -0.00014 \\
21	 & ABABCACBC 	& 0.68503	& -0.41978	& 0.14613	& -0.01604	& -0.19405	& -0.02998	& -0.50967	& -0.14314	& -0.18712	& -0.04060	& -0.00685	& -0.00014 \\
22	 & ABACBCABC	& 0.34273	& -0.10502	& 0.07317	& -0.00402	& -0.09693	& -0.00748	& -0.25463	& -0.03574	& -0.09351	& -0.01014	& -0.00342	& -0.00004 \\
23	 & ABABABABAC   & 1.23183	& -1.35852	& 0.26226	& -0.05169	& -0.35000	& -0.09751	& -0.91873	& -0.46487	& -0.33710	& -0.13176	& -0.01233	& -0.00046 \\
24	 & ABABABACAC   & 1.23183	& -1.35852	& 0.26226	& -0.05169	& -0.35000	& -0.09751	& -0.91873	& -0.46487	& -0.33710	& -0.13176	& -0.01233	& -0.00046 \\
25	 & ABABABACBC   & 0.92436	& -0.76461	& 0.19681	& -0.02911	& -0.26239	& -0.05481	& -0.68873	& -0.26133	& -0.25277	& -0.07409	& -0.00925	& -0.00026 \\
26	 & ABABABCBAC   & 0.92438	& -0.76465	& 0.19698	& -0.02916	& -0.26223	& -0.05475	& -0.68853	& -0.26117	& -0.25271	& -0.07405	& -0.00925	& -0.00026 \\
27	 & ABABACACBC   & 0.92439	& -0.76465	& 0.19699	& -0.02916	& -0.26223	& -0.05475	& -0.68853	& -0.26117	& -0.25271	& -0.07405	& -0.00925	& -0.00026 \\
28	 & ABABACABAC   & 0.92441	& -0.76469	& 0.19716	& -0.02921	& -0.26208	& -0.05468	& -0.68832	& -0.26101	& -0.25266	& -0.07402	& -0.00925	& -0.00026 \\
29	 & ABABACBCBC	& 0.92436	& -0.76461	& 0.19682	& -0.02911	& -0.26239	& -0.05481	& -0.68873	& -0.26133	& -0.25277	& -0.07409	& -0.00925	& -0.00026 \\
30	 & ABABCBABAC	& 0.92441	& -0.76469	& 0.19716	& -0.02921	& -0.26207	& -0.05468	& -0.68832	& -0.26101	& -0.25266	& -0.07402	& -0.00925	& -0.00026 \\
31	 & ABABABCABC   & 0.61655	& -0.34001	& 0.13118	& -0.01293	& -0.17496	& -0.02437	& -0.45908	& -0.11614	& -0.16851	& -0.03293	& -0.00617	& -0.00011 \\
32	 & ABABACBABC   & 0.61658	& -0.34003	& 0.13135	& -0.01296	& -0.17480	& -0.02433	& -0.45888	& -0.11604	& -0.16846	& -0.03291	& -0.00617	& -0.00011 \\
33	 & ABABCBCABC   & 0.61658	& -0.34004	& 0.13135	& -0.01296	& -0.17480	& -0.02433	& -0.45888	& -0.11604	& -0.16846	& -0.03291	& -0.00617	& -0.00011 \\
34	 & ABABCBACBC   & 0.61660	& -0.34006	& 0.13152	& -0.01300	& -0.17464	& -0.02429	& -0.45867	& -0.11593	& -0.16840	& -0.03289	& -0.00616	& -0.00011 \\
35	 & ABABCACABC   & 0.61658	& -0.34004	& 0.13135	& -0.01296	& -0.17480	& -0.02433	& -0.45888	& -0.11604	& -0.16846	& -0.03291	& -0.00617	& -0.00012 \\
36	 & ABACABACBC   & 0.61662	& -0.34009	& 0.13169	& -0.01303	& -0.17449	& -0.02424	& -0.45847	& -0.11583	& -0.16834	& -0.03286	& -0.00616	& -0.00011 \\
37	 & ABACABCABC   & 0.30847	& -0.08507	& 0.06586	& -0.00326	& -0.08723	& -0.00606	& -0.22916	& -0.02895	& -0.08416	& -0.00822	& -0.00308	& -0.00003 \\
38	 & ABACBACABC   & 0.30847	& -0.08507	& 0.06586	& -0.00326	& -0.08723	& -0.00606	& -0.22916	& -0.02895	& -0.08416	& -0.00822	& -0.00308	& -0.00003 \\
39	 & ABABABABABC	& 1.12003	& -1.12293	& 0.23828	& -0.04267	& -0.31831	& -0.08066	& -0.83530	& -0.38431	& -0.30650	& -0.10892	& -0.01121	& -0.00038 \\
40	 & ABABABABABAC & 1.28303	& -1.47393	& 0.27310	& -0.05606	& -0.36465	& -0.10585	& -0.95715	& -0.50453	& -0.35118	& -0.14299	& -0.01284	& -0.00050 \\
\hline
\end{tabular}
  \caption{Difference $\Delta \gamma=\frac{1}{N}\gamma_T^\text{opt} - \gamma_0$ values (in units of $10^{-4}$) with $\gamma_0=\sqrt{\frac{2}{3}}$ and corresponding change in the cohesive energy $\Delta E_\text{coh}=E_\text{coh}(\gamma_T)-E_\text{coh}(\gamma_0)$ (in units of $10^{-6}$) for case 2 (see main text). For case 3 see supporting information.}
\label{tab:barlowdeltagamma}
\end{table}

\bibliography{references}

\end{document}